\documentclass[
PhD, 
DIC=off,
12pt,
bibliography=totoc, 
listof = totoc, 
index= totoc, 
onehalfspacing, 
a4paper 
]
{icldt}
\usepackage{amsmath}
\usepackage{amssymb}
\usepackage{subcaption}
\usepackage{caption}
\usepackage{cite}

\usepackage{float}
\usepackage{graphicx}

\usepackage{hyperref}
\usepackage[title]{appendix}

\hypersetup{colorlinks = true, linkcolor=blue, citecolor={blue},urlcolor={blue}}
            
\makeatletter
\newcommand\appendix@numberline[1]{}
\g@addto@macro\appendix{%
  \addtocontents{toc}{
    \let\protect\numberline\protect\appendix@numberline}%
}
\makeatother

\begin{document}
\setlength{\tabcolsep}{0in}
\newcommand{\isep}{-2 pt}
\newcommand{\lsep}{-0.5cm}
\newcommand{\psep}{-0.6cm}
\renewcommand{\labelitemii}{$\circ$}

\college{Queen Mary}
\department{Physics and Astronomy}
\supervisor{Dr Karim A. Malik}
\title{Relativistic effects in the power spectrum of the
large scale structure}
\author{Rebeca Mart\'inez Carrillo}
\date{November 2021}
%
%
\declaration{I, Rebeca Mart\'inez Carrillo, confirm that the research included within this thesis is my own work or that where it has been carried out in collaboration with, or supported by others, that this is duly acknowledged below and my contribution indicated. Previously published material is also acknowledged below.
\newline
\newline
I attest that I have exercised reasonable care to ensure that the work is original, and does not to the best of my knowledge break any UK law, infringe any third party's copyright or other Intellectual Property Right, or contain any confidential material.
\newline
\newline
I accept that the College has the right to use plagiarism detection software to check the electronic version of the thesis.
\newline
\newline
I confirm that this thesis has not been previously submitted for the award of a degree by this or any other university. The copyright of this thesis rests with the author and no quotation from it or information derived from it may be published without the prior written consent of the author.
\newline
\newline
Details of collaboration and publications: 
\newline
Part of this work is done in collaboration with Karim A. Malik, Juan Carlos Hidalgo, Josue De-Santiago and Alkistis Pourtsidou. I have made a major contribution to the original research presented in this thesis. It is based on the following papers, all of which have been published or submitted for publication:
\newline
\begin{itemize}
\item 
\underline{Relativistic and non-Gaussianity contributions to the one-loop power}
\newline \underline{spectrum}
\newline
R. Martinez-Carrillo, J. De-Santiago, J. C. Hidalgo and K. A. Malik,
\newline
\textit{JCAP 2004 (2020) no.04, 028,}
arXiv: 1911.04359 [astro-ph.CO]
\newpage
\item 
\underline{Contributions from primordial non-Gaussianity and General Relativity}
\newline
\underline{to the galaxy power spectrum}\newline
R. Martinez-Carrillo, J. C. Hidalgo, K. A. Malik and A. Pourtsidou,
\newline
\textit{JCAP 2112 (2021) no.12, 025,} 
arXiv: 2107.10815 [astro-ph.CO]
\newline
\end{itemize}
\vfill
\noindent Signature: Rebeca Mart\'inez Carrillo 
\newline
Date: November 10, 2021}
\maketitle

\chapter*{Abstract}
\label{ch:abstract}
\addcontentsline{toc}{chapter}{Abstract}
\section*{}
\singlespacing

The forthcoming Stage-IV experiments aim to map the large scale structure of the Universe at high precision. The scales explored require a relativistic description, in addition to statistical tools for their analysis. In this thesis, we study the effects of adding relativistic and primordial non-Gaussianity contributions to the power spectrum. We begin by reviewing the standard cosmology, then we present the cosmological and Newtonian perturbation theory, which are necessary mathematical tools in the computation of our main results. Afterwards we present the main contributions to this thesis. First, we present solutions to the Einstein equations in the long-wavelength approximation, this allow us to obtain expressions for the relativistic density power spectrum at second and third order, these expressions also include contributions from primordial non-Gaussianity, in terms of the parameters $f_{\mathrm{NL}}$ and $g_{\mathrm{NL}}$. These results are complemented with the well known Newtonian solutions for the density contrast and are used in the computation of the total (relativistic + Newtonian) one-loop power spectrum. For completeness we also calculate the bispectrum at tree-level. We discuss the possibility of these relativistic effects being detectable with the future surveys considering different limiting values for $f_{\mathrm{NL}}$ and $g_{\mathrm{NL}}$. Subsequently, we compute the real space galaxy power spectrum, including relativistic and primordial non-Gaussianity effects. These effects come from the relativistic one-loop power spectrum terms and from factors of the non-linear bias parameter $b_{\mathrm{NL}}$. We use our modelling to assess the ability of Stage-IV surveys to constrain primordial non-Gaussianity. Finally, we show how this non-linear bias parameter can effectively renormalize diverging relativistic contributions at large scales.


\chapter*{Acknowledgements}
\label{ch:acknowledgements}
\addcontentsline{toc}{chapter}{Acknowledgements}

The work presented here would not have been possible without the help and support of many people that I wish to thank here. 

First, I would like to thank to my supervisor Karim Malik, for all his support and advice during my PhD, that as a result, made this thesis possible. 
I would also like to thank Juan Carlos Hidalgo for his hospitality during my visits to ICF-UNAM and all his help and advice during my PhD. 

I would like to give special thanks to my collaborators Josu\'e De-Santiago and Alkistis Pourtsidou for all their comments and suggestions that helped to improve my work. 

During my time in Queen Mary I have met really nice and friendly people,
I would like to particularly thank to Jorge Fuentes, for helping me during my PhD and to settle in London. To Shailee Imrith and John Ronayne for their long and sincere friendship. To Pedro Carrilho for his friendship and advice to improve my papers. To Clark Baker, Charlambos Pittordis, Viraj Sanghai and  Maritza Soto, for all the fun times. Jack Skinner and Fraser Kennedy for the very interesting conversations. 

I would like to give special thanks to Sanson Poon for his patience and keeping me company these four years, making them more enjoyable.

Thanks to my office colleagues and friends, Francesco Lovascio, Domenico Trotta, Christopher Gallagher, Sandy Zeng, Jessie Durk, Louis Coates, Paula Soares, Jesse Coburn, Alice Giroul, Eline De Weerd, Callum Boocock, Md G. Shah, Ali Barlas, Paul Hallam, John Strachan, Ahdab Althukair, Theo Anton, Kevin Chan, Pedro Fernandes, Anson Chen, George Turpin and Lin Qiao.

I would also like to thank to all the lecturers and researchers in the Astronomy Unit, particularly to Bernard Carr, Tessa Baker, Phil Bull, Juan Valiente-Kroon, Tim Clifton, David Mulryne, Chris Clarkson, Richard Nelson, Craig Agnor, Guillem Anglada-Escud\'e, Julian Adamek, Colin McNally and Tommi Tenkanen.

I would like to thank to my friends, Sonia Marrocu, Neha Narvekar and Cindy Tsai, which I meet at the beginning of my PhD and made many of my evenings so enjoyable. And also I want to thank to Pamela Robles, Andr\'es Ram\'irez and Dibya Chakraborty with whom I had very great times during my PhD time.

Finally, I would like to give massive thanks to my parents Ana Mar\'ia and Ram\'on and my sister Brenda, without their support I would not have been able to be here.

I acknowledge the support of a studentship funded by Queen Mary University of London and CONACyT grant No. 661285.


\tableofcontents

\newpage
\thispagestyle{plain} 

\null\vfill 

\textit{``There are two ways of becoming wise. One way is to travel out into the world and see as much as possible of God's creation. The other is to put down roots in one spot and to study everything that happens there in as much detail as you can. The trouble is that it's impossible to do both at the same time."}

\begin{flushright}
---Joistein Gaarder.
\end{flushright}
\vfill\vfill\vfill\vfill\vfill\vfill\null 

\clearpage 

\onehalfspacing

\chapter{Introduction}
\label{Introduction}

The current understanding of the Universe is the product of hundreds of years of scientific work. In the last decades, thanks to all the scientific and technological developments, we have been able to achieve observations with precision at percentage level, allowing us to probe theoretical predictions. A recent example of this, is the detection of the gravitational waves by the LIGO collaboration \cite{2016PhRvL.116f1102A} in 2015, a prediction from Albert Einstein's General Relativity (GR) Theory published in 1916.

The Universe as we observe it nowadays, is the result of 13.8 billion years of evolution. At this moment the best model to describe its evolution and composition, is the Cosmological Standard  model, also known as ``Lambda Cold Dark Matter" ($\Lambda$CDM). This model considers four main components in the Universe: radiation, baryonic matter, cold dark matter (CDM) and dark energy (DE), the latter one associated with the cosmological constant $\Lambda$. 

Observations have played an essential role in the establishment of this model. One of the most important sources of information has been the Cosmic Microwave Background (CMB), giving us access to the picture of a 380,000 years old universe, the oldest light that we have from the early Universe. 

Missions like the Cosmic Background Explorer (COBE) \cite{1999AIPC..476....1S}, Wilkinson Microwave Anisotropy Probe (WMAP) \cite{2013ApJS..208...19H} and more recent ones like Planck \cite{2020A&A...641A...1P} have measured and mapped the anisotropies in the temperature of the  Cosmic Microwave Background, nowadays we know these anisotropies are of the order  of $10^{-5}$ K. Among its results, we have the densities of the components in the Universe, with 69.35\% of dark energy, 25.81\% of cold dark matter and 4.83\% of baryonic matter. These values correspond to the parameter $68\%$ confidence limits of Planck 2015 presented in Ref.~\cite{2016A&A...594A..13P} for TT+lowP+lensing+ext (BAO+JLA+$H_0$).\footnote{In Cosmology, protons, neutrons and electrons are considered as baryonic matter.}\textsuperscript{,}\footnote{TT represents the temperature power spectrum, lowP is Planck polarisation data in the low-$\ell$ likelihood, lensing is CMB lensing reconstruction, ext is the external data from BAO (Baryonic Acoustic Oscillations), JLA (Joint Light-curve Analysis) and $H_0$ (Hubble constant).} It is also thanks to these observations, that we have constraints on some simple models of inflation \cite{2016A&A...594A..20P} and we think the Universe has a zero curvature \cite{2016A&A...594A..13P}. This information has been decisive in our understanding of the formation process of the large scale structure (LSS) in the Universe. 
However, despite all the theoretical and observational progress, some unanswered questions remain, for example, we have yet to explain the nature of the dark matter and dark energy (for a review, see e.g.~Refs.~\cite{2003RvMP...75..559P,2018RvMP...90d5002B,2005PhR...405..279B}).

\begin{figure}[ht]
\centering
\includegraphics[width=105mm]{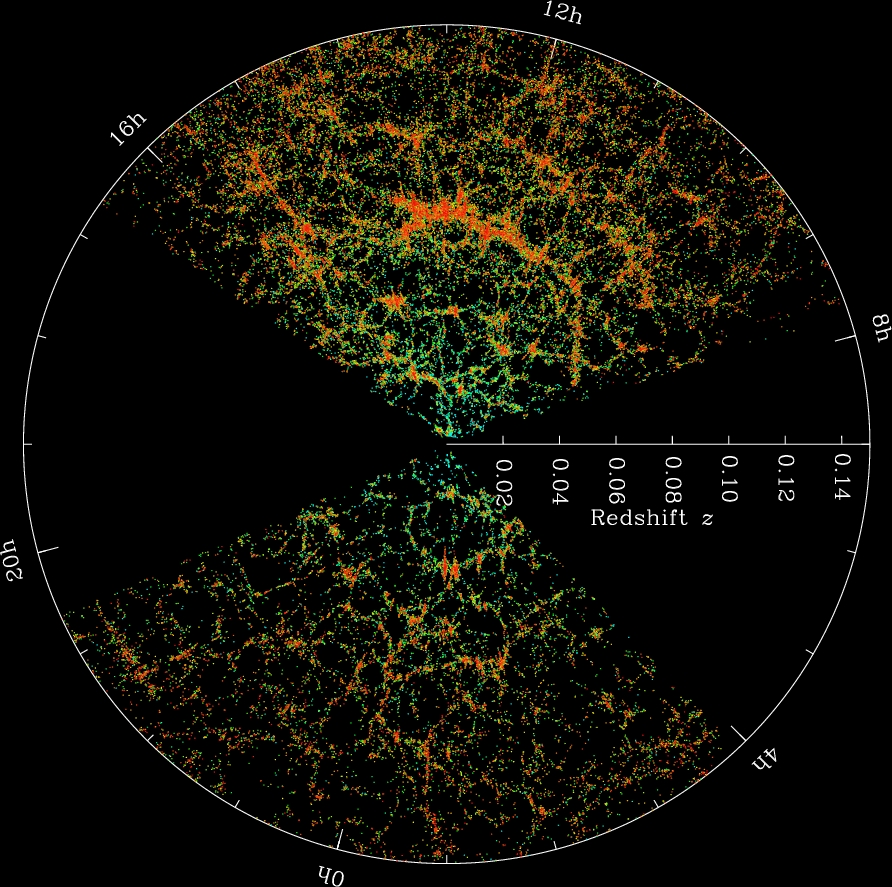}
\caption[The SDSS’s map of the Universe.]{The 2-dimensional SDSS's map of the Universe. Each dot in the image represents a galaxy, with us located at the centre. The green-red colour of the galaxies represents their actual colour. Image Credit: M. Blanton and SDSS.}
\label{sdss}
 \end{figure}

In addition to the CMB experiments, we also have surveys like the 2dF Galaxy Redshift Survey (2dfGRS) \cite{2005MNRAS.362..505C} and more recently the Sloan Digital Sky Survey (SDSS) \cite{2017AJ....154...28B}, that have mapped the distribution of galaxies. In Fig.~\ref{sdss} we show a 2-dimensional SDSS's map of the distribution of galaxies. Each dot on the image represents a galaxy, with us observing at the centre.  This map shows that galaxies do not follow a random, uncorrelated distribution, instead there is structure on the large scales \cite{2003moco.book.....D}. In this map the outer edges are at a distance of $\sim1.5$ billion lightyears. Although not shown in this map, beyond $2$ billion lightyears we do not find larger structures. The results of these surveys are an observational demonstration of the Cosmological principle \cite{2003imc..book.....L}, that states that the Universe is homogeneous and isotropic at large scales.

From a theoretical point of view, the formation of the large scale structure is modelled using perturbation theory by considering small departures from a smooth background. The Newtonian standard perturbation theory (SPT) \cite{1992PhRvD..46..585M,1994ApJ...431..495J,2002PhR...367....1B} has been a widely used description and although for many years the SPT  was sufficient to study the evolution of the matter content of the Universe, the Newtonian description of the Universe is only adequate for small scales and non-relativistic matter.

In recent years our  understanding of the evolution of the Universe has greatly benefited from observations. With the forthcoming Stage-IV experiments such as Euclid\footnote{\url{http://euclid-ec.org}} \cite{2011arXiv1110.3193L}, the Dark Energy Spectroscopic Instrument (DESI)\footnote{\url{https://www.desi.lbl.gov/}} \cite{2016arXiv161100036D} and the Vera C. Rubin Observatory’s Legacy Survey of Space and Time (LSST)\footnote{\url{https://www.lsst.org/}} \cite{2019ApJS..242....2C}, we are expecting to obtain high-precision measurements in order to improve our comprehension of the large scale structure (LSS) of the Universe, offering the possibility to give us a better understanding e.g.~of the nature of dark matter and dark energy. The progress achieved over the last few years demands a comprehensive description of observables from the full theory, a general relativistic description, in order to take advantage of the detail and scales these surveys aim to map.

One of the quantities that the community aims to constrain with the data released by the forthcoming surveys, is the primordial non-Gaussianity (PNG) of the density fluctuations, considered as seed of the large scale structure and generated during inflation.

The inflationary epoch is crucial
for early structure formation and the later evolution of the LSS \cite{2005JCAP...10..010B,2015JCAP...10..024D}, which makes constraining the PNG such an exciting prospect (see e.g.~Ref.~\cite{2010CQGra..27l4011D} for a review), as this offers the possibility to probe different inflationary scenarios (see also Refs.~\cite{2004PhR...402..103B,2012MNRAS.422.2854G} for prospects of detection).

Most of the cosmological information is encoded in the 2-point correlation function or its Fourier space equivalent, the
power spectrum \cite{2008cmbg.book.....D,2003moco.book.....D,1986ApJ...304...15B}, which, among other contributions, includes  those of primordial non-Gaussianity. The possibility to constrain PNG through the power spectrum has been reported e.g.~in Refs.~\cite{2008ApJ...684L...1C,2008JCAP...08..031S,2011PhRvD..84h3509H,2008PhRvD..78l3534T,2009MNRAS.396...85D,2013MNRAS.428.1116R,2015PhRvD..91d3506F,2017PhRvD..95l3513D,2018MNRAS.478.1341K,2019JCAP...09..010C,2020JCAP...11..052K,2020JCAP...12..031B,2021arXiv210706887B,2021arXiv210208315P}. Another statistical quantity widely used in the study of PNG is the bispectrum (the three-point correlation function in Fourier space), see e.g.~Refs.~\cite{2009ApJ...703.1230J,2015JCAP...07..004T,2016JCAP...06..014T,2017JCAP...03..006D,2018JCAP...07..050K,2021JCAP...04..013M}. These works show that the late-time statistics contain crucial information of the physics prevalent in the early universe, e.g.~the bispectrum is sensitive to the non-Gaussianity parameters $f_{\mathrm{NL}}$ and $g_{\mathrm{NL}}$, thus the bispectrum can be used in the study of PNG models and estimation of these parameters, which in turn can be used to test inflation models.

One of the main strategies to estimate primordial non-Gaussianity effects in the current matter distribution is to analyse the corresponding galaxy power spectrum, given by the correlator of the galaxy density contrast. This galaxy density contrast is related to the density contrast of underlying dark matter distribution through a set of bias parameters, which depend e.g.~on the kind of galaxies considered. In the past years galaxy bias has been included mostly in Newtonian descriptions of the LSS \cite{2008PhRvD..77l3514D,2010PhRvD..81f3530G,2015JCAP...09..029A,2015JCAP...12..043A,2020PhRvD.102j3530E,2020JCAP...08..007D}. An extensive review, presenting some of the main studies of large scale galaxy bias can be found in Ref.~\cite{2018PhR...733....1D}.

The ever-increasing precision in the measurement of the power spectrum that we have seen over the last decades (see e.g.~Refs~\cite{1994ARA&A..32..319W,1995Sci...268..829S,2002PhRvD..66j3508T,2004ApJ...606..702T,2020A&A...641A...1P}), where a wide range of different cosmological probes have been used to infer the power spectrum allows us to consider that, in addition to primordial non-Gaussianity contributions, general relativistic effects may also be observable at the large scales of the evolved matter distribution. This is due to the non-linear nature of the theory \cite{2014ApJ...794L..11B,2016PDU....13...30B}.

In addition, relativistic effects in the galaxy clustering have been studied for example in Refs.~\cite{2009PhRvD..80h3514Y,2010PhRvD..82h3508Y,2011PhRvD..84d3516C,2011PhRvD..84f3505B,2012PhRvD..85b3504J,2014PhRvD..90l3507Y,2014JCAP...09..037B,2014JCAP...11..013B,2014PhRvD..90b3513Y,2014JCAP...12..017D,2015PhRvD..91d3507C,2016JCAP...09..046Y,2017JCAP...03..010T,2017JCAP...03..034U,2017JCAP...09..040J,2018JCAP...09..037F,2019arXiv190808400F,2019MNRAS.486L.101C,2019arXiv190501293M,2020JCAP...03..065M,2018PhRvD..97b3531B,2020arXiv201215326F}. Furthermore, N-body simulations using General Relativity (see e.g.~Refs.~\cite{2016JCAP...07..053A,2016PhRvL.116y1302B, 2020JCAP...01..007B}) are also studying these relativistic effects in the LSS from a numerical point of view.

Galaxy bias in a relativistic context has been previously explored including primordial non-Gaussianities, e.g.~in Refs.~\cite{2011JCAP...10..031B,2011JCAP...04..011B}, and the most suitable gauge to define the bias in, namely the comoving-synchronous gauge, has been discussed in Refs.~\cite{2012PhRvD..85d1301B,2015CQGra..32q5019B}.

A derivation of the local bias at second order in cosmological perturbation theory, for Gaussian and non-Gaussian initial conditions is studied in Refs.~\cite{2019JCAP...05..020U,2019JCAP...12..048U}, where it is argued that general relativistic effects affect local clustering predominantly through the distortion of the volume element of the local patch, and that modulations of the short wavelength modes are possible only through primordial non-Gaussianity.

At one-loop order under the weak field approximation, relativistic corrections have been included for the power spectrum and bispectrum at intermediate scales in Ref.~\cite{2019JCAP...07..030C}, with results extended to obtain the galaxy power spectrum using a Lagrangian bias expansion in General Relativity in Ref.~\cite{2019arXiv191213034C}.

More recently, methods to avoid the infrared divergences of relativistic contributions, in the galaxy power spectrum, have been proposed in Refs.~\cite{2020JCAP...11..064G,2021arXiv210608857C}.\footnote{The integrals in $q$ for the power spectrum which have limits of $[0,\infty]$, have integrands which depend on terms like $1/q^n$, this integration diverges in the infrared, i.e.~for the lower limit of integration.}

Our aim is to move a step forward towards precision cosmology by studying the relativistic effects in the statistics of the large scale structure. Our focus is on the power spectrum, and to this end we structured this thesis as follows:

Chapter~\ref{Introduction} introduces the Standard Cosmology, and its mathematical formalism, the basic framework of the description of our work. In addition in this chapter we also present how cosmological distances are measured in Cosmology and how the Cosmic Microwave Background is measured and analysed to constrain cosmological parameters.

In chapter~\ref{Perturbation-Theory} we present a brief review of the perturbation theory formalism, the standard perturbation theory will be key to study the non-linear regime, while the cosmological perturbation theory is essential to describe the large scales. In this chapter we also introduce the statistical quantities that are the main subject of study of this thesis, the power spectrum and bispectrum, as well as a brief introduction to primordial-non Gaussianity.

In chapter~\ref{chapter:Relativistic} we compute the one-loop density power spectrum including Newtonian and relativistic contributions, as well as the primordial non-Gaussianity contributions from $f_{\mathrm{NL}}$ and $g_{\mathrm{NL}}$ in the local configuration. To this end we take solutions to the Einstein equations in the long-wavelength approximation and provide expressions for the matter density perturbation at second and third order. These solutions have shown to be complementary to the usual Newtonian cosmological perturbations. For completeness, we present the matter bispectrum at the tree-level including the mentioned contributions.

In chapter~\ref{paper2} we compute the real space galaxy power spectrum, including the leading order effects of General Relativity and primordial non-Gaussianity from the $f_{\mathrm{NL}}$ and $g_{\mathrm{NL}}$ parameters. Such contributions come from the one-loop matter power spectrum terms dominant at large scales, computed in chapter~\ref{chapter:Relativistic}, and from the factors of the non-linear bias parameter $b_{\mathrm{NL}}$. We use our modelling to assess the ability of Stage-IV surveys to constrain primordial non-Gaussianity. In addition, we show how this non-linear bias parameter can effectively renormalize diverging relativistic contributions at large scales.

Finally in chapter~\ref{Conclusions} we present our conclusions and possible future work.

\section{Notation}

In this thesis we use natural units where the convention for the gravitational constant $G$ and the speed of light $c$ is $G=c=1$.

When Greek indices ($\mu$, $\nu$,...) are used, these range from 0 to 3, lower case Latin indices, like $i$, $j$, and $k$, have range 1, 2, 3.

Vectors are denoted in two different ways. Using the tensor notation e.g.~$v^i$ and 
using bold letters e.g.~$\mathbf{v}$. 

Derivatives with respect to the cosmic time $t$ are denoted by a dot
\begin{equation}
\frac{d\phi}{dt}=\dot\phi~.
\end{equation}

Derivatives with respect to conformal time $\eta$ are denoted by a prime 
\begin{equation}
\frac{d\phi}{d\eta}=\phi'~.
\end{equation}

Partial derivatives are denoted by the following
\begin{equation}
\frac{\partial\phi}{\partial x^i}=\partial_i\phi=\phi_{,i}~.
\end{equation}

Covariant derivatives are denoted by a semi-colon e.g.
\begin{equation}
\nabla_\mu u^\nu=\frac{\partial u^\nu}{\partial x^\mu}+\Gamma^{\nu}_{\mu\alpha}u^\alpha=u^\nu_{;\mu}~,
\end{equation}

\noindent where $\Gamma^{\nu}_{\mu\alpha}$ are the Christoffel symbols, that are defined in Eq.~\eqref{Christoffel}.

Unless stated otherwise, throughout this thesis we follow the following Fourier convention, the Fourier transform of a function $A(\mathbf{x})$ is defined as 
\begin{equation}
{A}(\mathbf k)=\int d^3 x A(\mathbf x)e^{-i\mathbf k\cdot\mathbf x}~,
\label{Fourierconv1}
\end{equation}
\noindent with inverse Fourier transform given by 
\begin{equation}
A(\mathbf x)=\int\frac{d^3 k}{(2\pi)^3} {A}(\mathbf k)e^{i\mathbf k\cdot\mathbf x}~.
\label{Fourierconv2}
\end{equation}

Note that we do not use special symbols for real and Fourier space quantities, instead we use arguments like ($\mathbf{x}$, $\mathbf{y}$, etc.) for real space quantities, and ($\mathbf{k}$, $\mathbf{p}$, etc.) for Fourier space quantities. From these expressions the Dirac Delta is given by 

\begin{equation}
\delta_D(\mathbf{x}-\mathbf{x'})\equiv\int\frac{d^3k}{(2\pi)^3}e^{i\mathbf{k}\cdot(\mathbf{x}-\mathbf{x'})}~,
\end{equation}

\noindent where $\mathbf{k}$ is the wavenumber vector.

Additional notation will be introduced as required.

\section{Standard Cosmology }

The Cosmological principle establishes that on large enough scales ($\sim$\,$100\,\mathrm{Mpc}$) the Universe is homogeneous and isotropic \cite{2003imc..book.....L}. An appropriate description of the standard cosmology requires the introduction of some basic concepts of General Relativity \cite{2012reco.book.....E,2009fcgr.book.....S}. Firstly, the geometry of the Universe is described by the metric tensor $g_{\mu\nu}$, defined through the line element $ds^2$ as  
\begin{equation}
ds^2=g_{\mu\nu}dx^\mu dx^\nu.
\end{equation}

In order to study the dynamics of the Universe it is necessary to define some quantities, the first one is the Riemann tensor $R_{\mu\beta\nu}^{\alpha}$, which quantifies the curvature of space-time
\begin{equation}
R_{\mu\beta\nu}^{\alpha}=\Gamma_{\mu\nu,\beta}^{\alpha}-\Gamma_{\mu\beta,\nu}^{\alpha}+\Gamma_{\lambda\beta}^{\alpha}\Gamma_{\mu\nu}^{\lambda}-\Gamma_{\lambda\nu}^{\alpha}\Gamma_{\mu\beta}^{\lambda}\,,
\end{equation}

\noindent where $\Gamma_{\beta\gamma}^{\alpha}$ are the Christoffel symbols or connection coefficients, defined by

\begin{equation}
\Gamma_{\beta\gamma}^{\alpha}=\frac{1}{2}g^{\alpha\lambda}(g_{\lambda\beta,\gamma}+g_{\lambda\gamma,\beta}-g_{\beta\gamma,\lambda}). 
\label{Christoffel}
\end{equation}
The Riemann tensor satisfies the Bianchi identities \cite{2012reco.book.....E}. The first identity is given by 
\begin{equation}
R_{\mu\beta\nu\alpha}+R_{\mu\nu\alpha\beta}+R_{\mu\alpha\beta\nu}=0,
\end{equation}
and the second identity is given as
\begin{equation}
R_{\mu\beta\nu\alpha;\gamma}+R_{\mu\beta\alpha\gamma;\nu}+R_{\mu\beta\gamma\nu;\alpha}=0.
\label{Bianchitwo}
\end{equation}

Derived from the Riemann tensor we calculate the Ricci tensor $R_{\mu\nu}$ as  

\begin{equation}
R_{\mu\nu}\equiv R^{\alpha}_{\mu\alpha\nu}\, ,
\end{equation}
by contracting the Ricci tensor with the metric tensor we obtain the Ricci scalar $R$
 
\begin{equation}
R\equiv g^{\mu\nu}R_{\mu\nu}\, .
\end{equation}

These quantities allow to define the Einstein tensor $G_{\mu\nu}$, that encodes the information of the spacetime curvature  
\begin{equation}
G_{\mu\nu}\equiv R_{\mu\nu}-\frac{1}{2}g_{\mu\nu}R\, .
\end{equation}
This tensor has the property of symmetry
\begin{equation}
G_{\mu\nu}=G_{\nu\mu}\, ,
\end{equation}
and is divergence free due to the second Bianchi identity in Eq.~\eqref{Bianchitwo}
\begin{equation}
\nabla_\nu G^{\mu\nu}=0.
\end{equation}

It is also important to define the stress-energy tensor $T_{\mu\nu}$, which describes the energy and momentum content of the spacetime. For a perfect fluid the stress-energy tensor is defined by \cite{2012reco.book.....E,2009fcgr.book.....S}
\begin{equation}
T_{\mu\nu}=(\rho+P)u_\mu u_{\nu}+Pg_{\mu\nu}\, ,
\label{perfectfluidstress}
\end{equation}
where $\rho$ is the density, $P$ is the pressure and  $u_{\mu}$  is the four-velocity that satisfies the relation
\begin{equation}
u^\mu u_\mu=-1.
\end{equation}

Finally, we define the Einstein equations as
\begin{equation}
G_{\mu\nu}=8\pi GT_{\mu\nu}\, ,
\label{GReinstein}
\end{equation}
from this equation follows that the stress-energy tensor $T_{\mu\nu}$ is also divergence free
\begin{equation}
\nabla_\nu T^{\mu\nu}=0\, ,
\label{massconservation}
\end{equation}
which means it obeys the conservation of energy and momentum.\footnote{For the perfect fluid the anisotropic stress tensor is $\Pi_{ij}=0$.}

The Einstein equations are a set of ten coupled, non-linear, second order partial differential equations \cite{2012reco.book.....E,2009fcgr.book.....S}, which relate the geometry of the spacetime with the matter content of the Universe. The Einstein equations are covariant, which means they do not depend on the gauge choice.  Due to their complexity there are only few exact solutions to these equations known. 

One of the known solutions is the Friedmann-Lema\^{i}tre-Robertson-Walker (FLRW) metric, which is also consistent with the Cosmological principle.
The line element in spherical coordinates is given by \cite{2003imc..book.....L,2008cmbg.book.....D}
\begin{equation}
ds^2=-dt^2+a(t)^2\left(\frac{dr^2}{1-kr^2}+r^2d\theta^2+r^2\sin^2\theta d\phi^2\right),
\label{frw}
\end{equation}
where $a(t)$ is the scale factor, that measures the universal expansion rate, and is related with the redshift $z$ by 
\begin{equation}
a(t)=\frac{1}{1+z},
\end{equation}
$k$ is the spatial curvature that can take values of $k<0$ for an open universe, $k>0$ for a closed universe and $k=0$ for a flat universe. Since recent observations from Planck \cite{2016A&A...594A..13P} are in agreement with a universe that has zero curvature we will only consider the flat FLRW metric $k=0$, that can also be defined by 
\begin{equation}
ds^2=-dt^2+a^2(t)\delta_{ij}dx^idx^j,
\label{metricafrwl}
\end{equation}
where $\delta_{ij}$ is the Kronecker delta. 

From the $0-0$ component of the Einstein equations for the FLRW metric, we obtain the first Friedmann equation
\begin{equation}
H^2=\left(\frac{\dot a}{a}\right)^2=\frac{8\pi G\rho}{3}+\frac{\Lambda}{3},
\end{equation}

\noindent and the second Friedmann equation is obtained from  $i-i$ component of the Einstein equations
\begin{equation}
\dot{H}+H^2=\frac{\ddot{a}}{a}=-\frac{4\pi G}{3}\left(\rho+3P\right)+\frac{\Lambda}{3},
\end{equation}

\noindent where we introduced the Hubble rate $H$ defined as
\begin{equation}
H=\frac{\dot{a}}{a},
\end{equation}
and $\Lambda$ is the cosmological constant. The present value of the Hubble rate is given by $H_0=100h\, \mathrm{km}\mathrm{s}^{-1}\mathrm{Mpc}^{-1}$, where $h$ is a dimensionless parameter and its value is obtained from observations. 

We also introduce the critical density $\rho_c$, which is defined as the density in a flat Universe, and  is given by 
\begin{equation}
\rho_c\equiv\frac{3H_0^2}{8\pi G}.
\end{equation}

From the conservation of mass given in Eq.~\eqref{massconservation} we find the fluid equation 
\begin{equation}
\dot\rho+3\frac{\dot a}{a}(\rho+P)=0.
\end{equation}

In order to find the evolution of the scale factor it is necessary to introduce a equation of state 
\begin{equation}
P=\omega \rho,
\end{equation}

\noindent where $\omega$ can take different values depending on the fluid component considered 

\begin{equation}
\omega= \left\lbrace
\begin{array}{ll}
\frac{1}{3} &\textup{ Radiation}\\
0 &\textup{Pressureless Matter}\\
-1 &\textup{Cosmological Constant ($\Lambda$).}
\end{array}
\right.
\end{equation}

\begin{figure}[htbp]
\centering
\includegraphics[width=125mm]{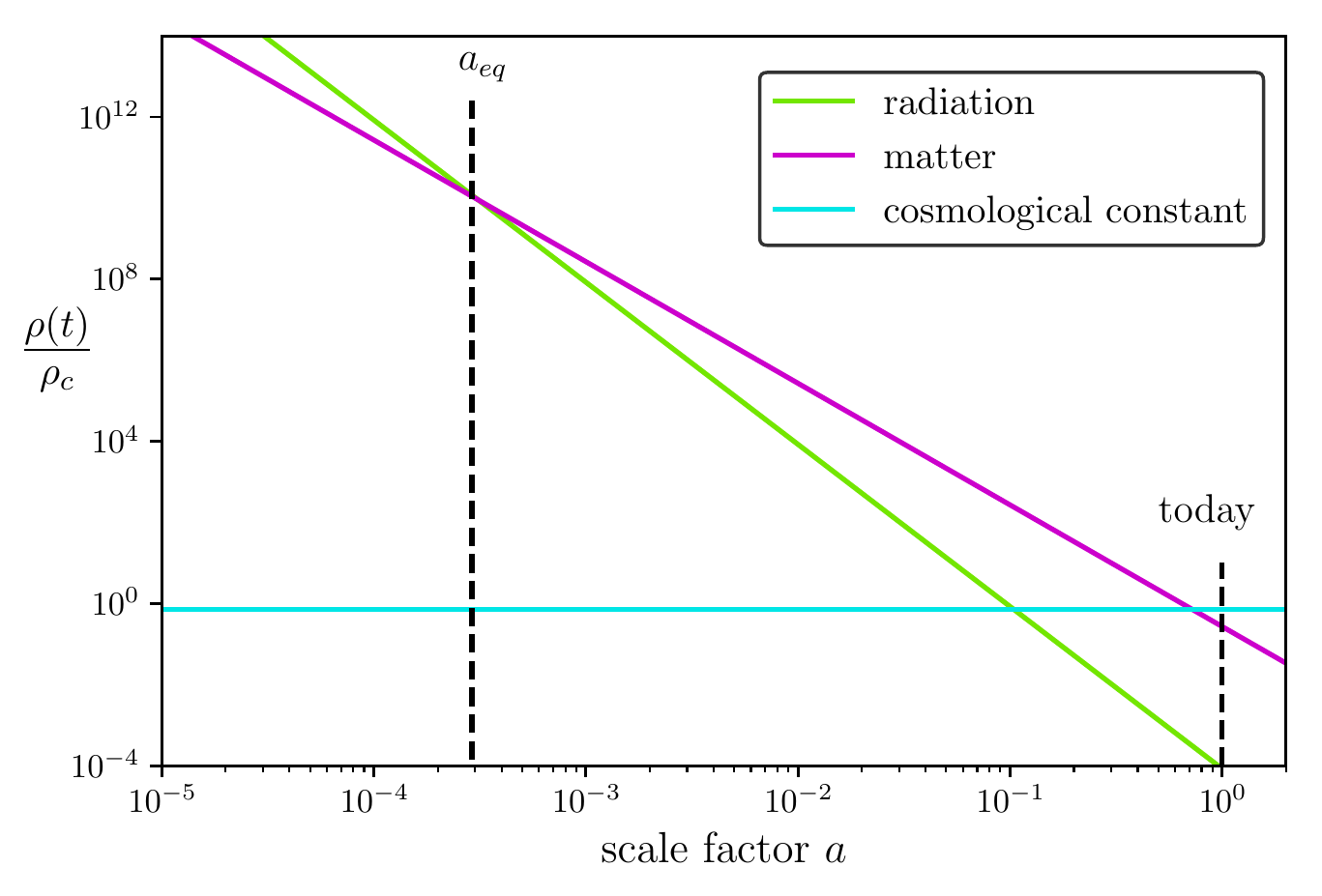}
\caption[Evolution of the different components of the Universe respect to the scale factor $a$.]{Evolution of the different components of the Universe respect to the scale factor $a$. At early times the Universe was radiation dominated ($\propto a^{-4}$). The time at which the radiation and matter content were equal is indicated with $a_{eq}$. After this the Universe was matter dominated ($\propto a^{-3}$). Nowadays the Universe is dominated by the cosmological constant.}
\label{densityevol}
 \end{figure}

If we consider each of the fluid components independently, the time evolution of the density for the different fluids in the Universe is given by \cite{2003imc..book.....L,2008cmbg.book.....D} 
\begin{equation}
\frac{\rho_i(t)}{\rho_{i0}}=\left(\frac{a_0}{a(t)}\right)^{3(1+\omega)},
\label{rhoevol}
\end{equation}

\noindent where the subscript $i$ denotes the different fluid components, throughout this chapter quantities with a subscript $0$ are quantities evaluated at their present value. These components dominate in different epochs as shown in Fig.~\ref{densityevol}. To give an idea of the evolution times, at the time were the matter and radiation were equal the Universe was $\sim50,000$ years old, after this, matter dominated for about $10$ billion years, and it was $\sim3.8$ billion years ago that the dark energy dominated epoch started. 

This definition allow us to define the density parameter $\Omega_i$, for the different components of the Universe,
\begin{equation}
\Omega_i(t)=\frac{\rho_i}{\rho_c}.
\end{equation}
In terms of the density parameter $\Omega_i$, the Friedmann equation can be rewritten as
\begin{equation}
\frac{H^2(a)}{H_0^2}=\frac{\Omega_{r}}{a^4}+\frac{\Omega_m}{a^3}+\Omega_{\Lambda}\,,
\label{Friedmanngral}
\end{equation}
where $r$, $m$ and $\Lambda$ denote radiation, matter and cosmological constant, respectively. According to the Planck results \cite{2016A&A...594A..13P} currently for the $\mathrm{\Lambda}$CDM model the matter content in the Universe is $30.65\%$ of the total, from which $4.83\%$ corresponds to baryonic matter and the other $25.81\%$ corresponds to dark matter, the remaining $69.35\%$ corresponds to cosmological constant,  usually associated to dark energy. 

The scale factor evolves as 
\begin{align}
\label{scaleevol}
\frac{a(t)}{a_0}&=\left(\frac{t}{t_0}\right)^{\frac{2}{3(\omega+1)}} \quad \mathrm{for}~\omega=\mathrm{constant}\neq-1,\\
\frac{a(t)}{a_0}&=e^{\sqrt{\frac{\Lambda}{3}}(t-t_0)} \quad \mathrm{for}~\omega=-1.
\label{scalefactorlambda}
\end{align}

In some of the chapters of this thesis we will use the conformal time denoted by the Greek letter $\eta$, the conformal time is  related with the cosmic time $t$ by \cite{2003moco.book.....D}

\begin{equation}
\eta(t)=\int_0^t\frac{dt'}{a(t')}.
\end{equation}

Using conformal time, the Friedmann equations take the form 

\begin{align}
\mathcal{H}^2=&\frac{8\pi G\rho a^2}{3}+\frac{\Lambda a^2}{3},\\
\mathcal{H}'=&-\frac{4\pi G a^2}{3}(\rho+3P)+\frac{\Lambda a^2}{3},
\end{align}
where $\mathcal{H}$ is the Hubble rate in conformal time defined as
\begin{equation}
\mathcal{H}=\frac{a'}{a}=aH.
\end{equation}

In conformal time the scale factor evolves as 
\begin{align}
\label{scaleconf}
\frac{a(\eta)}{a_0}=&\left(\frac{\eta}{\eta_0}\right)^{\frac{2}{1+3\omega}}\quad \mathrm{for}~\omega=\mathrm{constant}\neq-1,\\
\frac{a(\eta)}{a_0}=&\left(\frac{\eta}{\eta_0}\right)^{-1}\quad \mathrm{for}~\omega=-1. 
\end{align}

\subsection{Einstein-de Sitter universe}
\label{Einsteindesitter}

As an example of the solutions to the Friedmann equation we present the solution to the so called Einstein-de Sitter universe, this model considers a matter only universe, i.e. 
\begin{equation}
\Omega_m=1,\quad \Omega_r=\Omega_\Lambda=0.
\end{equation}
Using Eq.~\eqref{rhoevol} with $\omega=0$, the density evolves as 
\begin{equation}
\frac{\rho_m(t)}{\rho_{m0}}=\left(\frac{a_0}{a(t)}\right)^3,
\end{equation}
the scale factor in  Eq.~\eqref{scaleevol} in cosmic time evolves as 
\begin{equation}
\frac{a(t)}{a_0}=\left(\frac{t}{t_0}\right)^{\frac{2}{3}},
\label{matterscalefactor}
\end{equation}
and in conformal time, the scale factor in Eq.~\eqref{scaleconf} evolves as 

\begin{equation}
\frac{a(\eta)}{a_0}=\left(\frac{\eta}{\eta_0}\right)^{\frac{1}{2}}.
\end{equation}

\subsection{Matter + cosmological constant universe}

If we consider a more realistic case, a flat universe with matter and cosmological constant, i.e.~
\begin{equation}
\Omega_m+\Omega_{\Lambda}=1, \quad \Omega_r=0,
\end{equation}
the Friedmann Eq.~\eqref{Friedmanngral} takes the form 

\begin{equation}
\frac{H^2(t)}{H_0^2}=\frac{\Omega_m}{a^3}+\Omega_{\Lambda},
\end{equation}

\noindent which can be rearranged as 
\begin{equation}
\left(\frac{da}{dt}\right)^2=H_0^2\left[\Omega_m a^{-1}+\Omega_{\Lambda}a^2\right],
\end{equation}

\noindent then 

\begin{equation}
H_{0}\int{dt}=\int\frac{1}{\sqrt{\Omega_m}}\frac{a^{1/2}da}{\sqrt{1+(\Omega_{\Lambda}/\Omega_m)a^3}},
\end{equation}

\noindent introducing a change of variable of the form $u^2=\Omega_{\Lambda}/\Omega_m a^3$,
the integral takes the form,

\begin{equation}
H_0\int dt=\int\frac{2/3}{\sqrt{\Omega_\Lambda}}\frac{du}{\sqrt{1+u^2}}.
\end{equation}

After integration, we found 
\begin{equation}
H_{0}t=\frac{2/3}{\sqrt{\Omega_\Lambda}}\sinh^{-1}{(u)}=\frac{2/3}{\sqrt{\Omega_\Lambda}}\sinh^{-1}{\left(\sqrt{\frac{\Omega_\Lambda}{\Omega_m}}a^{3/2}\right)},
\end{equation}
then, the scale factor for a universe with matter and cosmological constant evolves as \cite{2008cosm.book.....W}
\begin{equation}
a(t)=\left(\frac{\Omega_m}{\Omega_{\Lambda}}\right)^{1/3}\sinh^{2/3}{\left(\frac{3\sqrt{\Omega_\Lambda}H_{0}}{2}t\right)}.
\end{equation}

If we consider this solution at early times, i.e.~for small $t$, we can use the fact that $\sinh(x)\approx x$ for small $x$

\begin{equation}
a(t)\approx\left(\frac{3}{2}\sqrt{\Omega_m}H_0t\right)^{2/3},
\end{equation}

\noindent which recovers the solution for a matter dominated universe, given in Eq.~\eqref{matterscalefactor}. 

On the other hand for late times, i.e.~large $t$, we can use that $\sinh(x)\rightarrow e^x/2$, then the scale factor is
\begin{equation}
a(t)\approx \exp\left({\sqrt{\Omega_\Lambda}H_0t}\right),
\end{equation}
which recovers the solution for a universe dominated by a cosmological constant given by Eq.~\eqref{scalefactorlambda}.

\section{Distances in Cosmology}
Our Universe is expanding, which means that at the moment of measuring distances in the sky, we need take into account this expansion and this is not always straightforward. In this section we present different methods used to obtain cosmological distances. 

\subsection{Comoving distance}
If we could stop the expansion of the Universe for an instant, the distance between two objects that we could  measure would be a proper distance, however this is not possible. A more useful distance is a comoving distance that takes into account the expansion in the Universe. Proper coordinates $\mathbf{x}_p$ are related to comoving coordinates $\mathbf{x}$ by 
\begin{equation}
\mathbf{x}_p=a(t)\mathbf{x}.
\end{equation}

To calculate the comoving distance $r(a)$ travelled by a light ray emitted by an object and received by an observer, we can make use of the fact that light rays obey the condition
\begin{equation}
ds=0,
\end{equation}
and if we consider a flat universe, then from the metric given by Eq.~\eqref{frw} we obtain a relation between comoving coordinates and time
\begin{equation}
cdt=\pm a(t)dr.
\end{equation}

\noindent Using the fact that $da/dt=aH$ and after some rearranging we obtain an expression for the comoving distance $r(a)$
\begin{equation}
r(a)=\int_0^rdr'=-\int_1^a\frac{cda'}{(a')^2H(a)},
\end{equation}

\noindent where $H(a)$ is defined in Eq.~\eqref{Friedmanngral}. The comoving distance is fixed for any given redshift, however it is a quantity that we can not measure directly. 

\subsection{Luminosity distance}
The luminosity distance $d_L$ is the distance that an object appears to have, assuming that the reduction of light intensity with distance follows an inverse square law \cite{2003imc..book.....L}. The luminosity distance is given by the relation
\begin{equation}
d_L=\sqrt{\frac{L}{4\pi f}},
\end{equation}

\noindent where $f$ is the measured flux of a luminous object, and $L$ is the luminosity of the object.

The luminosity distance requires the knowledge of not only the flux, but the luminosity, and it is important to note that there are objects for which we know their luminosity through the study of other properties, these objects are called standard candles. Cepheid variable stars and the Type Ia supernovae are types of standard candles, and are used to determine cosmological parameters like the Hubble constant (see e.g.~Ref.~\cite{2016inco.book.....R} for a pedagogical explanation).  

The luminosity distance can be also related to the comoving distance by 
\begin{equation}
d_L(z)=(1+z)r(z).
\end{equation}

\subsection{Angular diameter distance}
The angular diameter distance $d_A$ is the distance that an object of known physical size $l$  appears to be at \cite{2003imc..book.....L}. The angular diameter distance  is given by the relation
\begin{equation}
d_A\equiv\frac{l}{\sin{\theta}}\simeq\frac{l}{{\theta}},
\label{diameterdistance}
\end{equation}

\noindent where $\theta$ is the angular distance, and the small-angle approximation $\sin{\theta}\approx\theta$, was used since it is valid for most astronomical objects.

The relation between the angular diameter and the comoving distance is given by
\begin{equation}
d_A(z)=\frac{r(z)}{(1+z)}, 
\end{equation}
and the angular diameter distance is related to the luminosity distance by
\begin{equation}
d_A(z)=\frac{d_L}{(1+z)^2}.
\end{equation}

\section{Cosmic Microwave Background}

As we briefly mentioned the study of the Cosmic Microwave Background has been crucial in our understanding  of the formation process of the large scale structure. In this section we discuss it in more detail.

The early Universe was filled by a hot and dense plasma of nuclei, electrons and photons. As the Universe expanded and cooled down, the conditions in the Universe became adequate for the electrons and nuclei to combine and form the first neutral hydrogen atoms, this is called recombination and occurred at a redshift of $z=1100$, when the Universe was $\sim 380,000$ years old. At this point photons stopped being scattered by electrons, i.e.~matter and radiation decoupled, which meant that photons were able to travel freely through space. This is the last scattering surface and these photons have been travelling since then, and we are able to observe them today. This remnant radiation from the early Universe is what we call the Cosmic Microwave Background. 
 
The Cosmic Microwave Background was first discovered by Penzias and Wilson in 1965 \cite{1965ApJ...142..419P}, and it is one of the most important sources of information for cosmologists, since this is the oldest light that we receive from the early Universe, the CMB contains relevant information for our understanding of the structure formation. 

\subsection{The angular power spectrum}

\begin{figure}[htbp]
\centering
\includegraphics[width=130mm]{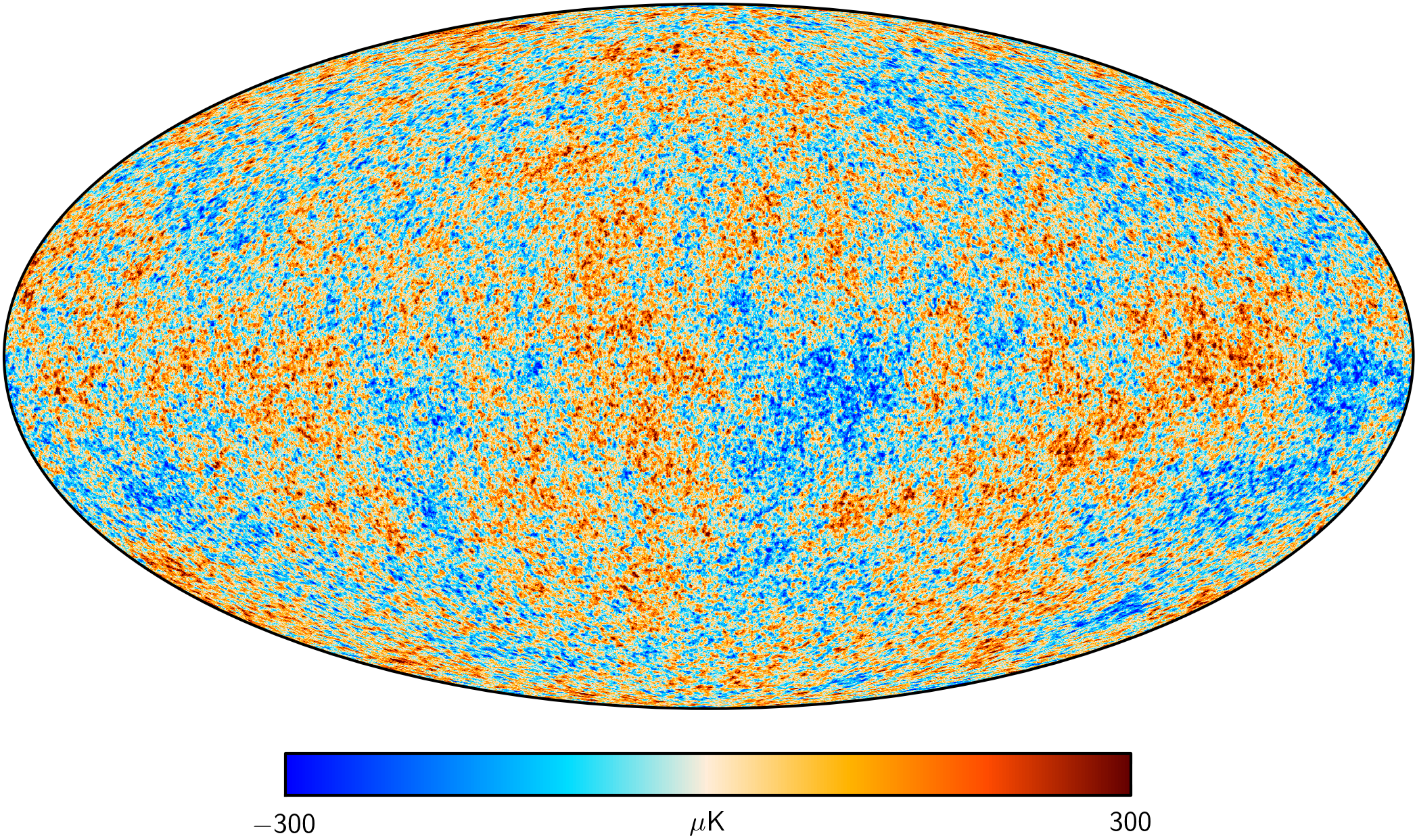}
\caption[Mollweide map of the temperature anisotropies of the CMB, as seen by the Planck satellite.]{Mollweide map of the temperature anisotropies of the CMB, as seen by the Planck satellite. Image Credit: Planck Collaboration.}
\label{CMB-map}
\end{figure}

Observations have measured that the Cosmic Microwave Background has a black-body mean temperature of $T=2.725$ K that is very uniform across the sky, except for tiny fluctuations in the temperature, the so called anisotropies, that are of the size $\sim 10^{-5}$ K. A map of these temperature fluctuations is shown in  Fig.~\ref{CMB-map}. In order to obtain information from the anisotropies, we require to study their statistics. 

First, we define the dimensionless temperature fluctuation at a given direction in the sky $\mathbf{\hat n}\equiv(\theta,\phi)$ as 
\begin{equation}
\Theta(\mathbf{\hat n})=\frac{T(\mathbf{\hat n})-\langle T \rangle}{\langle T \rangle}.
\end{equation}
The temperature fluctuations are projected in a 2D spherical surface sky, spherical harmonics are used to describe them. The spherical harmonics $Y_{\ell m}(\theta,\phi)$ form an orthonormal basis and are defined by 
\begin{equation}
Y_{\ell m}(\theta,\phi)=\sqrt{\frac{2\ell+1}{4\pi}\frac{(\ell-m)!}{(\ell+m)!}}P_\ell^m(\cos{\theta})e^{im\phi},
\end{equation}
\noindent  where $P^m_\ell$ are the Legendre polynomials, where the indices  are $\ell=0,...\infty$ and $-\ell\leq m\leq \ell$. Thus, we expand the temperature in terms of the spherical harmonics as
\begin{equation}
\Theta(\mathbf{\hat n})=\sum_{\ell=0}^{\ell=\infty}\sum_{m=-\ell}^{\ell}a_{\ell m}Y_{\ell m}(\mathbf{\hat n}),
\end{equation}

\noindent $\ell$ is the multipole and encodes the angular information with a characteristic scale of $\Delta\theta\simeq\pi/\ell$. The spherical harmonic coefficients  $a_{\ell m}$ can be calculated by
\begin{equation}
a_{\ell m}=\int_{\theta=-\pi}^{\pi}\int_{\phi=0}^{2\pi}\Theta(\mathbf{\hat n})Y_{\ell m}^{*}(\mathbf{\hat n})d\Omega,
\end{equation}

\noindent and give information of the size of the irregularities on different scales \cite{2003imc..book.....L}.

\begin{figure}[htbp]
\centering
\includegraphics[width=130mm]{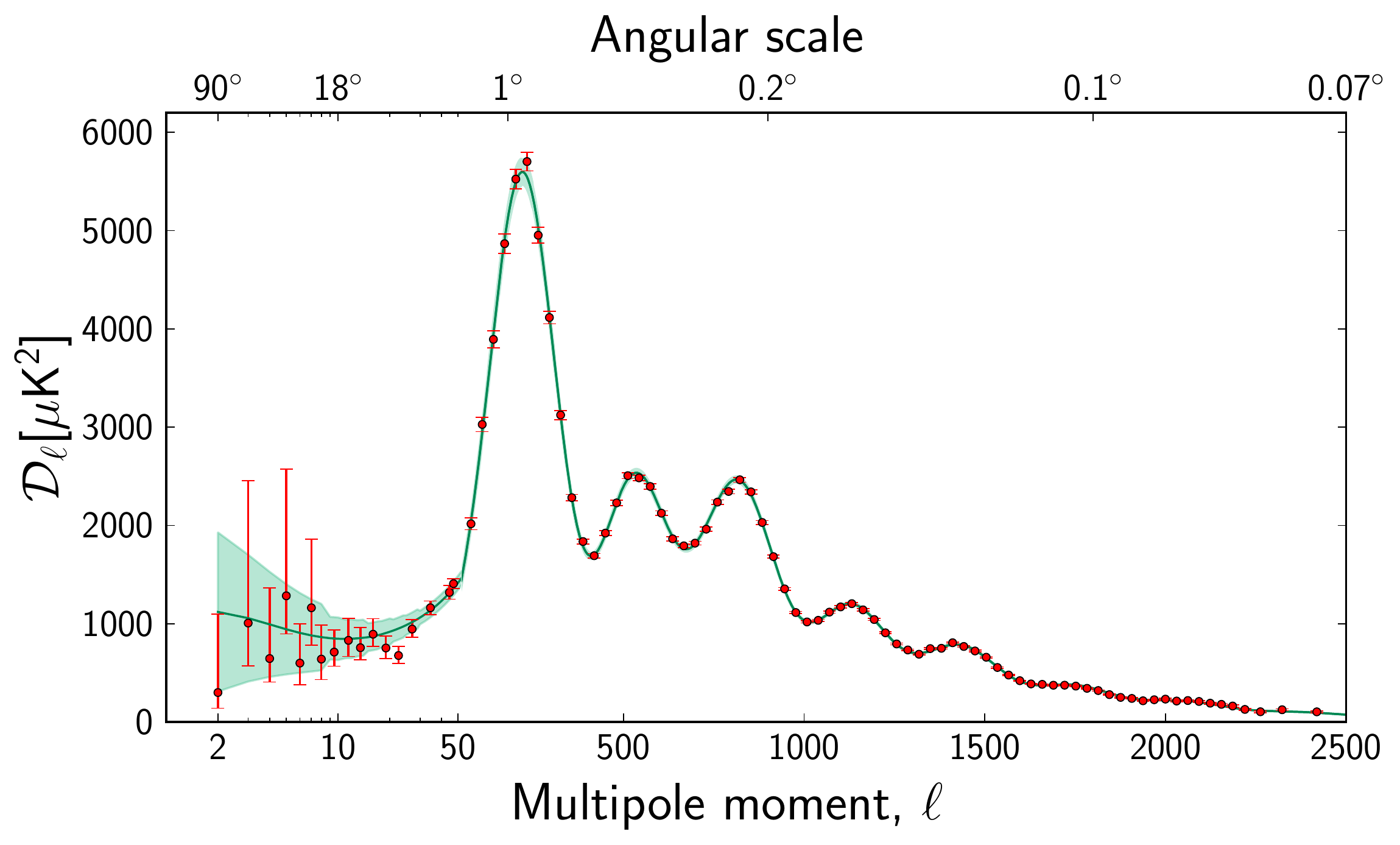}
\caption[Temperature angular power spectrum of the CMB from Planck.]{Temperature angular power spectrum of the CMB from Planck. The vertical scale is $\mathcal{D}_\ell=\ell(\ell+1)C_\ell/2\pi$. Image Credit: Planck Collaboration \cite{2014A&A...571A...1P}.}
\label{CMB-power}
 \end{figure}

The angular power spectrum $C_\ell$ of the temperature fluctuations is the variance of the spherical harmonics coefficients
\begin{equation}
\langle a_{\ell m} a_{\ell' m'}^{*}\rangle=\delta_{\ell \ell'}\delta_{mm'}C_{\ell},
\end{equation}
where the average is taken over many ensembles. The fact that the power spectrum $C_\ell$ is a function of the multipole $\ell$ only is a consequence of the isotropy of the Universe \cite{2008cmbg.book.....D,2013arXiv1302.4640L}. We have a limited number of independent $m$-modes, since we have only one Universe. There are only $(2\ell+1)$ $m$-modes for each multipole $\ell$, hence the angular power spectrum can be written as an average of the variance of the $m$-modes as
\begin{equation}
C_{\ell}=\frac{1}{2\ell+1}\sum_{m=-\ell}^{\ell}\langle|a_{\ell m}|^2\rangle,
\end{equation}

\noindent and in Fig.~\ref{CMB-power} the temperature angular power spectrum of the CMB as measured by Planck is shown. 

\subsection{Cosmological parameters}

The CMB angular power spectrum shape depends on different cosmological parameters \cite{2003moco.book.....D}. We now briefly describe how the CMB power spectrum is used to constrain cosmological parameters.\footnote{All the values for the cosmological parameters presented in this section correspond to the parameter $68\%$ confidence limits of Planck 2015 presented in Ref.~\cite{2016A&A...594A..13P} for TT+lowP+lensing+ext (BAO+JLA+$H_0$).}

First, if we change the amplitude of primordial fluctuations $\mathcal{A}_s$, this will shift the CMB power spectrum up and down, in other words, it is a factor that multiplies all $C_\ell$ \cite{Baumann, 2003moco.book.....D}. The value for the amplitude of primordial fluctuations according to Planck \cite{2016A&A...594A..13P} is $\ln({10^{10}\mathcal{A}_s})=3.064\pm0.024$.

Changing the spectral index $n_s$, tilts the primordial scalar power spectrum. If $n_s=1$, this means that the power spectrum $\ell(\ell+1)C_\ell$ has the same power on all scales, having $n_s<1$ means more power on large scales, whereas $n_s>1$ means there is more power on small scales \cite{Baumann}. The value for the spectral index according to Planck \cite{2016A&A...594A..13P} is  $n_s=0.9681\pm0.0044$.

The optical depth $\tau$ is a parameter that measures the probability that a photon from the last scattering surface scattered from a free electron travelling to us \cite{2003imc..book.....L}. The Universe at low redshifts is highly ionised. Ionised electrons are able to scatter photons and as a result of this, the pattern of the CMB anisotropies gets smooth out. The fact that we observe anisotropies down to small scales in the CMB, means that the optical depth is $\tau<1$, a value of $\tau=0$ implies that reionisation did not occur, and a value of $\tau>1$ means that each CMB photon would be scattered many times, losing all the information of its direction \cite{2016inco.book.....R}. The value for the optical depth according to Planck \cite{2016A&A...594A..13P} is $\tau=0.067\pm0.013$.

In the CMB power spectrum (see Fig.~\ref{CMB-power}) we observe several peaks, these are the acoustic peaks. The first peak in the Cosmic Microwave Background power spectrum at about $1^\circ$, corresponds to the sound horizon $r_s$, which is the distance sound waves travelled before recombination \cite{2002astro.ph..9504R}. The angular size of the temperature fluctuations in the CMB is related to the sound horizon $r_s$ by 
\begin{equation}
\Delta\theta=\frac{r_s}{d_A(z_{dec})},
\end{equation}
where $d_A$ is the angular diameter distance (see Eq.\eqref{diameterdistance}) and $z_{dec}$ is the redshift of decoupling. 
The size of the sound horizon can be determined using properties of the photon-baryon fluid, the angular diameter distance depends strongly on the geometry (curvature) of the Universe \cite{2008cmbg.book.....D,2008arXiv0802.3688H}, thus determining $d_A$ can give us information of the curvature of the Universe. The curvature is given by the parameter $\Omega_k$. The current observations from Planck are consistent with a flat universe. 

The second peak in the CMB power spectrum is related to the baryon density $\Omega_b$. If the baryon-photon ratio changes, the sound horizon $r_s$ is changed, this will not only cause a shift in the peak locations, but a modification of the ratio of the heights of odd to even peaks \cite{2003moco.book.....D,2016ASSP...45....3W}. The baryon density according to Planck \cite{2016A&A...594A..13P} is given by  $\Omega_bh^2=0.02227\pm0.00020$.

Finally, the third peak in the CMB power spectrum is related to the cold dark matter density $\Omega_c$. If the matter-radiation ratio changes, this will have an effect on the amplitude of the acoustic peaks. Also, the decay of the gravitational potential (that is dominated by the cold dark matter) in the radiation era, enhances the third and higher peaks \cite{2008arXiv0802.3688H,2016ASSP...45....3W}. According to Planck \cite{2016A&A...594A..13P} the cold dark matter density is $\Omega_ch^2=0.1184\pm0.0012$.
 
This chapter is the basic framework of Cosmology that will help us to describe the evolution of the Universe and we will use it as a background in the description of the Cosmological Perturbation Theory in the next chapter. 
\chapter{Perturbation Theory}
\label{Perturbation-Theory}
Perturbation theory is a widely used tool in the study of non-linear problems in physics. As part of our study of the structures in the Universe, in this chapter we review the perturbation theory techniques that we will use in chapter~\ref{chapter:Relativistic} for our main calculations. In section~\ref{CPTsection}, we introduce the  Cosmological Perturbation Theory, this relativistic formalism is necessary for the description of the large scale structure. In section~\ref{section:SPT}, we present the Newtonian perturbation theory, this formalism is the adequate to describe the small scales. Finally, in section~\ref{section:statistics} we present the statistical quantities that we use in the analysis of our results. 

\section{Cosmological Perturbation Theory}
\label{CPTsection}
In this section, we review the main points of Cosmological Perturbation Theory. 

We begin by briefly reviewing some of the pioneering work on the topic. Lifshitz \cite{2017GReGr..49...18L,1963AdPhy..12..185L} and Tomita \cite{1967PThPh..37..831T} first studied the evolution of density perturbations at first and second order respectively. In Ref.~\cite{1980PhRvD..22.1882B} Bardeen tackled the gauge problem, by defining the first gauge invariant quantities, known as ``Bardeen potentials".  Relevant reviews on Cosmological Perturbation Theory are the ones by Kodama and Sasaki \cite{1984PThPS..78....1K} and 
Mukhanov et al \cite{1992PhR...215..203M}. Other important works are cited throughout the chapter. However, a complete review on the work in cosmological perturbation theory is beyond the scope of this thesis, a  more recent and extensive review of this topic can be found in Ref.~\cite{2009PhR...475....1M}, which we will also follow in the description of the following sections.

As discussed in chapter~\ref{Introduction}, our Universe at large scales is considered to be homogeneous and isotropic. However, on smaller scales the Universe is inhomogeneous and anisotropic. In order to study the inhomogeneous universe it is possible to use N-body simulations (see e.g.~Refs.~\cite{2016JCAP...07..053A,2016PhRvL.116y1302B, 2020JCAP...01..007B} ), which is one of the most complete approaches, however the amount of resources and time required may not be always accessible. As an alternative, we can use an analytical approach like perturbation theory.

The key idea of perturbation theory is to split quantities into a homogeneous background and inhomogeneous perturbations 
\begin{equation}
\mathbf{T}(x^i,\eta)= \mathbf{T}_0(\eta)+\delta\mathbf{T}(x^i,\eta).
\end{equation}

The homogeneous background can be described with the usual FLRW solutions that we presented in chapter~\ref{Introduction}, the background quantities are denoted with the subscript ${(0)}$, the first or linear order perturbations, with the subscript ${(1)}$, etc. The inhomogeneous universe is studied through perturbations around this homogeneous background, the perturbation terms can be expressed as a power series expansion of the form 
\begin{equation}
\delta\mathbf{T}(x^i,\eta)=\sum_{n=1}^\infty\frac{\epsilon^n}{n!}\delta\mathbf{T}_n(x^i,\eta),
\end{equation}
\noindent where $n$ gives the order of the perturbation, and $\epsilon$ is a small parameter of the expansion. This allow us to set up equations at different orders, which can be individually solved, order by order. 

Our starting point is the perturbed metric,
\begin{equation}
g_{\mu\nu}=g^{(0)}_{\mu\nu}+\delta g_{\mu\nu},
\end{equation}

\noindent where we consider as a background a flat FLRW metric, already defined in Eq.~\eqref{metricafrwl}. 

The perturbed metric is defined as 
\begin{equation}
\delta g_{\mu\nu}=a^2\begin{bmatrix}
-2\phi&\hat\omega_i\\
\hat\omega_i&\gamma_{ij}
\end{bmatrix},
\end{equation}

\noindent with 
\begin{equation}
\gamma_{ij}=-2\psi\delta_{ij}+\hat\chi_{ij},
\end{equation}
and the perturbed line element is thus \cite{1995ApJ...455....7M,1997CQGra..14.2585B}
\begin{equation}
ds^2=a^2(\eta)\{-(1+2\phi)d\eta^2+2\hat\omega_id\eta dx^i+[(1-2\psi)\delta_{ij}+\hat\chi_{ij}]dx^idx^j\},
\label{perturbedlinelement}
\end{equation}

\noindent where $\phi$ is the lapse perturbation, $\hat\omega_i$ is the shift perturbation and $\psi$ and $\hat\chi_{ij}$ are the spatial metric perturbations.

By means of the  Helmholtz decomposition theorem (see e.g.~Ref.~\cite{1990CQGra...7.1169S}) these components of the metric can be decomposed into their irreducible form of scalars, vectors and tensors. This decomposition is useful at linear order as it allow us to have decoupled governing equations, that can be solved separately.

The lapse perturbation $\phi$ and the spatial metric perturbation $\psi$ are scalars. In the case of vectors these can be decomposed into a scalar and a vector part. This decomposition of a vector consists of a curl-free part (also known as longitudinal part), and divergence-free part (also known as solenoidal part). The shift perturbation $\hat\omega_{i}$ can be decomposed in its scalar and vector components as
\begin{equation}
\hat\omega_{i}=\omega_{,i}-\omega_i,
\end{equation}

\noindent where $\omega_i$ is the solenoidal part, i.e.~$\partial^i\omega_{i}=0$.

For tensors we can also decompose into scalar, vector and tensor components. 
The spatial metric perturbation $\hat\chi_{ij}$ is traceless i.e.,~$\chi_i^i=0$ and transverse i.e.,~$\partial^i\chi_{ij}=0$. Its scalar, vector and tensor components are given by 
\begin{equation}
\hat\chi_{ij}=D_{ij}\chi+\chi_{i,j}+\chi_{j,i}+\chi_{ij},
\end{equation}
\noindent where, $\chi_i$ is a solenoidal vector field and $D_{ij}$ is defined by
\begin{equation}
D_{ij}=\partial_i\partial_j-\frac{1}{3}\delta_{ij}\nabla^2.
\end{equation}

The perturbed metric allow us to find the left hand side of the Einstein equations as defined in Eq.~\eqref{GReinstein}. The perturbations for the perfect fluid stress energy momentum tensor defined in Eq.~\eqref{perfectfluidstress} can be found by taking the density $\rho$, pressure $p$ and fourth-velocity $u_\mu$ perturbations and substituting them in Eq.~\eqref{perfectfluidstress}, these are given by 
\begin{equation}
\delta T_{\mu}^{\nu}=\begin{bmatrix}
-\delta\rho&(\rho_0+p_0)(v_i-\omega_i)\\
-(\rho_0+p_0)v_i&\delta p\delta^i_j
\end{bmatrix},
\end{equation}

\noindent where the quantities $\rho_0$ and $p_0$ denote the density and pressure in the background respectively. And we have defined $\delta\rho=\rho-\rho_{0}$ and $\delta p=p-p_{0}$. This allow us to write the right hand side of the Einstein equations.

\subsection{Gauge problem and gauge transformations}
Using General Relativity to study physical problems should be covariant, i.e.~it should not depend on the choice of coordinates. After splitting quantities into background and perturbations, the perturbations are no longer covariant, and depend on the gauge choice, this is the so called gauge problem. The splitting of background and perturbation quantities, introduces non physical gauge modes, meaning that we will obtain different results depending on the coordinate choice  \cite{2009PhR...475....1M}.  

However, it is possible to overcome this problem by studying the gauge transformations of the perturbations and defining gauge invariant quantities. An appropriate gauge choice can also simplify the  equations of our problem. 

It is important to establish a method to relate quantities in different gauges. In the literature (see e.g.~Ref.~\cite{2009PhR...475....1M}) we find two methods, the active and the passive approach. The passive approach specifies the relation between two coordinate systems, then the change in the perturbations is calculated using this coordinate transformation. The transformation in the passive approach is taken at the same physical point. On the other hand, in the active approach the perturbations are changed under a mapping, this map induces the transformation of the perturbed quantities. In the active approach the transformation is evaluated at the same coordinate point. Since both approaches have the same results, we use the active approach in the following.

The active approach relates a tensor $\widetilde{\mathbf{T}}$ with a tensor $\mathbf{T}$ using a exponential map
\begin{equation}
\widetilde{\mathbf{T}}=e^{\pounds_{\xi^\mu}}\mathbf{T},
\label{activetransformation}
\end{equation}
where $\xi^\mu$ is the gauge generator, and the   ${\pounds_{\xi^\mu}}$ is the Lie derivative with respect to $\xi^\mu$, defined as follows for scalars, vectors and tensors respectively,
\begin{align}
\pounds_{\xi^\mu}\rho&=\rho_{,\mu}\xi^{\mu}\, ,\\
\pounds_{\xi^\mu}v_{\mu}&=v_{\mu,\lambda}\xi^{\lambda}+v_{\lambda}\xi^{\lambda}_{,\mu}\, ,\\
\pounds_{\xi^{\mu}}T_{\mu\nu}&=T_{\mu\nu,\lambda}\xi^{\lambda}+T_{\mu\lambda}\xi_{,\nu}{^\lambda}+T_{\lambda\nu}\xi_{,\nu}^{\lambda}\, .
\label{lietensor}
\end{align}
Using the expanded gauge generator $\xi^{\mu}$ up to second order,
\begin{equation}
\xi^\mu\equiv\epsilon\xi_1+\frac{1}{2}\epsilon^2\xi_2^\mu+\mathcal{O}(\epsilon^3),
\end{equation}
then the exponential map expanded up to second order takes the form 
\begin{equation}
\mathrm{exp}(\pounds_\xi)=1+\epsilon\pounds_{\xi_1}+\frac{1}{2}\epsilon^2\pounds^2_{\xi_1}+\frac{1}{2}\epsilon^2\pounds_{\xi_2}+...
\end{equation}
and splitting the expansion by order, the tensor $\mathbf{T}$ transforms as 
\begin{align}
\widetilde{\mathbf{T}_0}=&\mathbf{T}_0,\\
\epsilon\widetilde{\delta\mathbf{T}_1}=&\epsilon{\delta\mathbf{T}_1}+\epsilon{\pounds}_{\xi_1}\mathbf{T}_0,\\
\epsilon^2\widetilde{\delta\mathbf{T}_2}=&\epsilon^2(\delta\mathbf{T}_2+\pounds_{\xi_2}\mathbf{T}_0+\pounds^2_{\xi_1}\mathbf{T}_0+2\pounds_{\xi_1}\delta\mathbf{T}_1).
\end{align}

Using the active approach given in Eq.~\eqref{activetransformation}, we can write the transformation metric perturbation at first order as 
\begin{equation}
\widetilde{\delta g_{\mu\nu}}^{(1)}=\delta g_{\mu\nu,\lambda}^{(0)}\xi^{\lambda}_1+g_{\mu\lambda}^{(0)}\xi_{1,\nu}+g_{\lambda\nu}^{(0)}\xi_{1,\mu}^{\lambda}\, .
\end{equation}

If we use the Eq.~\eqref{activetransformation} for a coordinate $x^{\mu}$ of a point $q$, this will transform for a point $p$ as
\begin{equation}
x^\mu(q)=e^{\xi^\lambda\frac{\partial}{\partial x^\lambda}|_p}x^\mu(p),
\end{equation}
expanding the terms up to second order, e.g. 
\begin{equation}
x^\mu(q)=x^\mu(p)+\epsilon\xi^\mu_1(p)+\frac{1}{2}\epsilon^2(\xi^\mu_{1,\nu}(p)+\xi_2^\mu(p)).
\end{equation}

\noindent At linear order the gauge transformation for the coordinates $\widetilde{\eta}$ and $\widetilde{x^i}$ is given by 
\begin{align}
\widetilde{\eta}=&\eta+\alpha\, ,\\
\widetilde{x^i}=&x^i+\partial^i\beta+d^i\, ,
\end{align}
where we used $\xi^\mu=(\alpha,\partial^i\beta+d^i)$ with $\partial_i d^i=0$.

Given the transformation rules, the perturbations at linear order take the form of \cite{1995ApJ...455....7M,1997CQGra..14.2585B}
\begin{align}
\widetilde{\phi}=&\phi+\mathcal{H}\alpha+\alpha'\, ,\\
\widetilde{\omega_i}=&\omega_i-\alpha_{,i}+\beta_{,i}'+d{_i}'\, ,
\\\widetilde{\gamma_{ij}}=&-2\left(\psi-\frac{1}{3}\nabla^2\beta-\mathcal{H}\alpha\right)\delta_{ij}\\
+&\chi_{ij}+2D_{ij}\beta+d_{i,j}+d_{j,i}\, .
\end{align}

Using the Helmholtz decomposition theorem we find that scalar metric variables transform as \cite{2016JCAP...01..030V,Cosper}
\begin{align}
\widetilde{\phi}=&\phi+\mathcal{H}\alpha+\alpha',\\
\widetilde{\omega}=&\omega-\alpha+\beta',\\
\widetilde{\psi}=&\psi-\frac{1}{3}\nabla^2\beta-\mathcal{H}\alpha,\\
\widetilde{\chi}=&\chi+\beta,
\end{align}
the vector metric variables transform as 
\begin{align}
\widetilde{\chi}_i&=\chi_i+d_i,\\
\widetilde{\omega}_i&=\omega_i+{d_i}',
\end{align}
and the tensor metric
\begin{equation}
\widetilde{\chi}_{ij}=\chi_{ij}.
\end{equation}

The stress energy momentum tensor also transforms using Eqs.~\eqref{activetransformation} and \eqref{lietensor}, while the temporal part of velocity is transformed as 
\begin{equation}
\widetilde{v}^{0}=v^{0}-\mathcal{H}\alpha-\alpha',
\end{equation}
the scalar part of the spatial velocity transforms as 
\begin{equation}
\widetilde{v}^i=v^i-{\beta^i}',
\end{equation}
and the vector part transforms as
\begin{equation}
\widetilde{v}^i=v^i-{d^i}'.
\end{equation}

The density perturbation, transforms as
\begin{equation}
\widetilde{\delta\rho}=\delta\rho+\rho_0'\alpha\, .
\end{equation}

An example of gauge invariant quantities are the Bardeen's potentials \cite{1980PhRvD..22.1882B} defined as
\begin{align}
\label{Bardeenpot}
\Phi&\equiv\phi+\mathcal{H}(\omega-\chi')+(\omega-\chi')',\\
\Psi&\equiv\psi-\mathcal{H}(\omega-\chi').
\end{align}

In this thesis we are only interested in scalar perturbations as these are responsible for the density perturbations which are the seeds for structure formation. At linear order only, the vector and tensor perturbations are decoupled from the scalar density perturbations and the latter ones can be treated independently, at higher order this is no the case anymore. The vector and tensor perturbations are used to describe other effects like vorticity, magnetic fields and gravitational waves respectively, see e.g.~Refs.~\cite{2011PhRvD..83l3512C,2014JCAP...09..023N,2001astro.ph..1009B,2003gr.qc.....3004M}.

\subsection{Gauge choice}
\label{Gaugechoice}
In this thesis we use two different gauge conditions, the synchronous gauge condition that is obtained when the conditions
\begin{equation}
\phi=\hat\omega_i=0,
\label{condition1}
\end{equation}
are imposed on the metric in Eq.~\eqref{perturbedlinelement}, and the comoving gauge given by the condition
\begin{equation}
v^i=0.
\label{condition2}
\end{equation}

These choices specify the gauge completely, and ensure the absence of spurious gauge modes \cite{2009PhR...475....1M}. The evolution equations resulting from these gauge choices are presented in chapter~\ref{chapter:Relativistic}.

\section{Newtonian standard perturbation theory}
\label{section:SPT}

In this section we review the key aspects of the Newtonian standard perturbation theory (SPT), following some of the pioneering work of Refs.~\cite{1992PhRvD..46..585M,1994ApJ...431..495J,1986ApJ...311....6G,1981MNRAS.197..931J,1983MNRAS.203..345V}.
An extensive review on SPT can be found in Ref.~\cite{2002PhR...367....1B}, we follow this work in this section.

The SPT is an adequate description for small scales, scales that are smaller than the Hubble radius \cite{2002PhR...367....1B}. The starting point of this description considers collisionless particles, interacting only gravitationally. The evolution equations in comoving coordinates are given by the continuity equation, which describes the conservation of mass
\begin{equation}
\delta'+\nabla\cdot[(1+\delta)\mathbf{v}]=0,
\label{continuityeq}
\end{equation}
the Euler equation, which describes the  conservation of momentum  
\begin{equation}
\mathbf{v}'+(\mathbf{v}\cdot\nabla)\mathbf{v}=-\mathcal{H}\mathbf{v}-\nabla\phi,
\label{eulereq}
\end{equation}
and the Poisson equation 
\begin{equation}
\nabla^2\phi=4\pi G a^2\bar\rho\delta,
\label{poissoneq}
\end{equation}

\noindent where $\bar\rho$ is the mean density in the background,  $\delta(\mathbf{x},\eta)$ is the density contrast given by 
\begin{equation}
\delta(\mathbf{x},\eta)=\frac{\rho(\mathbf{x},\eta)-\bar{\rho}(\eta)}{\bar\rho(\eta)}.
\end{equation}
 Here, $\mathbf{v}(\mathbf{x},\eta)=\frac{\partial \mathbf{x}}{\partial\eta}$ is the peculiar velocity and $\phi$ is the peculiar gravitational potential field sourced by the density fluctuations, which can be identified with the metric perturbation in the conformal Newtonian gauge \cite{2009PhR...475....1M}.

We can describe the peculiar velocity $\mathbf{v}(\mathbf{x},\eta)$ by its divergence $\theta=\nabla\cdot\mathbf{v}$, the vorticity can be neglected as it will decay with the expansion of the Universe \cite{1992PhRvD..46..585M}. 

Using $\theta$ as our new velocity variable, we can write in Fourier space (see Eqs.~\eqref{Fourierconv1}, \eqref{Fourierconv2} for convention) the Eqs. \eqref{continuityeq}, \eqref{eulereq} and \eqref{poissoneq} as 
\begin{align}
 \frac{\partial \delta(\mathbf{k},\eta)}{\partial \eta}&+\theta(\mathbf{k},\eta)=-\int\frac{ d^3\mathbf{k}_1d^3\mathbf{k}_2}{(2\pi)^3}\delta_D(\mathbf{k}-\mathbf{k}_{12})\frac{\mathbf{k}_{12}\cdot\mathbf{k}_1}{k_1^2}\theta(\mathbf{k}_1,\eta)\delta(\mathbf{k}_2,\eta),
 \label{deltaink}\\
\nonumber
\frac{\partial\theta(\mathbf{k},\eta)}{\partial \eta}&+\mathcal{H}(\eta)\theta(\mathbf{k},\eta)+\frac{3}{2}\Omega_m\mathcal{H}^2(\eta)\delta(\mathbf{k},\eta)=\\
&-\int\frac{d^3\mathbf{k}_1 d^3\mathbf{k}_2}{(2\pi)^3}\delta_D(\mathbf{k}-\mathbf{k}_{12})\frac{k_{12}^2(\mathbf{k}_1\cdot\mathbf{k}_2)}{2k_1^{2}k_2^{2}}\theta(\mathbf{k}_1,\eta)\theta(\mathbf{k}_2,\eta),
\label{thetaink}
\end{align}

\noindent where we used the short-hand notation $\mathbf{k}_{12}=\mathbf{k}_1+\mathbf{k}_2$, this notation will be used throughout this thesis where convenient. The right hand side of Eqs.~\eqref{deltaink} and \eqref{thetaink} describes the non-linear evolution of $\delta(\mathbf{k},\eta)$ and $\theta({\mathbf{k},\eta})$, which is specified by the coupling of the linear wavevectors $\mathbf{k_1}$ and $\mathbf{k_2}$.

\subsection{Linear solutions}
First we will review the linear solutions for $\delta(\mathbf{k},\eta)$. If we keep only the linear terms, the Eqs.~\eqref{deltaink} and \eqref{thetaink} take the form 
\begin{align}
 \frac{\partial \delta(\mathbf{k},\eta)}{\partial \eta} +&\theta(\mathbf{k},\eta)=0,
 \label{lineardelta}\\
\frac{\partial\theta(\mathbf{k},\eta)}{\partial \eta}+&\mathcal{H}(\eta)\theta(\mathbf{k},\eta)+\frac{3}{2}\Omega_m\mathcal{H}^2(\eta)\delta(\mathbf{k},\eta)=0.
\label{lineartheta}
\end{align}

Taking the derivative of Eq.~\eqref{lineardelta} and using Eq.~\eqref{lineartheta}, we can write 
\begin{equation}
\frac{\partial^2\delta(\mathbf{k},\eta)}{\partial\eta^2}+\mathcal{H}(\eta)\frac{\partial\delta(\mathbf{k},\eta)}{\partial\eta}-\frac{3}{2}\Omega_m\mathcal{H}^2(\eta)\delta(\mathbf{k},\eta)=0.
\label{secondorderdelta}
\end{equation}

\noindent This equation can be solved with an ansatz of the form $\delta(\mathbf{k},\eta)\propto D(\eta)\delta(\mathbf{k})$, where $D(\eta)$ is the linear growth factor that describes the growth of matter perturbations at late times \cite{2003moco.book.....D}. The solution for Eq.~\eqref{secondorderdelta} is composed by a linear combination of a growing mode $(+)$ and a decaying mode $(-)$
\begin{equation}
\delta(\mathbf{k},\eta)={D}_{+}(\eta)A(\mathbf{k})+{D}_{-}(\eta)B(\mathbf{k}).
\end{equation}

For example, for an Einstein-de Sitter universe (see section~\ref{Einsteindesitter}), the growing and decaying modes are given by 
\begin{equation}
{D}_{+}=\eta^{2},\quad {D}_{-}=\eta^{-3}.
\end{equation}

 In further calculations we will keep the growing mode solution only, as the decaying mode vanishes at late times.

\subsection{Non-linear solutions}

The Eqs.~\eqref{deltaink} and \eqref{thetaink} describe the non-linear behaviour of $\delta(\mathbf{k},\eta)$ and $\theta(\mathbf{k},\eta)$. Since these equations are coupled, it is difficult to give an exact solution. However, a formal solution can be written using a perturbation expansion. For simplicity we will only consider the solutions for an Einstein de-Sitter universe. We expand $\delta(\mathbf{k},\eta)$ and $\theta(\mathbf{k},\eta)$ as \cite{1994ApJ...431..495J,1986ApJ...311....6G}
\begin{equation}
\delta(\mathbf{k},\eta)=\sum_{n=1}^{\infty}a^{(n)}(\eta)\delta^{(n)}(\mathbf{k}),\quad \theta(\mathbf{k},\eta)=-\mathcal{H}(\eta)\sum_{n=1}^{\infty}a^{(n)}(\eta)\theta^{(n)}(\mathbf{k}),
\end{equation}

\noindent where 
\begin{equation}
\delta^{(n)}(\mathbf{k})=\int\frac{d^3q_1...d^3q_n}{(2\pi)^{3n}}(2\pi)^3\delta_D\left(\sum\mathbf{q}_i-\mathbf{k}\right)F^{(n)}(\{\mathbf{q}_i\})\delta^{(1)}(\mathbf{q}_1)...\delta^{(1)}(\mathbf{q}_n),
\end{equation}
\begin{equation}
\theta^{(n)}(\mathbf{k})=\int\frac{d^3q_1...d^3q_n}{(2\pi)^{3n}}(2\pi)^3\delta_D\left(\sum\mathbf{q}_i-\mathbf{k}\right)G^{(n)}\left(\{\mathbf{q}_i\}\right)\delta^{(1)}(\mathbf{q}_1)...\delta^{(1)}(\mathbf{q}_n),
\end{equation}

\noindent with the recursion relations $F^{(n)}$ and $G^{(n)}$ given by 
\begin{align}
\nonumber
F^{(n)}(\mathbf{q}_1,...,\mathbf{q}_n)=&\sum_{m=1}^{n-1}\frac{G^{(m)}(\mathbf{q}_1,...,\mathbf{q}_m)}{(2n+3)(n-1)}\left[(1+2n)\frac{\mathbf{k}\cdot\mathbf{k}_1}{k_1^2}F^{(n-1)}(\mathbf{q}_{m+1},...,\mathbf{q}_n)\right.\\
+&\left.\frac{k^2(\mathbf{k}_1\cdot\mathbf{k}_2)}{k_1^2k_2^2}G^{(n-m)}(\mathbf{q}_{m+1},...,\mathbf{q}_n)\right],
\end{align}
\begin{align}
\nonumber
G^{(n)}(\mathbf{q}_1,...,\mathbf{q}_n)=&\sum_{m=1}^{n-1}\frac{G^{(m)}(\mathbf{q}_1,...,\mathbf{q}_m)}{(2n+3)(n-1)}\left[3\frac{\mathbf{k}\cdot\mathbf{k}_1}{k_1^2}F^{(n-m)}(\mathbf{q}_{m+1},...,\mathbf{q}_n)\right.\\
+&\left.n\frac{k^2(\mathbf{k}_1\cdot\mathbf{k}_2)}{k_1^2k_2^2}G^{(n-m)}(\mathbf{q}_{m+1},...,\mathbf{q}_n)\right],
\end{align}

\noindent where $F^{(1)}=G^{(1)}=1$, $\mathbf{k}=\mathbf{q}_1+...+\mathbf{q}_n$, and  $\mathbf{k}\equiv\mathbf{k}_1+\mathbf{k}_2$.
 
For second order, $n=2$, the recursion relations allow us to obtain the kernels
\begin{equation}
\mathcal{F}_{N}^{(2)}(\mathbf k_1,\mathbf k_2,\eta)=\frac{5}{7}+\frac{2}{7}\frac{(\mathbf k_1\cdot\mathbf{ k}_2)^2}{k_1^2k_2^2}+\frac{\mathbf{k}_1\cdot\mathbf {k}_2(k_1^2+k_2^2)}{2k_1^2k_2^2},
\label{KernelFF2N}
\end{equation}
\begin{equation}
\mathcal{G}_{N}^{(2)}(\mathbf k_1,\mathbf k_2,\eta)=\frac{3}{7}+\frac{4}{7}\frac{(\mathbf k_1\cdot\mathbf {k}_2)^2}{k_1^2k_2^2}+\frac{\mathbf k_1\cdot\mathbf {k}_2(k_1^2+k_2^2)}{2k_1^2k_2^2}.
\end{equation}
These are symmetrized kernels, this results from the sum of ${F}^{(n)}$ and ${G}^{(n)}$ with all possible permutations of $\mathbf{q}_i$, the symmetrized kernels are written in calligraphic font. 

In this work we will require the recursion relations up to $n=3$, in this case the kernels are given by \cite{2010PhDT.........4J} 
\begin{align}
 \label{Kernel3Nchap2}
\nonumber
\mathcal{F}^{(3)}_{N}(\mathbf k_1,\mathbf k_2,\mathbf k_3,\eta)=&\frac{2k^2}{54}\left[\frac{\mathbf k_1\cdot\mathbf k_{23}}{k_1^2k_{23}^2}\mathcal{G}_N^{(2)}(\mathbf k_2,\mathbf k_3)+(2\ \textrm{cyclic})\right]\\
           +&\frac{7}{54}\mathbf k\cdot\left[\frac{\mathbf k_{12}}{k_{12}^2}\mathcal{G}_N^{(2)}(\mathbf k_1,\mathbf k_2)+(2\ \textrm{cyclic})\right]\\ 
\nonumber           
            +&\frac{7}{54}\mathbf k\cdot\left[\frac{\mathbf k_{1}}{k_{1}^2}\mathcal{F}_N^{(2)}(\mathbf k_2,\mathbf k_3)+(2\ \textrm{cyclic})\right],\\
            \nonumber
\end{align}
\begin{align}
 \label{Kernel3GNchap2}
\nonumber
\mathcal{G}^{(3)}_{N}(\mathbf k_1,\mathbf k_2,\mathbf k_3,\eta)=&\frac{k^2}{9}\left[\frac{\mathbf k_1\cdot\mathbf k_{23}}{k_1^2k_{23}^2}\mathcal{G}_N^{(2)}(\mathbf k_2,\mathbf k_3)+(2\ \textrm{cyclic})\right]\\
           +&\frac{1}{18}\mathbf k\cdot\left[\frac{\mathbf k_{12}}{k_{12}^2}\mathcal{G}_N^{(2)}(\mathbf k_1,\mathbf k_2)+(2\ \textrm{cyclic})\right]\\ 
\nonumber           
            +&\frac{1}{18}\mathbf k\cdot\left[\frac{\mathbf k_{1}}{k_{1}^2}\mathcal{F}_N^{(2)}(\mathbf k_2,\mathbf k_3)+(2\ \textrm{cyclic})\right].\\
            \nonumber
\end{align}

Higher orders can be calculated, however their symmetrization is not straightforward, and are not needed here. 

\section{Statistics in Cosmology}
\label{section:statistics}

So far we have described the evolution of the density contrast $\delta(\mathbf{k},\eta)$ in the Universe. However, the density contrast on its own does not provide information about the structure in the Universe as a whole. In order to be able to compare theoretical results with observations, we require to analyse the statistics of both.

\begin{figure}[tbp]
\centering
\includegraphics[width=115mm]{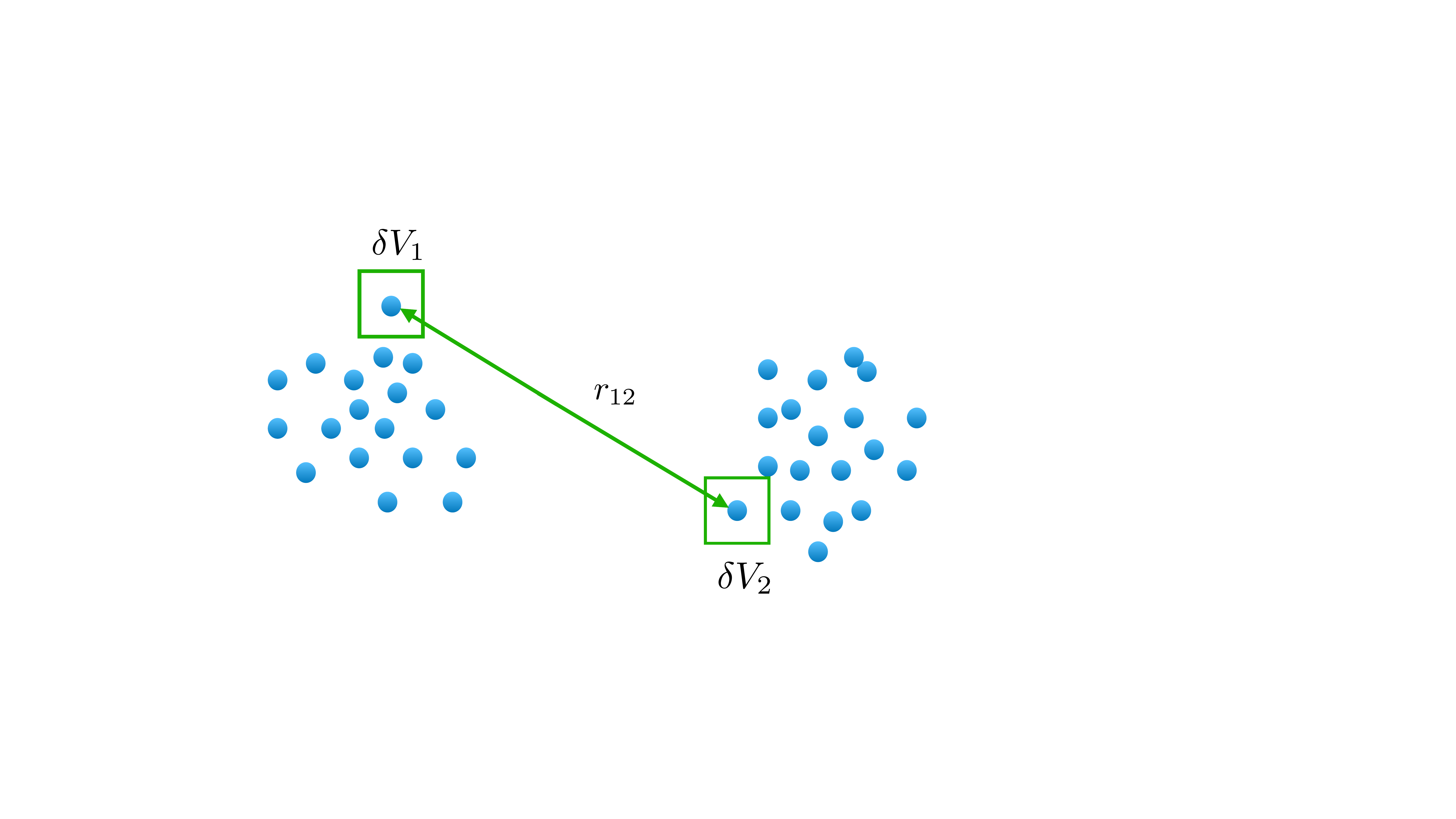}
\caption[Diagrammatic representation of the separation of the volume elements of 2-point correlation function.]{Diagrammatic representation of the volume elements $\delta V_1$ and $\delta V_2$, separated at a distance $r_{12}$ in the definition of 2-point correlation function.}
\label{correlationfunctioncartoon}
\end{figure}

One of the simplest statistical quantities is the galaxy 2-point correlation function $\xi(r)$ defined by the joint probability $\delta P$ of finding a galaxy in both of the volume elements $\delta V_1$  and $\delta V_2$ at a given separation $r_{12}$ (see Fig.~\ref{correlationfunctioncartoon})
\begin{equation}
\delta P=n^2\delta V_1\delta V_2 [1+\xi(r_{12})],
\end{equation}
where $n$ is the mean number density of galaxies \cite{Peebles1980}.\footnote{$\xi$ should not to be confused with the gauge generator defined in previous sections.}

Another definition of the correlation function is given by the spatial average of the product of the density contrast at two different points 

\begin{equation}
\langle{\delta(\mathbf{x}_1)\delta{(\mathbf{x}_2)}\rangle}=\xi(|\mathbf{r}_{12}|),
\end{equation}
where $|\mathbf{r}_{12}|=|\mathbf{x}_1-\mathbf{x}_2|$, is due to the isotropy and homogeneity that the correlation function depends only on the magnitude of the difference of the positions $\mathbf{x}_1$ and $\mathbf{x}_2$ \cite{Baldauf,2015daen.book.....A}.

Having defined the correlation function, we can define the power spectrum $P(k)$ as the Fourier transform of the 2-point correlation function
\begin{equation}
\langle{\delta(\mathbf{k})\delta(\mathbf{k}')\rangle}=(2\pi)^3P(\mathbf{|k|})\delta_D(\mathbf{k}+\mathbf{k}'),
\label{powerspectdef}
\end{equation}

\noindent in terms of calculations  the power spectrum turns out to be more convenient than the correlation function as the modes in Fourier space evolve independently at linear order.

When we consider Gaussian random fields, which is the case for the density contrast, Wick's theorem is useful. This theorem establishes that the correlator of the product of an odd number of Gaussian random fields vanishes 
\begin{equation}
\langle{\delta(\mathbf{k}_1)...\delta(\mathbf{k}_{2p+1})\rangle}=0,
\end{equation}

\noindent and if we have the product of an even number of Gaussian random fields, the correlator will be given by 

\begin{equation}
\langle{\delta(\mathbf{k}_1)...\delta(\mathbf{k}_{2p})\rangle}=\underset{\mathrm{all\,pairs\,associations}}\sum\quad \underset{p\,\mathrm{pairs}\,(i,j)}\prod \langle{\delta(\mathbf{k}_i)\delta(\mathbf{k}_j)\rangle},
\label{Wickstheorem}
\end{equation}
where all pairs associations refers to the sum of all possible pairings of $\delta(\mathbf{k}_i)$ with $\delta({\mathbf{k}_j})$, i.e. products of 2-point correlators, where $p$ is an integer number \cite{2002PhR...367....1B,2008cmbg.book.....D}. For the case of Gaussian random fields, the 2-point correlation function (or in Fourier space, the power spectrum) describes in full their statistics.

\begin{figure}[ht]
\centering
\includegraphics[width=115mm]{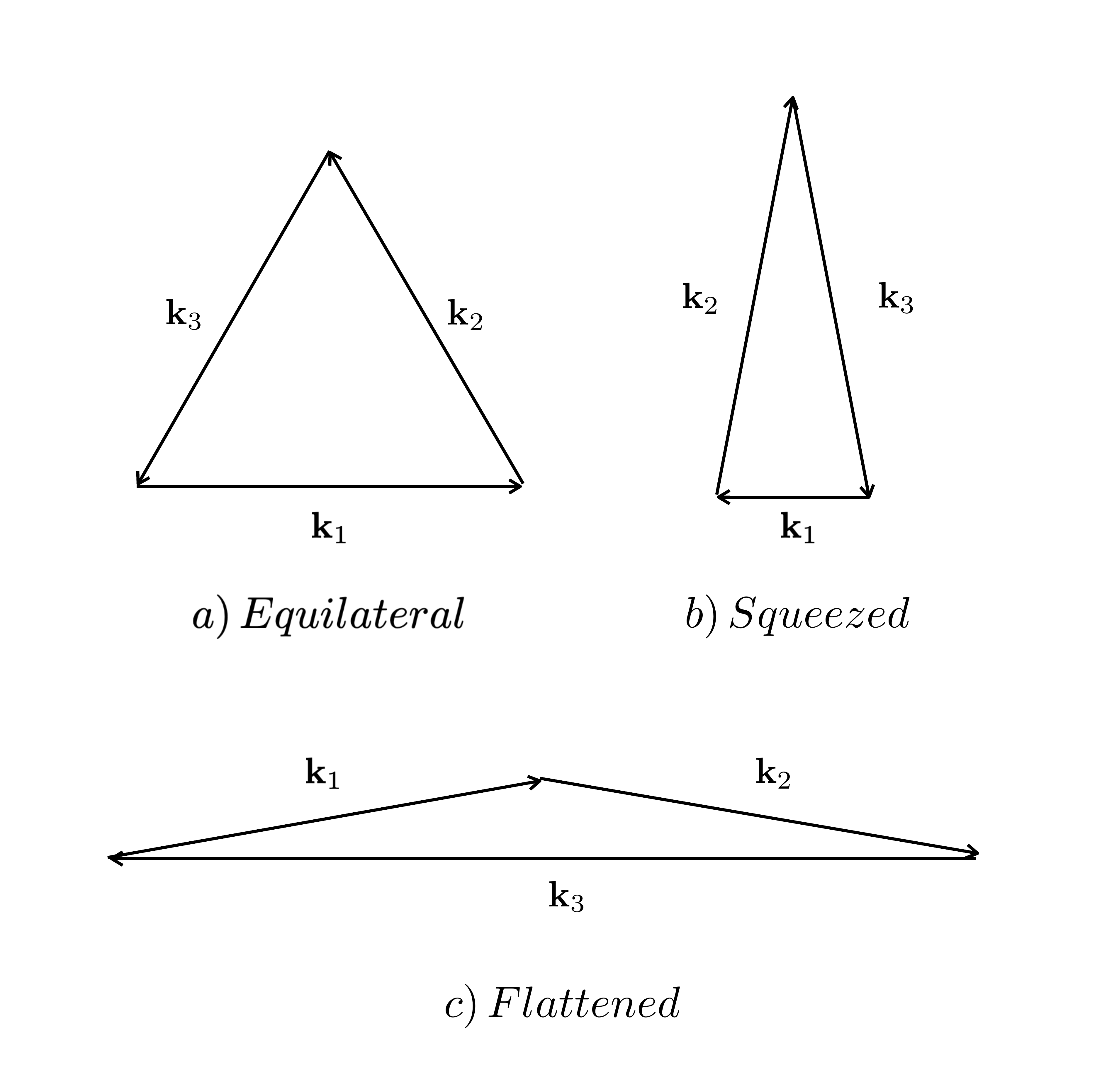}
\caption[Different configurations for the bispectrum triangles.]{Different configurations for the bispectrum triangles. a) Equilateral configuration where  $\mathbf{k}_1=\mathbf{k}_2=\mathbf{k}_3$. b) Squeezed configuration where $\mathbf{k}_2, \mathbf{k}_3\gg\mathbf{k}_1$. c) Flattened configuration where $\mathbf{k}_1=\mathbf{k}_2=\frac{1}{2}\mathbf{k}_3$.}
\label{triangles}
\end{figure}

The next order statistics that give us information that is not captured by the 2-point correlation function, is the three-point correlation function or in Fourier space, the bispectrum $B(\mathbf{k}_1,\mathbf{k}_2,\mathbf{k}_3)$ defined as \cite{1984ApJ...279..499F}
\begin{equation}
\langle{\delta(\mathbf{k}_1)\delta(\mathbf{k}_2)\delta(\mathbf{k}_3)\rangle}=(2\pi)^3B(\mathbf{k}_1,\mathbf{k}_2,\mathbf{k}_3)\delta_{D}(\mathbf{k}_1+\mathbf{k}_2+\mathbf{k}_3).
\end{equation}
Due to homogeneity and isotropy the three wave vectors have to sum to zero, as a result they form a triangle \cite{Baldauf,2011JCAP...10..026L}. The configurations for these triangles that are usually considered in the literature are shown in Fig.~\ref{triangles}. 

The bispectrum is necessary if we are dealing with non-Gaussian random fields or we treat with non-linear orders of Gaussian random fields we will have to consider higher order statistics \cite{Baldauf,2011JCAP...10..026L,2018arXiv180300070P}. For example, if we expand the Gaussian density contrast to non-linear orders
\begin{equation}
\delta=\delta^{(1)}+\frac{\delta^{(2)}}{2}+\frac{\delta^{(3)}}{6}+...,
\end{equation}
the correlation of three copies of the linear order density contrast
\begin{equation}
\langle{\delta^{(1)}(\mathbf{k}_1)\delta^{(1)}(\mathbf{k}_2)\delta^{(1)}(\mathbf{k}_3)\rangle}=0,
\end{equation}
due to Wick's theorem. However, the correlation between two linear order density contrast $\delta^{(1)}(\mathbf{k}_1)$ and $\delta^{(1)}(\mathbf{k}_2)$ with a non-linear order  density contrast $\delta^{(2)}(\mathbf{k}_3)$ will be 
\begin{equation}
\frac{1}{2}\langle{\delta^{(1)}(\mathbf{k}_1)\delta^{(1)}(\mathbf{k}_2)\delta^{(2)}(\mathbf{k}_3)\rangle}=(2\pi)^3B(\mathbf{k}_1,\mathbf{k}_2,\mathbf{k}_3)\delta_{D}(\mathbf{k}_1+\mathbf{k}_2+\mathbf{k}_3).
\end{equation}

\subsection{Primordial non-Gaussianity}
Since we are studying the large scale structure of the Universe, it is of great importance to consider the effects of primordial non-Gaussianities in it. It is believed that the primordial density fluctuations, seeds of the LSS, were generated during the inflation period. These primordial density fluctuations leave an imprint on the LSS that we observe nowadays. The simplest models of inflation tell us that the generation of these primordial density fluctuations follow a nearly Gaussian distribution, with small non-Gaussian deviations \cite{2016A&A...594A..17P,2003JHEP...05..013M,2003NuPhB.667..119A}.

 Studying and constraining primordial non-Gaussianities is a useful tool to study the early universe as it can help to distinguish between different inflation models (see e.g.~Refs.~\cite{2004PhR...402..103B,2007JCAP...01..002C}) being a key in our understanding of the physics of the early universe and the process of structure formation. For a review on primordial non-Gaussianity see e.g.~Refs.~\cite{2004PhR...402..103B,2018arXiv181208197C}.

As mentioned in the previous section in order to study non-Gaussianities it is necessary to consider the bispectrum statistics. For the 
Bardeen’s curvature potential $\Phi$ in Eq.~\eqref{Bardeenpot} 
we have 
\begin{equation}
\langle\Phi(\mathbf{k}_1)\Phi(\mathbf{k}_2)\Phi(\mathbf{k}_3)\rangle=(2\pi)^3B_\Phi(\mathbf{k}_1,\mathbf{k}_2,\mathbf{k}_3)\delta_{D}(\mathbf{k}_1+\mathbf{k}_2+\mathbf{k}_3).
\end{equation}

\noindent Following Ref.~\cite{2016A&A...594A..17P} the general the bispectrum $B_{\Phi}$ can be expressed as 
\begin{equation}
B_\Phi({k}_1,{k}_2,{k}_3)=f_{\mathrm{NL}}F(k_1,k_2,k_3),
\end{equation}
where $f_{\mathrm{NL}}$ is the non-Gaussian parameter, and measures the amplitude of the non-Gaussianity, the function $F(k_1,k_2,k_3)$ will be given in terms of the shape of bispectrum considered. 

As an example, we present the local model of PNG, for this $\Phi$ can be expanded as
\begin{equation}
\Phi=\phi+f_{\mathrm{NL}}(\phi^2-\langle\phi^2\rangle)+g_{\mathrm{NL}}(\phi^3)+...,
\end{equation}
where $\phi$ is a linear Gaussian random field, $f_{\mathrm{NL}}$ and $g_{\mathrm{NL}}$ are the non-Gaussian parameters at first and second order respectively.\footnote{In the literature, the local model is also defined through the curvature perturbation in the uniform-density gauge $\zeta$, where at large scales $\Phi=\frac{3}{5}\zeta$.}

The local model peaks in the squeezed limit and using Wick's theorem, $B_{\Phi}$ is given by \cite{2000MNRAS.313..141V,2001PhRvD..63f3002K}

\begin{equation}
B^{\mathrm{local}}_{\Phi}(\mathbf{k}_1,\mathbf{k}_2,\mathbf{k}_3)=2f_{\mathrm{NL}}^{\mathrm{local}}[P_{\Phi}(k_1)P_{\Phi}(k_2)+P_{\Phi}(k_2)P_{\Phi}(k_3)+P_{\Phi}(k_3)P_{\Phi}(k_1)],
\end{equation}
where the power spectrum $P_{\Phi}(k)$, is defined by the Eq.~\eqref{powerspectdef}
\begin{equation}
\langle\Phi(\mathbf{k}_1)\Phi(\mathbf{k}_2)\rangle=(2\pi)^3\delta(\mathbf{k}_1+\mathbf{k}_2)P_{\Phi}(k_1),
\end{equation}
with \cite{2016A&A...594A..17P,2010CQGra..27l4010K}
\begin{equation}
P_{\Phi}(k)=Ak^{n_s-4},
\end{equation}
where $A$ is the normalisation and  $n_s$ is the scalar spectral index.

Current constraints for non-Gaussianity parameters reported by the Planck collaboration \cite{2020A&A...641A...9P} are $f_{\mathrm{NL}}^{\mathrm{local}}=-0.9\pm5.1$ and  $g_{\mathrm{NL}}^{\mathrm{local}}=-5.8\times10^4\pm6.5\times10^4.$ These constraints will be used in the analysis in the next chapters.

\chapter{Relativistic and non-Gaussianity contributions to the one-loop power spectrum}
\label{chapter:Relativistic}

\section{Introduction}

The work presented in this chapter is based on the paper in Ref.~\cite{2020JCAP...04..028M}. The goal of this chapter is to include contributions from the scalar sector of the full relativistic theory at second and third order, as well as the primordial non-Gaussianity at the same orders (which can be easily included as an additional term to the density contrast in the chosen gauge, the synchronous-comoving gauge), and analyse their effects on the power spectrum at one-loop and the  tree-level bispectrum. This chapter is organised as follows:

In section~\ref{sec:Evolutioneqs} we review work previously done and present the evolution equations for the density contrast in synchronous-comoving gauge. We present its solutions up to third order using the gradient expansion. These solutions assume an Einstein-de Sitter universe (a matter only universe, see section \ref{Einsteindesitter}) and are necessary for the computation of the one-loop power spectrum. In section~\ref{sec:Fouriersol}, we present the Newtonian and relativistic solutions for the density contrast in Fourier space. Section~\ref{sec:One-loop} is dedicated to the one-loop power spectrum, which is our main result. We provide complete analytical expressions for the one-loop power spectrum, along with numerical integrations, including the contributions to the one-loop power spectrum for the allowed values of $f_{\textrm{NL}}$ and $g_{\textrm{NL}}$ reported by Planck \cite{2020A&A...641A...9P}. For completeness, in section~\ref{sec:Treebis} we present the tree-level bispectrum, along with numerical solutions. Finally, in section~\ref{sec:conclusions} we discuss our results in light of the forthcoming galaxy surveys.

\section{Evolution equations and relativistic density contrast solutions}
\label{sec:Evolutioneqs}

In this section we present the evolution equations for the density contrast which are given in synchronous-comoving gauge, based on work previously done in Refs.~\cite{2018JCAP...06..016G,2014ApJ...794L..11B,2014ApJ...785....2B}. The choice of this gauge provides a Lagrangian frame in General Relativity, which is also suitable for defining local Lagrangian galaxy bias up to second order \cite{2015CQGra..32q5019B}. Our starting point is the 
synchronous comoving line element, defined in subsection~\ref{Gaugechoice},
\begin{equation}
ds^2=a^2(\eta)[-d\eta^2+\gamma_{ij}dx^idx^j],
\end{equation}

\noindent where $a$ is the scale factor, $\eta$ is the conformal time and $\gamma_{ij}$ is the spatial metric.

As the matter content we consider an irrotational, pressureless fluid. Observers are comoving with the fluid, and as a consequence the four-velocity in the synchronous comoving gauge is $u_\mu=(-a,0,0,0)$. 

For the following fluid description, we define the deformation tensor,
\begin{equation}
{\vartheta}_\nu^\mu\equiv a{u^\mu}_{;\nu}-\mathcal{H}\delta_\nu^\mu,
\label{lineelement}
\end{equation}

\noindent where the isotropic background expansion was removed. In the chosen gauge, the deformation tensor has only spatial components and is proportional to the extrinsic curvature $K_j^i$ of the conformal spatial metric $\gamma_{ij}$,
\begin{equation}
\vartheta_j^i=-K_j^i,
\end{equation}

\noindent where $K_j^i$ is given by
\begin{equation}
K_j^i\equiv-\frac{1}{2}\gamma^{ik}\gamma_{kj}'.
\end{equation}

The density field $\rho$ is defined as 
\begin{equation}
\rho(\mathbf x,\eta)=\bar\rho(\eta)+\delta\rho(\mathbf x,\eta)=\bar\rho(\eta)(1+\delta(\mathbf x,\eta)),
\end{equation}

\noindent where $\bar\rho(\eta)$ is the density in the background, $\delta\rho(\mathbf x,\eta)$ is a small perturbation and $\delta(\mathbf x,\eta)$ is the density contrast. 
The evolution of the density contrast $\delta(\mathbf x,\eta)$ is given by the continuity equation 
\begin{equation}
\delta'+(1+\delta)\vartheta=0,
\label{continuity}
\end{equation}

\noindent where $\vartheta=\vartheta_\alpha^\alpha$ is the trace of $\vartheta_\nu^{\mu}$.

The evolution for $\vartheta$ is given by the Raychaudhuri equation (more details of the derivation can be found in Refs.~\cite{2014ApJ...785....2B,Meures})
\begin{equation}
\vartheta'+\mathcal{H}\vartheta+\vartheta_j^i\vartheta_i^j+4\pi G a^2 \bar\rho\delta=0.
\label{Ray}
\end{equation}

The energy constraint is given by 
\begin{equation}
\vartheta^2-\vartheta_j^i\vartheta_i^j+4\mathcal{H}\vartheta+{}^{3}R=16\pi Ga^2\bar\rho\delta,
\label{energycons}
\end{equation}

\noindent where ${}^{3}R$ is the spatial Ricci scalar of the spatial metric $\gamma_{ij}$. In the following subsections we use two approaches to find solutions to the evolution equations. 

\subsection{Cosmological perturbation theory}
\label{perturbation}
In order to show how cosmological perturbation theory is used to find the evolution of the density contrast, we present in this section the solutions to first order. The line element in Eq.~\eqref{lineelement} is equivalent to a spatially flat FLRW background with a perturbed spatial metric as it was defined in Eq.~\eqref{perturbedlinelement} (in synchronous-comoving gauge), and hence we can expand $\gamma_{ij}$ as in terms of the scalar metric potentials $\psi$ and $\chi$ as
\begin{align}
\label{pert_metric}  
\nonumber
\gamma_{ij}&=\delta_{ij}+\gamma_{ij}^{(1)}+\frac{1}{2}\gamma_{ij}^{(2)}+...\\
 &=(1-2\psi^{(1)}-\psi^{(2)})\delta_{ij}+\chi_{ij}^{(1)}+\frac{1}{2}\chi_{ij}^{(2)}+...
\end{align}

\noindent where 
\begin{equation}
\chi_{ij}=D_{ij}\chi,
\end{equation}
and 
\begin{equation}
D_{ij}=\left(\partial_i\partial_j-\frac{1}{3}\delta_{ij}\nabla^2\right).
\end{equation}

The density contrast is decomposed as
\begin{equation}
\delta=\delta^{(1)}+\frac{1}{2}\delta^{(2)}+\frac{1}{6}\delta^{(3)}+...
\end{equation}

For the case of the first order solutions for the density contrast, we combine the first order of the continuity equation \eqref{continuity} and the Raychaudhuri equation \eqref{Ray} at first order,  to obtain the first order density contrast evolution equation
\begin{equation}
\delta^{(1)''}+\mathcal{H}\delta^{(1)'}-\frac{3}{2}\mathcal{H}^2\Omega_m\delta^{(1)}=0.
\label{Secondorder}
\end{equation}

 From the first order energy constraint equation \eqref{energycons}, combined with the first order continuity equation \eqref{continuity} we obtain
 \begin{equation}
 4\mathcal{H}\delta^{(1)'}+6\mathcal{H}^2\Omega_m\delta^{(1)}-{}^{3}R^{(1)}=0,
 \label{deltaprime}
 \end{equation}
 
\noindent combining the time derivative of the Eq.~\eqref{deltaprime} and using the first order of Eqs.~\eqref{continuity} and \eqref{Ray} we find an equation for $R^{(1)}$ given by
 \begin{equation}
 {}^{3}R^{(1)'}=0.
 \end{equation}
 
 The general solution for a second order differential equation as is Eq.~\eqref{Secondorder}, will be composed of a linear combination of a growing mode  and a decaying mode 
 \begin{equation}
 \delta^{(1)}(\mathbf x,\eta)=C_{+}(\mathbf x)D_{+}(\eta)+C_{-}(\mathbf x)D_{-}(\eta).
 \end{equation}
 
 Since we choose to work in an Einstein-de Sitter universe (see section~\ref{Einsteindesitter}), the decaying mode solution is negligible and from now on we take a solution of the form
 \begin{equation}
\delta^{(1)}(\mathbf x,\eta)=C(\mathbf x)D_{+}(\eta),
\label{soldelta}
 \end{equation}

\noindent where $C(\mathbf x)$ will be given by \cite{2014ApJ...785....2B}
 \begin{equation}
C(\mathbf x)=\frac{{}^{3}R^{(1)}}{10\mathcal{H}_{IN}^2D_{+IN}},
\end{equation}
 
 \noindent $D_{+}$ is the growth factor, and the subscript ``$IN$" denotes a time early in the matter dominated era.

At first order in perturbation, for an unspecified gauge, the spatial Ricci scalar, is 
\begin{equation}
{}^{3}R^{(1)}=4\nabla^2\left(\psi^{(1)}+\frac{1}{6}\nabla^2\chi^{(1)}\right).
\label{Riccist}
\end{equation}
We can define the gauge invariant comoving curvature perturbation $\mathcal{R}_c$ by
\begin{equation}
\mathcal{R}_c=\psi^{(1)}+\frac{1}{6}\nabla^2\chi^{(1)}-\mathcal{H}(v+\omega).
\label{Riccinew}
\end{equation}
\noindent We can then evaluate Eq.~\eqref{Riccinew} in the comoving gauge, where $(v+\omega)=0$, and get \cite{2009PhR...475....1M,2016JCAP...02..021C}
\begin{equation}
\mathcal{R}_c=\psi^{(1)}_c+\frac{1}{6}\nabla^2\chi^{(1)}_c.
\end{equation}

The comoving curvature perturbation is related to the curvature perturbation on  uniform-density hypersurfaces as (see for example Ref.~\cite{2009PhR...475....1M})
\begin{equation}
\zeta^{(1)}\equiv -\psi^{(1)}-\frac{1}{6}\nabla^2\chi^{(1)}-\frac{\mathcal{H}}{\rho'}\delta\rho^{(1)}=-\mathcal{R}_c+\frac{1}{3}\delta^{(1)},
\end{equation}

\noindent and at early times and large scales $\zeta^{(1)}$ and $\mathcal{R}_c$ are approximately equal:
\begin{equation}
\zeta^{(1)}\simeq-\mathcal{R}_c.
\label{Rczeta}
\end{equation}
Substituting Eq.~\eqref{Rczeta} into Eq.~\eqref{Riccist}, we write the first order solution for  the density contrast as 
\begin{equation}
\delta^{(1)}=\frac{D_{+}(\eta)}{10\mathcal{H}^2_{IN}D_{+IN}}\left(-4\nabla^2\zeta^{(1)}\right),
\label{delta1N}
\end{equation}

\noindent where the growth factor in Einstein-de Sitter is
\begin{equation}
D_{+}=\frac{D_{+IN}\mathcal{H}_{IN}^2}{\mathcal{H}^2},
\end{equation}

\noindent with  $D_{+IN}=1$ and $\mathcal{H}_{IN}=\mathcal{H}_0$, where $\mathcal{H}_0$ is the conformal Hubble parameter at present time.\footnote{The order by order correspondence between the density contrast and the curvature perturbation means that $\delta^{(1)}$ represents a Gaussian field.} These choices are made to recover the standard Newtonian solutions. 

\subsection{Gradient expansion approach}

In section~\ref{perturbation} we presented the first order equations and solutions for the density contrast using cosmological perturbation theory, in this section we present the solutions for the second and third order equations using a different approach, the gradient expansion, that leads to the same equations and solutions obtained using the perturbative treatment. Instead of using the expansion Eq.~\eqref{pert_metric}, we can also write the spatial metric as \cite{1990PhRvD..42.3936S,2005JCAP...05..004L}
\begin{equation}
g_{ij}=a^2\gamma_{ij}=a^2e^{2\zeta}\check\gamma_{ij},
\end{equation}
\noindent where $\zeta$ is the curvature perturbation on uniform density hypersurfaces. 

The initial conditions for perturbations are set in the inflationary epoch. After this period, the curvature perturbation $\zeta$ is almost scale-invariant and remains constant (see for example Ref.~\cite{2009pdp..book.....L}). As a consequence is it possible to consider small initial inhomogeneities on large scales, allowing for a gradient expansion \cite{1963AdPhy..12..185L, 1975PThPh..54..730T, 2013MNRAS.430L..54R, 2013PhRvD..87l3525R, 1995PhRvD..52.2007D}. In this long-wavelength approximation the spatial gradients are small compared to time derivatives. Using this approximation we find  
\begin{equation}
\delta\sim \vartheta\sim {}^{3}R\sim \nabla^2,
\end{equation}

\noindent and using this approximation with the continuity \eqref{continuity} and energy constraint equations \eqref{energycons}, lead us back to the Eq.~\eqref{deltaprime}.

On large scales, and only considering scalars,  the conformal metric can be approximated as $\check\gamma_{ij}\simeq \delta_{ij}$. As a consequence of this simplified spatial metric, the Ricci scalar $R$ is a nonlinear function of the curvature perturbation $\zeta$ only, taking the form \cite{{2018JCAP...06..016G},{2014ApJ...794L..11B},{1984ucp..book.....W}}
\begin{equation}
{}^{3}R=-4\nabla^2\zeta+\sum_{m=0}^{\infty}\frac{\left(-2\right)^{m+1}}{(m+1)!}\left[(m+1)(\nabla\zeta)^2-4\zeta\nabla^2\zeta\right]\zeta^m.
\end{equation}

This expansion for $R$ will allow us to obtain solutions for the density contrast to higher orders. In this chapter we are interested in solutions up to third order. The third order corrections are obtained after expanding $R$ up to $m=1$ and are given by
\begin{equation}
{}^{3}R=-4\nabla^2\zeta+(-2)[(\nabla\zeta)^2-4\zeta\nabla^2\zeta]+2[2(\nabla\zeta)^2-4\zeta\nabla^2\zeta]\zeta.
\label{Rm1}
\end{equation}

 The curvature perturbation can be expanded in terms of a Gaussian random field $\zeta^{(1)}$ as
\begin{equation}
\zeta=\zeta^{(1)}+\frac{3}{5}f_{\textrm{NL}}\zeta^{(1)2}+\frac{9}{25}g_{\textrm{NL}}\zeta^{(1)3},
\label{zetaexpansion}
\end{equation}

\noindent where $f_{\textrm{NL}}$ and $g_{\textrm{NL}}$ are the non-Gaussian parameters at first and second order respectively \cite{2010CQGra..27l4002W}.
 After substituting Eq.~\eqref{zetaexpansion} into Eq.~\eqref{Rm1}, we get an expression for the Ricci scalar, that will allow us to find the density contrast solutions
\begin{align}
 \label{Riccifnlexpansion}   
\nonumber
{}^{3}R\simeq-4\nabla^2\zeta^{(1)}&+\left(\nabla\zeta^{(1)}\right)^2\left[-2-\frac{24}{5}f_{\textrm{NL}}\right]+\zeta^{(1)}\nabla^2\zeta\left[-\frac{24}{5}f_{\textrm{NL}}+8\right]\\
              &+\zeta^{(1)}\left(\nabla\zeta^{(1)}\right)^2\left[-\frac{216}{25}g_{\textrm{NL}}+\frac{24}{5}f_{\textrm{NL}}+4\right]\\
              &+\zeta^{(1)2}\nabla^2\zeta^{(1)}\left[-\frac{108}{25}g_{\textrm{NL}}+\frac{72}{5}f_{\textrm{NL}}-8\right]+\mathcal{O}(\zeta^{(1)4}).
              \nonumber
\end{align}

From Eq.~\eqref{Riccifnlexpansion} it is straightforward to see that solutions to first order in the gradient expansion agree with the ones produced using the perturbation theory treatment. 

In a similar way to the first order, using the continuity equation \eqref{continuity}, along  with the energy constraint equation \eqref{energycons}, the  second order  evolution equation of $\delta$ will be given by
\begin{equation}
4\mathcal{H}\delta^{(2)'}+6\mathcal{H}^2\Omega_m\delta^{(2)}-{}^{3}R^{(2)}=2\vartheta^{(1)2}-2\vartheta^{(1)i}_j\vartheta_i^{(1)j}-8\mathcal{H}\delta^{(1)}\vartheta^{(1)},
\end{equation}

\noindent using
\begin{equation}
{}^{3}R^{(2)'}=-4\vartheta^{(1)i}_jR_i^{(1)j}.
\end{equation}

As shown in the Ref.~\cite{2014ApJ...785....2B}, the solution for these equations is composed of an homogeneous and a particular solution (labelled with subscripts ``$h$" and ``$p$" respectively) of the form
\begin{equation}
\label{partplushom}
\delta^{(2)}=\delta_h^{(2)}+\delta_p^{(2)}, \quad {}^{3}R^{(2)}={}^{3}R_h^{(2)}+{}^{3}R_p^{(2)}, 
\end{equation}

\noindent where the particular solution recovers the Newtonian density contrast obtained within the Newtonian standard perturbation theory formalism and the homogeneous solution corresponds to the relativistic contributions to the density contrast also presented in Ref.~\cite{2014ApJ...785....2B}.\footnote{Expressions for the relativistic contributions in the Lagrangian perturbation formalism have also been reported in \cite{2016PhRvD..93d3539C}.}

Thus, using the expansion for the Ricci scalar given in Eq.~\eqref{Riccifnlexpansion} up to second order ($m=0$), the homogeneous solution for the  second order of the density contrast is 
\begin{equation}
\frac{1}{2}\delta^{(2)}=\frac{D_{+}(\eta)}{10\mathcal{H}^2_{IN}D_{+IN}}\frac{24}{5}\Bigg[-(\nabla\zeta^{(1)})^2\bigg(\frac{5}{12}+f_{\textrm{NL}}\bigg)+\zeta^{(1)}\nabla^2\zeta^{(1)}\bigg(\frac{5}{3}-f_{\textrm{NL}}\bigg)\Bigg],
\label{Reldelta2}
\end{equation}

\noindent in analogous way the homogeneous  third order solution for the density contrast is 
\begin{align}
\label{Reldelta3}
\nonumber
\frac{1}{6}\delta^{(3)}=\frac{D_{+}(\eta)}{10\mathcal{H}^2_{IN}D_{+IN}}\frac{108}{25}\Bigg[&2\zeta^{(1)}(\nabla\zeta^{(1)})^2\bigg(-g_{\textrm{NL}}+\frac{5}{9}f_{\textrm{NL}}+\frac{25}{54}\bigg)\\
+&\zeta^{(1)2}\nabla^2\zeta^{(1)}\bigg(-g_{\textrm{NL}}+\frac{10}{3}f_{\textrm{NL}}-\frac{50}{27}\bigg)\Bigg].
\end{align}

\noindent The expression in Eq.~\eqref{Reldelta3} slightly differs from the 
expression provided in Ref.~\cite{2018JCAP...06..016G} (Eq.~(5.9)), where the authors missed the negative sign in the $-2\zeta^{(1)}(\nabla\zeta^{(1)})^2 g_{\mathrm{NL}}$ term and they have missed a factor of $10$ in $\frac{9}{27}\zeta^{(1)2}\nabla^2\zeta^{(1)}f_{\mathrm{NL}}$ term, that if included leads to obtain the $\frac{10}{3}\zeta^{(1)2}\nabla^2\zeta^{(1)}f_{\mathrm{NL}}$ term. We are interested in the new effects to the one-loop power spectrum due to Newtonian and relativistic contributions focusing on the derivation of the relativistic solutions for the density contrast, since the Newtonian solutions are well known(see e.g.~Refs.~\cite{1992PhRvD..46..585M,1994ApJ...431..495J,2002PhR...367....1B,2009PhRvD..80d3531C}).\footnote{It is important to note that our expressions for the relativistic density contrast are derived in the synchronous-comoving gauge, i.e.~from a Lagrangian formalism. In our analysis we are also including the Newtonian density contrast, given instead in the Eulerian frame. Since the relativistic corrections only affect the large scales and the Lagrangian variables only present differences with respect to the Eulerian frame in the small scales, it is safe to state that our result is valid in the Eulerian frame at the one-loop order \cite{2015CQGra..32q5019B}.}

\section{Complete density contrast solutions in Fourier space}
\label{sec:Fouriersol}
In this section we present the complete solutions for the density contrast in Fourier space, these solutions consider both Newtonian and relativistic contributions.

As presented in section~\ref{section:SPT}, in Fourier space the second order density contrast is defined by 
\begin{equation}
\frac{\delta^{(2)}(\mathbf k,\eta)}{2}=\int\frac{d^3\mathbf k_1d^3\mathbf k_2}{(2\pi)^3}\delta_D(\mathbf k-\mathbf k_1-\mathbf k_2)\mathcal{F}^{(2)}(\mathbf k_1,\mathbf k_2,\eta)\delta^{(1)}(\mathbf k_1,\eta)\delta^{(1)}({\mathbf k_2},\eta),
\label{delta2N}
\end{equation}

\noindent the kernel $\mathcal{F}^{(2)}(\mathbf k_1,\mathbf k_2,\eta)$ is given by
\begin{equation}
\mathcal{F}^{(2)}(\mathbf k_1,\mathbf k_2,\eta)=\mathcal{F}_{N}^{(2)}(\mathbf k_1,\mathbf k_2,\eta)+\mathcal{F}_{R}^{(2)}(\mathbf k_1,\mathbf k_2,\eta),
\label{Kernels2}
\end{equation}

\noindent where  $\mathcal{F}_{N}^{(2)}(\mathbf k_1,\mathbf k_2,\eta)$  is the Newtonian contribution, corresponding to the particular solution in Eq.~\eqref{partplushom}, previously defined in Eq.~\eqref{KernelFF2N}
\begin{equation}
\mathcal{F}_{N}^{(2)}(\mathbf k_1,\mathbf k_2,\eta)=\Bigg\{\frac{5}{7}+\frac{2}{7}\frac{(\mathbf k_1\cdot\mathbf k_2)^2}{k_1^2k_2^2}+\frac{\mathbf k_1\cdot\mathbf k_2(k_1^2+k_2^2)}{2k_1^2k_2^2}\Bigg\},
\label{Kernel2N}
\end{equation}

\noindent the relativistic corrections $\mathcal{F}_{R}^{(2)}(\mathbf k_1,\mathbf k_2,\eta)$, obtained from Eqs.~\eqref{delta1N} and \eqref{Reldelta2}, in Fourier space are given by 
\begin{equation}
\mathcal{F}_{R}^{(2)}(\mathbf k_1,\mathbf k_2,\eta)=3\mathcal{H}^2\Bigg\{\left(f_{\textrm{NL}}-\frac{5}{3}\right)\frac{k_1^2+k_2^2}{2k_1^2k_2^2}+\left(f_{\textrm{NL}}+\frac{5}{12}\right)\frac{\mathbf k_1\cdot \mathbf k_2}{k_1^2k_2^2}\Bigg\},
\label{Kernel2R}
\end{equation}

\noindent the relativistic kernel is subdominant with respect to the Newtonian kernel at small scales, due to the factor $\mathcal{H}^2/k^2$ that for large values of $k$ is small.

Similarly, the third order density contrast is defined as 
\begin{align}
\label{delta3N}
\frac{\delta^{(3)}(\mathbf k,\eta)}{6}=\int\frac{d^3\mathbf k_1d^3\mathbf k_2d^3\mathbf k_3}{(2\pi)^6}&\delta_D(\mathbf k-\mathbf k_1-\mathbf k_2-\mathbf k_3 )\mathcal{F}^{(3)}(\mathbf k_1,\mathbf k_2,\mathbf k_3,\eta)\\
\times&\delta^{(1)}(\mathbf k_1,\eta)\delta^{(1)}(\mathbf k_2,\eta)\delta^{(1)}({\mathbf k_3},\eta),
\nonumber
\end{align}

\noindent where the kernel $\mathcal{F}^{(3)}(\mathbf k_1,\mathbf k_2,\mathbf k_3,\eta)$ is also composed by Newtonian and relativistic contributions
\begin{equation}
\mathcal{F}^{(3)}(\mathbf k_1,\mathbf k_2,\mathbf k_3,\eta)=\mathcal{F}_{N}^{(3)}(\mathbf k_1,\mathbf k_2,\mathbf k_3,\eta)+\mathcal{F}_{R}^{(3)}(\mathbf k_1,\mathbf k_2,\mathbf k_3,\eta),
\label{Kernels3}
\end{equation}

\noindent with the third order Newtonian kernel previously defined in Eq.~\eqref{Kernel3Nchap2} and given by
\begin{align}
 \label{Kernel3N}
\nonumber
\mathcal{F}^{(3)}_{N}(\mathbf k_1,\mathbf k_2,\mathbf k_3,\eta)=&\frac{2k^2}{54}\left[\frac{\mathbf k_1\cdot\mathbf k_{23}}{k_1^2k_{23}^2}\mathcal{G}_N^{(2)}(\mathbf k_2,\mathbf k_3)+(2\ \textrm{cyclic})\right]\\
           +&\frac{7}{54}\mathbf k\cdot\left[\frac{\mathbf k_{12}}{k_{12}^2}\mathcal{G}_N^{(2)}(\mathbf k_1,\mathbf k_2)+(2\ \textrm{cyclic})\right]\\ 
\nonumber           
            +&\frac{7}{54}\mathbf k\cdot\left[\frac{\mathbf k_{1}}{k_{1}^2}\mathcal{F}_N^{(2)}(\mathbf k_2,\mathbf k_3)+(2\ \textrm{cyclic})\right],\\
            \nonumber
\end{align}

\noindent and the relativistic contribution from Eqs.~\eqref{delta1N} and \eqref{Reldelta3}
\begin{align}
\nonumber
\mathcal{F}_{R}^{(3)}(\mathbf k_1,\mathbf k_2,\mathbf k_3,\eta)=\frac{27}{2}\mathcal{H}^4\Bigg[-&\frac{\mathbf k_1\cdot \mathbf k_2+\mathbf k_1\cdot\mathbf k_3+\mathbf k_2\cdot\mathbf k_3}{k_1^2k_2^2k_3^2}
\bigg(-g_{\textrm{NL}}+\frac{5}{9}f_{\textrm{NL}}+\frac{25}{54}\bigg)\\
-&\frac{1}{6}\frac{ k_1^2+k_2^2+k_3^2}{k_1^2k_2^2k_3^2}\bigg(-g_{\textrm{NL}}+\frac{10}{3}f_{\textrm{NL}}-\frac{50}{27}\bigg)\Bigg].
\label{Kernel3R}
\end{align}

\section{One-loop power spectrum}
\label{sec:One-loop}
The $n^{th}$ order contribution to the density power spectrum $P^{(n)}(\mathbf{k}, \eta)$ \cite{2002PhR...367....1B} is defined as, 
\begin{equation}
\label{powdefoneloop}
(2\pi)^3\delta^{D}(\mathbf k+\mathbf k')P^{(n)}(\mathbf k,\eta)=\sum_{m=1}^{2n-1}\frac{1}{m!{(2n-m)!}}\langle \delta^{(m)}(\mathbf k,\eta)\delta^{(2n-m)}(\mathbf k',\eta)\rangle.
\end{equation}
\noindent From this expression we find that the first order power spectrum $P^{(1,1)}(k,\eta)$, is also known as the tree-level power spectrum, corresponding to the linear power spectrum $P_{L}(k,\eta)$. Writing all the contributions up to second order ($n=2$) for the density power spectrum we obtain \cite{1996ApJ...473..620S}:
\begin{equation}
P(k,\eta)=P_L(k,\eta)+2P^{(1,3)}(k,\eta)+P^{(2,2)}(k,\eta),
\label{Oneloop}
\end{equation}

\noindent where  $P^{(1,3)}(k,\eta)$ and $P^{(2,2)}(k,\eta)$ corrections are known as the one-loop corrections to the density power spectrum. Since $\delta^{(1)}(k,\eta)$ is a Gaussian field, correlations of the order $P^{(1,2)}(k,\eta)$ are null (in contrast with the expansions presented in e.g.~Refs.~\cite{2008PhRvD..78l3534T,2009MNRAS.396...85D}).
 
\subsection[Second order density power spectrum correction \texorpdfstring{$P^{(2,2)}(k,\eta)$}{n}]{Second order density power spectrum correction \texorpdfstring{$\boldsymbol{P^{(2,2)}(k,\eta)}$}{n}}

The second order contribution to the density power spectrum $P^{(2,2)}(k,\eta)$ (see Appendix~\ref{one-loopderivation}) is defined as 
\begin{equation}
P^{(2,2)}(k,\eta)=2\int\frac{d^3q}{(2\pi)^3}P_L(q,\eta)P_L(|\mathbf k-\mathbf q|,\eta)[\mathcal{F}^{(2)}(\mathbf q,\mathbf k-\mathbf q,\eta)]^2.
\end{equation}

\noindent After substituting the expressions for $\mathcal{F}^{(2)}(\mathbf q,\mathbf k-\mathbf q,\eta)$ defined in Eq.~\eqref{Kernels2} and using the following variable transformation \cite{1992PhRvD..46..585M} 

 \begin{equation}
 x=\frac{\mathbf k\cdot\mathbf q}{|\mathbf k||\mathbf q|}=\cos{\theta}, \quad r=\frac{|\mathbf q|}{|\mathbf k|},
 \label{variablechange}
 \end{equation} 
 
\noindent we can write the total second order power spectrum correction $P^{(2,2)}(k,\eta)$ as a sum of a Newtonian density power spectrum $P_{NN}^{(2,2)}(k,\eta)$, a cross term  $P_C^{(2,2)}(k,\eta)$ that includes Newtonian and relativistic terms, and a purely relativistic term $P_{RR}^{(2,2)}(k,\eta)$
\begin{equation}
P^{(2,2)}(k,\eta)=P_{NN}^{(2,2)}(k,\eta)+P_C^{(2,2)}(k,\eta)+P_{RR}^{(2,2)}(k,\eta).
\end{equation}
 
 \noindent Altogether this is 
 \begin{align}
 \nonumber
 P^{(2,2)}(k,\eta)=&\frac{k^3}{2\pi^2}\int_{0}^{\infty}r^2dr  P_L(kr,\eta)\int_{-1}^{1}dxP_L(k\sqrt{1+r^2-2rx},\eta)\\
  \times&\Bigg\{  \bigg[ \frac{3r+7x-10rx^2}{14r(1+r^2-2rx)}\bigg]^2 \label{integral22}\\
  \nonumber
+& \left(\frac{\mathcal{H}^2}{k^2}\right)\frac{(6f_{\textrm{NL}}-10-25r(r-x))(3r+7x-10rx^2)}{28r^3(1+r^2-2rx)^2}\\
 +&  \left(\frac{\mathcal{H}^4}{k^4}\right)\bigg[\frac{6f_{\textrm{NL}}-10-25r^2+25rx}{4r^2(1-2rx+r^2)}\bigg]^2\Bigg\},
 \nonumber
 \end{align}

\noindent where the first and second lines correspond to $P_{NN}^{(2,2)}(k,\eta)$, while the  third and fourth lines correspond to  $P_C^{(2,2)}(k,\eta)$ and $P_{RR}^{(2,2)}(k,\eta)$ respectively.

\subsection[Second order density power spectrum correction \texorpdfstring{$P^{(1,3)}(k,\eta)$}{n}]{Second order density power spectrum correction \texorpdfstring{$\boldsymbol{P^{(1,3)}(k,\eta)}$}{n}}

The second order contribution $P^{(1,3)}(k,\eta)$ (see Appendix~\ref{one-loopderivation}) is defined as 

\begin{equation}
P^{(1,3)}(k,\eta)=3\mathcal{F}^{(1)}(\mathbf k)P_L(k,\eta)\int\frac{d^3q}{(2\pi)^3}P_L(q,\eta)\mathcal{F}^{(3)}(\mathbf k,\mathbf q,-\mathbf q,\eta),
\end{equation}
 
\noindent where $\mathcal{F}^{(1)}(\mathbf k)=1$ and $\mathcal{F}^{(3)}(\mathbf k,\mathbf q,-\mathbf q,\eta)$ is defined by Eq.~\eqref{Kernels3} and is written in terms of the variables defined in Eq.~\eqref{variablechange}.  For the total second order contribution $P^{(1,3)}(k,\eta)$ we have the sum of a Newtonian contribution $P_{NN}^{(1,3)}(k,\eta)$ and a relativistic contribution $P_{RR}^{(1,3)}(k,\eta)$. Note that, strictly speaking $P_{RR}^{(1,3)}(k,\eta)$ is a cross term, since it comes from the combination of $\mathcal{F}^{(1)}(\mathbf k)$ and $\mathcal{F}_R^{(3)}(\mathbf k,\mathbf q,-\mathbf q,\eta)$, and $\mathcal{F}^{(1)}(\mathbf k)$ does not have relativistic corrections. However, since $\mathcal{F}^{(1)}(\mathbf k)=1$, the product of $\mathcal{F}^{(1)}(\mathbf k)$  with $\mathcal{F}^{(3)}(\mathbf k,\mathbf q,-\mathbf q,\eta)$ will not modify $\mathcal{F}^{(3)}(\mathbf k,\mathbf q,-\mathbf q,\eta)$. Thus, in this thesis we will consider $P_{RR}^{(1,3)}(k,\eta)$ as a relativistic term. Then
\begin{equation}
P^{(1,3)}(k,\eta)=P_{NN}^{(1,3)}(k,\eta)+P_{RR}^{(1,3)}(k,\eta),
\end{equation}

\noindent using the change of variables in \eqref{variablechange} and integrating over the variable $x$, we obtain
\begin{align}
\nonumber
P^{(1,3)}(k,\eta)=\frac{k^3}{4\pi^2}P_L(k,\eta)\int_{0}^{\infty} drP_L(kr,\eta)&\Bigg\{ \frac{1}{504}\bigg[\frac{12}{r^2}-{158}+100r^2-42r^4\\
\nonumber
+&\frac{3}{r^3}(r^2-1)^3(7r^2+2)\ln{\left(\frac{r+1}{|r-1|}\right)}\bigg] \label{integral13}\\
\nonumber
+&81\left(\frac{{\mathcal{H}^4}}{k^4}\right)
\bigg[\Big(-g_{\textrm{NL}}+\frac{5}{9}f_{\textrm{NL}}+\frac{25}{54}\Big)\\
+&\frac{1+2r^2}{6r^2}\Big(g_{\textrm{NL}}-\frac{10}{3}f_{\textrm{NL}}+\frac{50}{27}\Big) \bigg]\Bigg\},
\end{align}

\noindent where the first and second line correspond to $P_{NN}^{(1,3)}(k,\eta)$ and third and fourth line to $P_{RR}^{(1,3)}(k,\eta)$.

We obtain numerical solutions for the different contributions to the density power spectrum presented in this section. All our integrations use as an input a linear power spectrum generated with the Boltzmann solver \texttt{CLASS} \cite{2011arXiv1104.2932L,2011JCAP...07..034B}, assuming a flat $\Lambda$CDM cosmology given by the Planck collaboration \cite{2016A&A...594A..13P} with a sharp cut-off in $P_L(k,\eta)$ at $k=10^{-5}h{\textrm{Mpc}}^{-1}$ due to the infrared behaviour of the purely relativistic terms (see the Appendix~\ref{IRL}), this cut-off is necessary for the numerical integration, that can not be performed using the lower limit of $0$ as stated in the analytical expressions, as these will diverge since we require as an input a numerical linear power spectrum $P_L(k,\eta)$ that is generated for a specific range of values in $k$. The large scales removed by this cut-off are not observable with the current surveys, meaning these are outside the horizon and can be considered a part of the background. To test the convergence of the numerical integration of the density power spectrum, we have computed these integrals with the Mathematica package and with a Python script independently.

In this way, the total Newtonian one-loop power spectrum is given as usual by
\begin{equation}
P_{NN}(k,\eta)=P_L(k,\eta)+2P_{NN}^{(1,3)}(k,\eta)+P_{NN}^{(2,2)}(k,\eta).
\end{equation}

\begin{figure}[htbp]
\centering
\includegraphics[width=130mm]{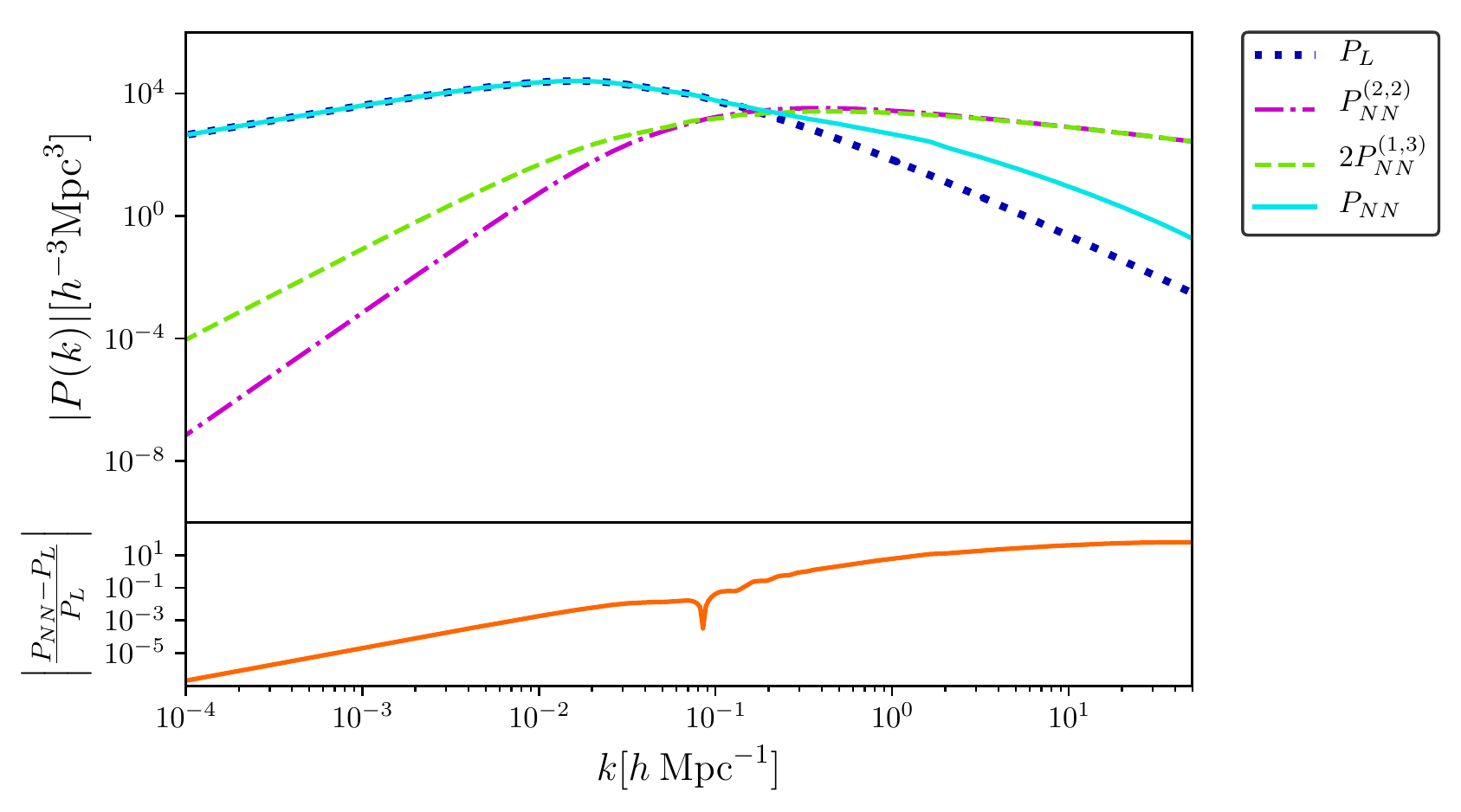}
\caption[The total Newtonian one-loop power spectrum $P_{NN}$ and the individual second order density power spectrum contributions to the total Newtonian one-loop power spectrum $P^{(2,2)}_{NN}$ and $P^{(1,3)}_{NN}$, at redshift
\newline
$z=0$.]{Upper panel: The total Newtonian one-loop power spectrum $P_{NN}$ and the individual second order density power spectrum contributions to the total Newtonian one-loop power spectrum $P^{(2,2)}_{NN}$ and $P^{(1,3)}_{NN}$, at redshift $z=0$. Bottom panel: The relative difference of the Newtonian one-loop power spectrum with respect to the linear power spectrum normalised with the linear power spectrum. 
 }
\label{NewtonOne}
\end{figure}

In Fig.~\ref{NewtonOne} we present the Newtonian standard perturbation theory results, showing the second order Newtonian contributions to the one-loop power spectrum, $P_{NN}^{(2,2)}(k,\eta)$ and $P_{NN}^{(1,3)}(k,\eta)$, along with the total Newtonian one-loop power spectrum $P_{NN}(k,\eta)$, for comparison we also plot the linear power spectrum $P_L(k,\eta)$ in all the Figures presented. The relative difference of the Newtonian one-loop power spectrum with respect to the linear power spectrum is also shown. The Newtonian contributions show a relevant effect only for the small scales.

\begin{figure}[htbp]
\centering
\includegraphics[width=130mm]{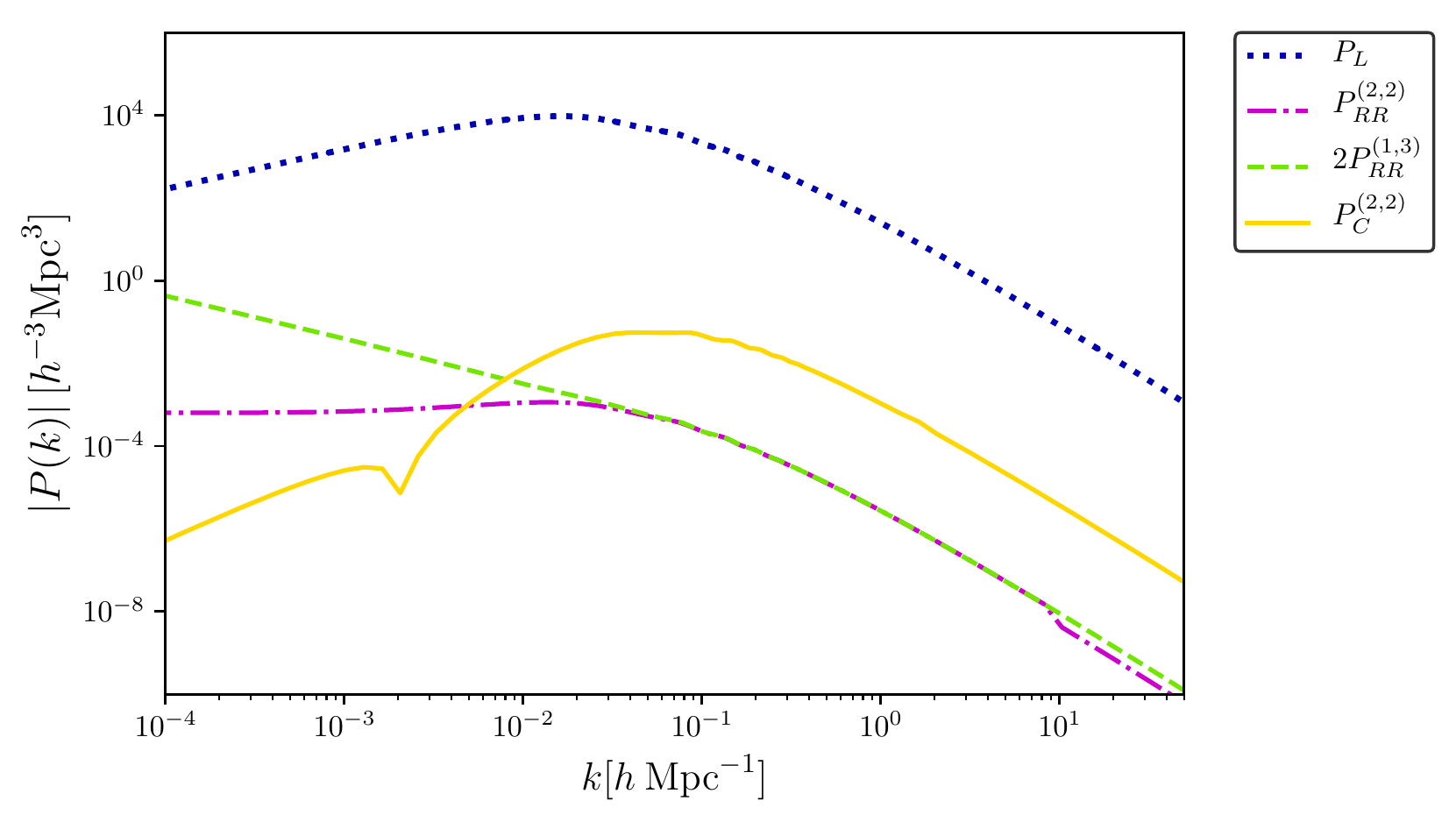}
\caption[Individual second order density power spectrum contributions to the total one-loop power spectrum $P^{(2,2)}_{RR}$, $P^{(2,2)}_{C}$ and $P^{(1,3)}_{RR}$, with $f_{\textrm{NL}}=g_{\textrm{NL}}=0$, at redshift $z=0$.]{Individual second order density power spectrum contributions to the total one-loop power spectrum $P^{(2,2)}_{RR}$, $P^{(2,2)}_{C}$ and $P^{(1,3)}_{RR}$ for $f_{\textrm{NL}}=g_{\textrm{NL}}=0$, at redshift $z=0$.}
\label{Crossed}
\end{figure}

In Fig.~\ref{Crossed} we present the individual relativistic contributions to the total one-loop power spectrum coming from, $P_{RR}^{(2,2)}(k,\eta)$ and $P_{RR}^{(1,3)}(k,\eta)$, along with the cross term $P_{C}^{(2,2)}(k,\eta)$. In this Figure we consider the case in where $f_{\textrm{NL}}=g_{\textrm{NL}}=0.$ The cross term is subdominant respect to the  second order density power spectrum relativistic contributions $P_{RR}^{(2,2)}(k,\eta)$ and $P_{RR}^{(1,3)}(k,\eta)$, the contribution of $P_{C}^{(2,2)}(k,\eta)$ is mainly in the small scales. 

\begin{figure}[htbp]
\centering
\includegraphics[width=135mm]{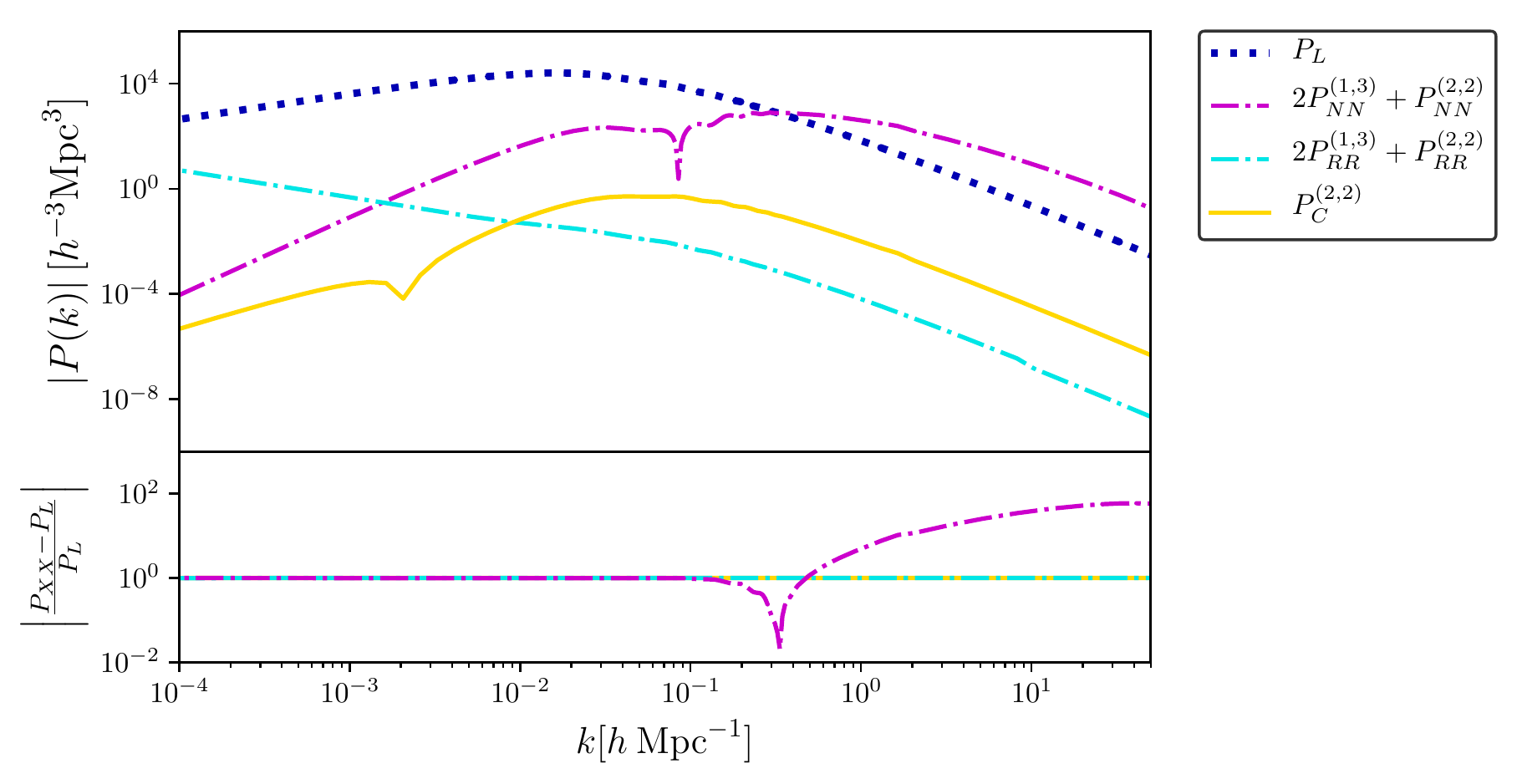}
\caption[Newtonian $(2P_{NN}^{(1,3)}+P_{NN}^{(2,2)})$, relativistic $(2P_{RR}^{(1,3)}+P_{RR}^{(2,2)})$ and cross term $P_C^{(2,2)}$ contributions to the one-loop power spectrum, at redshift $z=0$.]{Upper panel: Newtonian $(2P_{NN}^{(1,3)}+P_{NN}^{(2,2)})$, relativistic $(2P_{RR}^{(1,3)}+P_{RR}^{(2,2)})$ and cross term $P_C^{(2,2)}$ contributions to the one-loop power spectrum, with $f_{\textrm{NL}}=g_{\textrm{NL}}=0$, at redshift $z=0$. Bottom panel: The relative difference of each term $P_{XX}$ (i.e.~the Newtonian, relativistic and cross term) contributions with respect to the linear power spectrum normalised with the linear power spectrum. 
 }
\label{Relativedifcomparison}
\end{figure}

In Fig.~\ref{Relativedifcomparison} we present the Newtonian $(2P_{NN}^{(1,3)}+P_{NN}^{(2,2)})$, relativistic $(2P_{RR}^{(1,3)}+P_{RR}^{(2,2)})$ and cross term $P_C^{(2,2)}$ contributions to the one-loop power spectrum. The relative difference for each of these contributions to the one-loop power spectrum with respect to the linear power spectrum is also shown. In this Figure we consider the case in where $f_{\textrm{NL}}=g_{\textrm{NL}}=0.$ 

The total relativistic one-loop power spectrum is defined as 
\begin{equation}
P_{RR}(k,\eta)=P_L(k,\eta)+2P_{RR}^{(1,3)}(k,\eta)+P_{RR}^{(2,2)}(k,\eta).
\end{equation}

\begin{figure}[htbp]
\centering
\includegraphics[width=130mm]{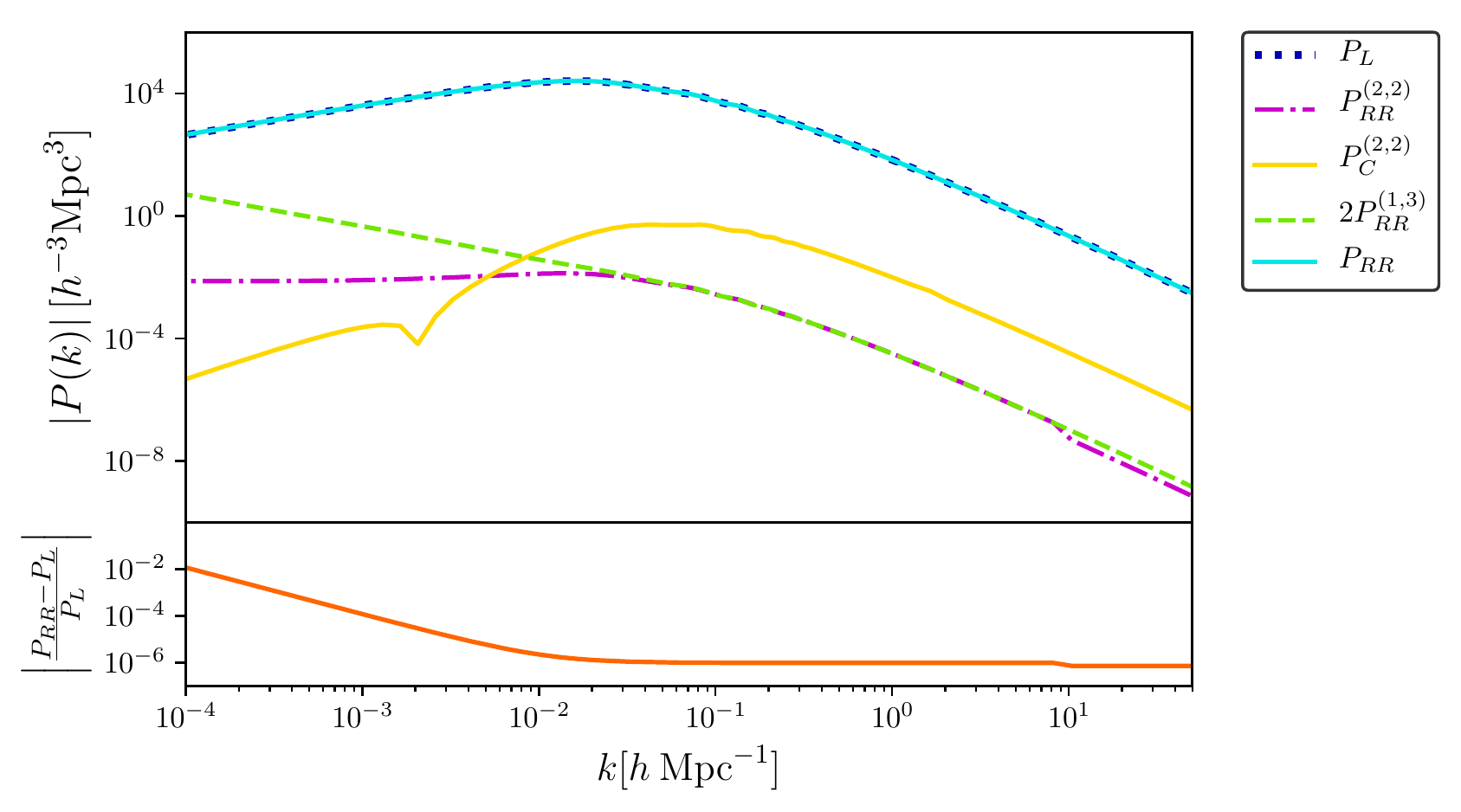}
\caption[The total relativistic one-loop power spectrum $P_{RR}$ and the individual second order density power spectrum contributions to the total relativistic one-loop power spectrum $P^{(2,2)}_{RR}$ and $P^{(1,3)}_{RR}$, with $f_{\textrm{NL}}=g_{\textrm{NL}}=0$, at redshift $z=0$.]{Upper panel: The total relativistic one-loop power spectrum $P_{RR}$ and the individual second order density power spectrum contributions to the total relativistic one-loop power spectrum $P^{(2,2)}_{RR}$, $P^{(2,2)}_C$ and $P^{(1,3)}_{RR}$ for $f_{\textrm{NL}}=g_{\textrm{NL}}=0$, at redshift $z=0$. Bottom panel: The relative difference of the relativistic one-loop power spectrum  with respect to the linear power spectrum normalised with the linear power spectrum. }
\label{GROne}
\end{figure}

In Fig.~\ref{GROne} we present the relativistic results, we show the relativistic contributions to the one-loop power spectrum coming from, $P_{RR}^{(2,2)}(k,\eta)$ and $P_{RR}^{(1,3)}(k,\eta)$, along with the total relativistic one-loop power spectrum $P_{RR}(k,\eta)$, we have also included the cross term $P_{C}^{(2,2)}(k,\eta)$, in this Figure we consider the case in where $f_{\textrm{NL}}=g_{\textrm{NL}}=0.$ The relative difference of the relativistic one-loop power spectrum  with respect to the linear power spectrum is also shown. We note that relativistic one-loop power spectrum  corrections are relevant in the large scales, the relativistic contributions are subdominant in smaller scales.

Finally, the total one-loop power spectrum defined in Eq.~\eqref{Oneloop} reads as
\begin{equation}
P_{RN}(k,\eta)=P_L(k,\eta)+2P^{(1,3)}(k,\eta)+P^{(2,2)}(k,\eta).
\label{Oneloopexp}
\end{equation}

\begin{figure}[htbp]
\centering
\includegraphics[width=130mm]{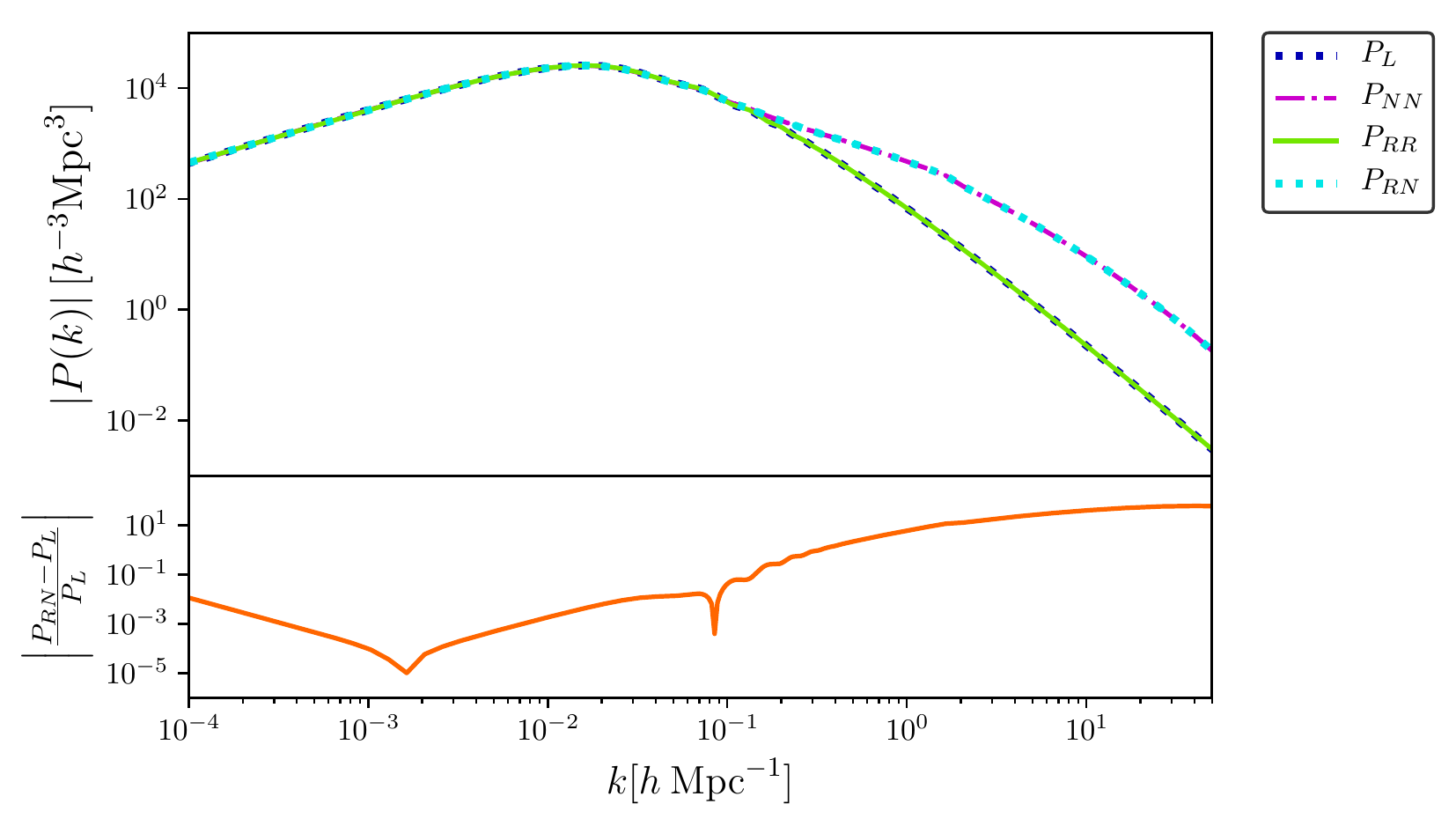}
\caption[Total Newtonian one-loop power spectrum $P_{NN}$ along with the total relativistic one-loop power spectrum $P_{RR}$ and total one-loop power spectrum $P_{RN}$, with $f_{\textrm{NL}}=g_{\textrm{NL}}=0$, at redshift $z=0$.]{Upper panel: Total Newtonian one-loop power spectrum $P_{NN}$ along with the total relativistic one-loop power spectrum $P_{RR}$ and total one-loop power spectrum $P_{RN}$, with $f_{\textrm{NL}}=g_{\textrm{NL}}=0$, at redshift $z=0$. Bottom panel: Relative difference of the total one-loop power spectrum  with respect to the linear power spectrum normalised with the linear power spectrum.}
\label{TotalOne}
\end{figure}

In Fig.~\ref{TotalOne}  we present a comparison of the total Newtonian one-loop power spectrum $P_{NN}(k,\eta)$, the total relativistic one-loop power spectrum $P_{RR}(k,\eta)$, along with the total one-loop power spectrum $P_{RN}(k,\eta)$, in this Figure we consider the case with no primordial non-Gaussianity $f_{\textrm{NL}}=g_{\textrm{NL}}=0.$ The difference of the total one-loop power spectrum $P_{RN}(k,\eta)$ respect to the linear power spectrum $P_L(k,\eta)$ lies in the large scales is due to the relativistic corrections, whereas the difference in the small scales is given purely by the Newtonian contributions.

\begin{figure}[htbp]
\centering
\includegraphics[width=150mm]{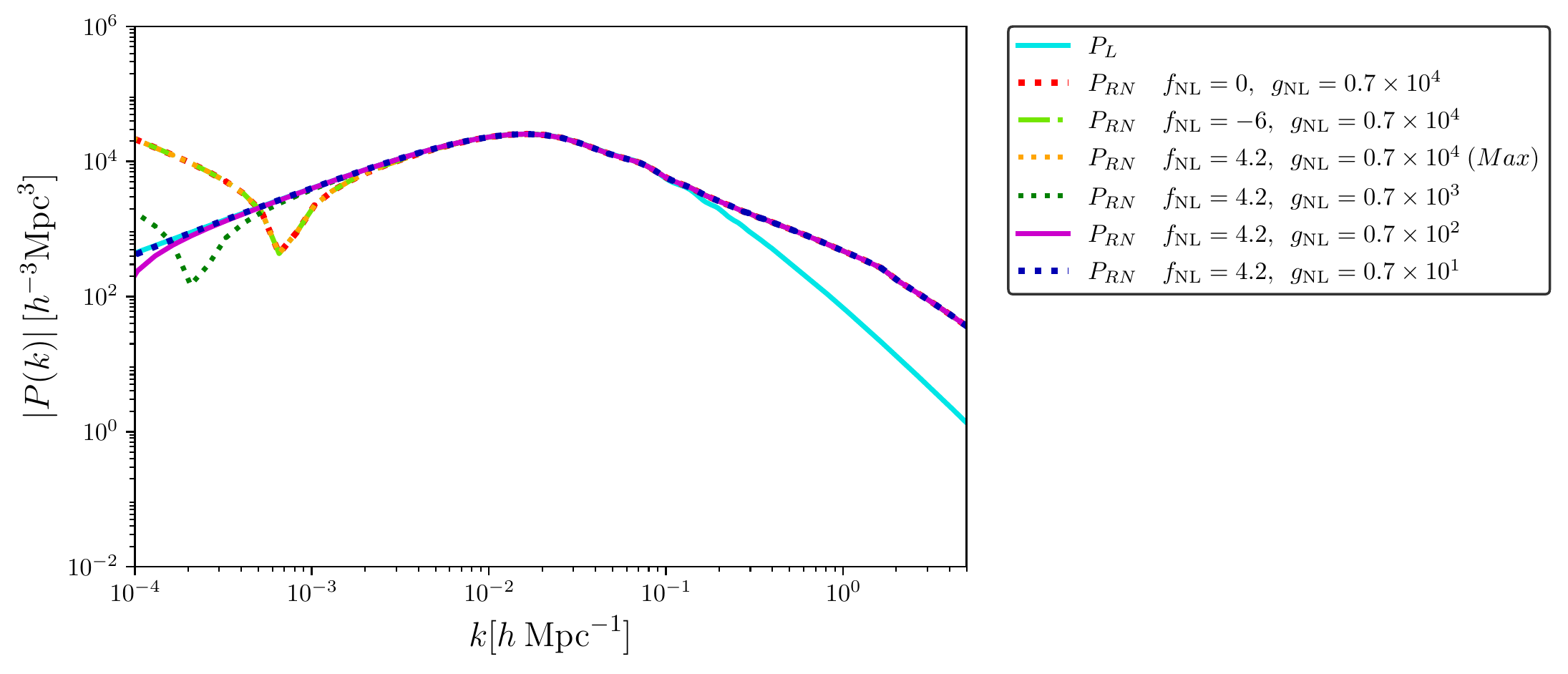}
\caption[Total one-loop power spectrum $P_{RN}$, at redshift $z=0$, for different values of  $g_{\textrm{NL}}$.]{Total one-loop power spectrum $P_{RN}$, at redshift $z=0$, for different values of  $g_{\textrm{NL}}$ reported by Planck \cite{2020A&A...641A...9P}. We observe negative contributions to the power spectrum for $g_{\mathrm{NL}}>7$.}
\label{Differentgnl}
\end{figure}

\begin{figure}[htbp]
\centering
\includegraphics[width=150mm]{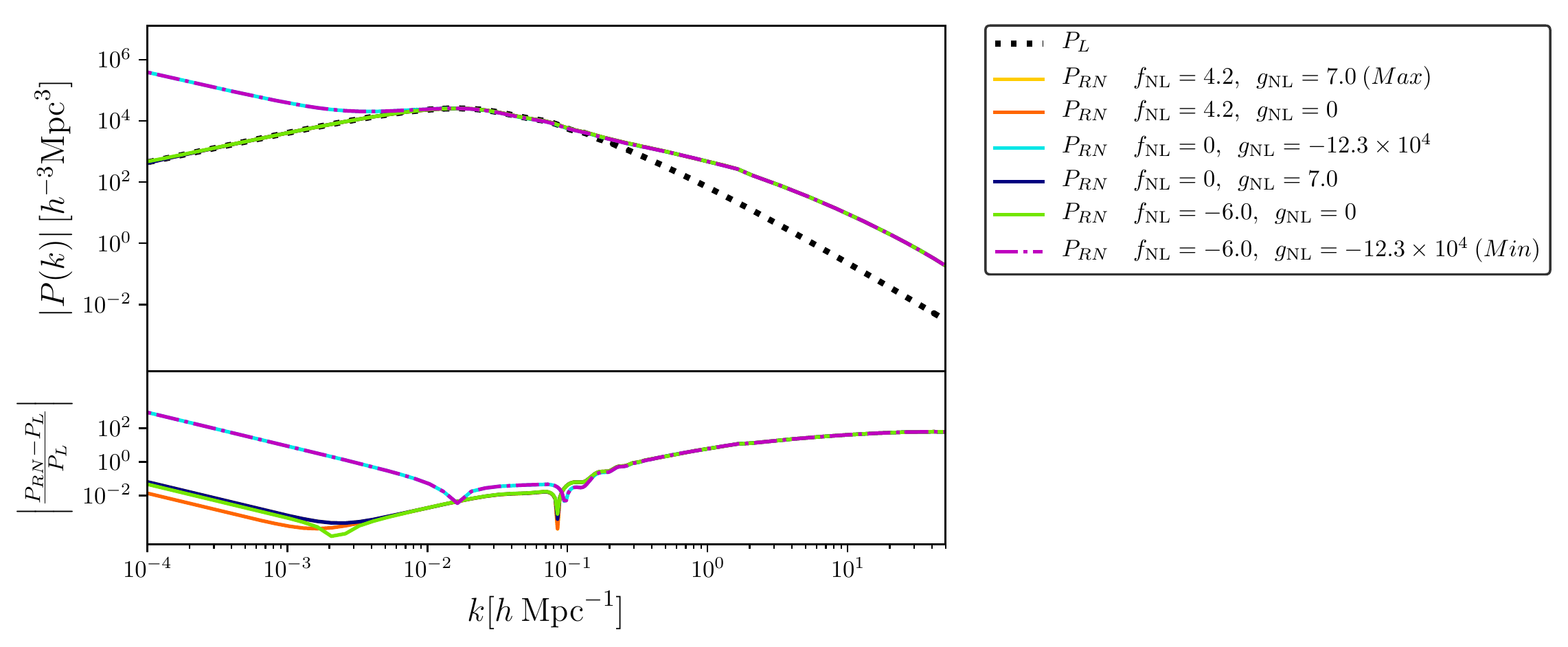}
\caption[Total one-loop power spectrum $P_{RN}$, at redshift $z=0$, for different limiting values of $f_{\textrm{NL}}$ and $g_{\textrm{NL}}$.]{Upper panel: Total one-loop power spectrum $P_{RN}$, at redshift $z=0$, for different limiting values of $f_{\textrm{NL}}$ and $g_{\textrm{NL}}$ reported by Planck \cite{2020A&A...641A...9P}. Note that the non-visible lines (yellow, orange and dark blue) are hidden behind the light green line as the have similar behaviour. Bottom panel: Relative difference of the relativistic one-loop power spectrum with respect to the linear power spectrum  normalised with the linear power spectrum.}
\label{Combinations}
\end{figure}

In Fig.~\ref{Combinations} we present the total one-loop power spectrum $P_{RN}(k,\eta)$, using different combinations of values of $f_{\textrm{NL}}$ and $g_{\textrm{NL}}$ reported in by the  Planck collaboration in Ref.~\cite{2020A&A...641A...9P}. The current constraints are given by $f^{\textrm{local}}_{\textrm{NL}}=-0.9\pm5.1$ and $g^{\textrm{local}}_{\textrm{NL}}=-5.8\pm6.5\times10^{4}$. For $f_{\textrm{NL}}$ we use the minimum and maximum values allowed by Planck i.e. $f_{\textrm{NL}}=-6.0$ and $f_{\textrm{NL}}=4.2$. In the case of $g_{\textrm{NL}}$, we use the minimum value allowed by Planck i.e.~$g_{\textrm{NL}}=-12.3\times10^4$, and the maximum value of $g_{\textrm{NL}}$ that we will use is $\sim 7$. Higher values for $g_{\textrm{NL}}$ are allowed by the results of the Planck collaboration \cite{2020A&A...641A...9P}, however might be in conflict with our perturbative expansion, giving negative contributions to the density power spectrum on large scales as shown in Fig.~\ref{Differentgnl} and it is not clear whether these features are physical. This does not mean that using these values for $g_{\textrm{NL}}$ is not allowed, however clarifying this issue  would require to calculate higher perturbative orders.

These values for $g_{\textrm{NL}}$ and $f_{\textrm{NL}}$ were chosen to show which values of $f_{\textrm{NL}}$ and $g_{\textrm{NL}}$ have a more significant contribution to the one-loop power spectrum. The relative difference with respect to the linear power spectrum shows that the largest corrections to the power spectrum in the large scales are present when $g_{\textrm{NL}}$ takes its minimum value, being this the dominant correction term as is not affected by the chosen value of $f_{\textrm{NL}}$. On the other hand, larger values of $g_{\textrm{NL}}$ present a similar behaviour for the different combinations with $f_{\textrm{NL}}$, having a small relative difference with respect to the linear power spectrum in comparison to the corrections given by minimum values of $g_{\textrm{NL}}$.

\begin{figure}[htbp]
\centering
\includegraphics[width=150mm]{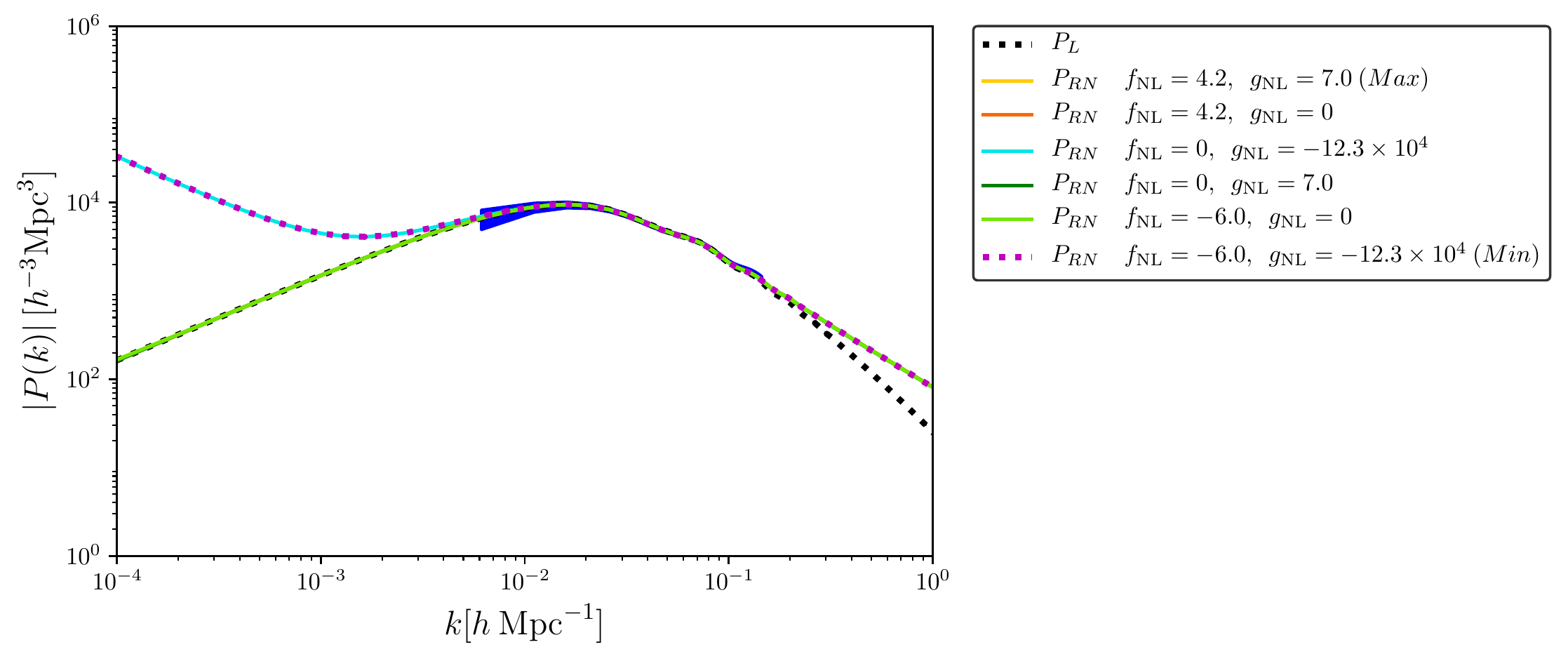}
\caption[Total one-loop power spectrum $P_{RN}$, at redshift $z=1$, for different limiting  values of $f_{\textrm{NL}}$ and $g_{\textrm{NL}}$.]{Total one-loop power spectrum $P_{RN}$, at redshift $z=1$, for different limiting  values of $f_{\textrm{NL}}$ and $g_{\textrm{NL}}$ reported by Planck \cite{2020A&A...641A...9P}. The blue shaded area corresponds to the measurement error of a typical Stage-IV-like survey redshift bin with $\Delta z = 0.2$, as detailed in the main text. We also used a $k$-binning $\Delta k = 0.006 \, h{\textrm{Mpc}^{-1}}$. Note that the non-visible lines (yellow, orange and dark green) are hidden behind the light green line as the have similar behaviour}
\label{Combinationsz1}
\end{figure}

In Fig.~\ref{Combinationsz1} we present the same set of total one-loop power spectrum $P_{RN}(k,\eta)$ plots as in Fig.~\ref{Combinations} but at a redshift $z=1$. In addition to the density power spectrum $P_{RN}(k,\eta)$ we also present in the blue shaded area the measurement errors assuming a cosmic variance limited Stage-IV galaxy survey like DESI \cite{2016arXiv161100036D}, Euclid \cite{2019arXiv191009273E}, or LSST \cite{2019ApJS..242....2C}. These errors are given by \cite{2008PhRvD..77l3514D}
\begin{equation}
\delta P_{gg}=\frac{1}{\sqrt{N_k}}\left[P_{gg}(k,z)+\frac{1}{\bar n_g}\right],
\label{error}
\end{equation}

\noindent where $P_{gg}(k,z)$ is the galaxy power spectrum, $\bar n_g$ is the mean number density and the term $1/\bar n_g$ is the shot noise, in our case we consider an idealised case, thus the shot noise is set to zero. The factor $N_k$ is the number of Fourier modes in a survey of volume $V_s$ is expressed as
\begin{equation}
N_k=\frac{k^2\Delta kV_s}{4\pi^2}.
\end{equation}
\noindent The galaxy power spectrum in this case is modelled as 
\begin{equation}
P_{gg}(k,z)=b^2D^2(z)P_{L}(k),
\end{equation} 
\noindent where $D(z)$ is the growth factor, and  $b$ is a linear bias parameter, following the parametrisation as\cite{2020A&A...642A.191E,2008arXiv0810.0003R}\footnote{Note that we calculated $P(k)$ not $P_{gg}(k)$, thus in order to compare $P(k)$ and the measurement errors in Fig.~\ref{Combinationsz1}, we rescale $P_{gg}(k)$ to remove the bias. A more complete treatment including the galaxy bias is presented in the next chapter.} 
\begin{equation}
b=\sqrt{1+z}.
\end{equation}

The Fisher matrix $F_{ij}$, which provides with the best possible error that we expect to achieve, is related to the errors as \cite{1997PhRvL..79.3806T,2007ApJ...665...14S,2011PhRvD..83j3527T}
\begin{equation}
F_{ij}=\int_{k_{min}}^{k_{max}}\frac{\partial P_{gg}}{\partial p_i}\frac{\partial P_{gg}}{\partial p_j}\left[P_{gg}+\frac{1}{\bar n_g}\right]^{-2}\frac{V_sk^2}{4\pi^2}dk,
\label{fisher}
\end{equation}
where $p_i$ are the parameters for which we want to forecast errors.

More specifically, we have assumed a sky area of $15,000 \, {\textrm{deg}}^2$ at $z=1$ with bin width $\Delta z = 0.2$. These numbers correspond to typical specifications of such surveys used in recent forecast and model validation studies at $z=1$ (see e.g.~Ref.~\cite{2019OJAp....2E..13M}). Note however that the measurement errors would decrease if we chose a wider redshift bin given the large \emph{total} redshift coverage of Stage-IV surveys. Similarly, we have defined the largest measurable scale as $k_{\textrm{min}} \simeq 2\pi / V^{1/3}_{\textrm{bin}} = 0.003 \, h{\textrm{Mpc}^{-1}}$, where $V_{\textrm{bin}}$ is the volume corresponding to $\Delta z = 0.2$; this volume would increase if we were to consider a wider redshift bin, allowing us to reach larger scales. The minimum values of $f_{\textrm{NL}}$ and $g_{\textrm{NL}}$ show the largest impact at the largest measured scales of the upcoming experiments, forecasting a detectability of PNG for values of $g_{\textrm{NL}}$ or $f_{\textrm{NL}}$.

\section{Tree-level bispectrum}
\label{sec:Treebis}
For completeness we calculate the tree-level bispectrum, which is  defined as 
\begin{equation}
B(k_1,k_2,k_3,\eta)\equiv2P_L(k_1,\eta)P_L(k_2,\eta)\mathcal{F}^{(2)}(\mathbf k_1,\mathbf k_2,\eta)+\textrm{(2\enspace cyclic)},
\label{bispe}
\end{equation}
\noindent the components to calculate the bispectrum at tree-level are given in  Eq.~\eqref{Kernel2N} and Eq.~\eqref{Kernel2R}.
We define the Newtonian tree-level bispectrum $B_{NN}(k_1,k_2,k_3,\eta)$ as
\begin{equation}
B_{NN}(k_1,k_2,k_3,\eta)\equiv2P_L(k_1,\eta)P_L(k_2,\eta)\mathcal{F}_{N}^{(2)}(\mathbf k_1,\mathbf k_2,\eta)+\textrm{(2\enspace cyclic)},
\end{equation}

\noindent and the  relativistic tree-level bispectrum as
\begin{equation}
B_{RR}(k_1,k_2,k_3,\eta)\equiv2P_L(k_1,\eta)P_L(k_2,\eta)\mathcal{F}_{R}^{(2)}(\mathbf k_1,\mathbf k_2,\eta)+\textrm{(2\enspace cyclic)},
\end{equation}

the total tree-level bispectrum $B_{RN}(k_1,k_2,k_3,\eta)$,  is defined by Eq.~\eqref{bispe}, where $\mathcal{F}^{(2)}(\mathbf{k}_1,\mathbf{k}_2,\eta)$ is given in Eq.~\eqref{Kernels2}.

\begin{figure}[htbp]
\centering
\includegraphics[width=100mm]{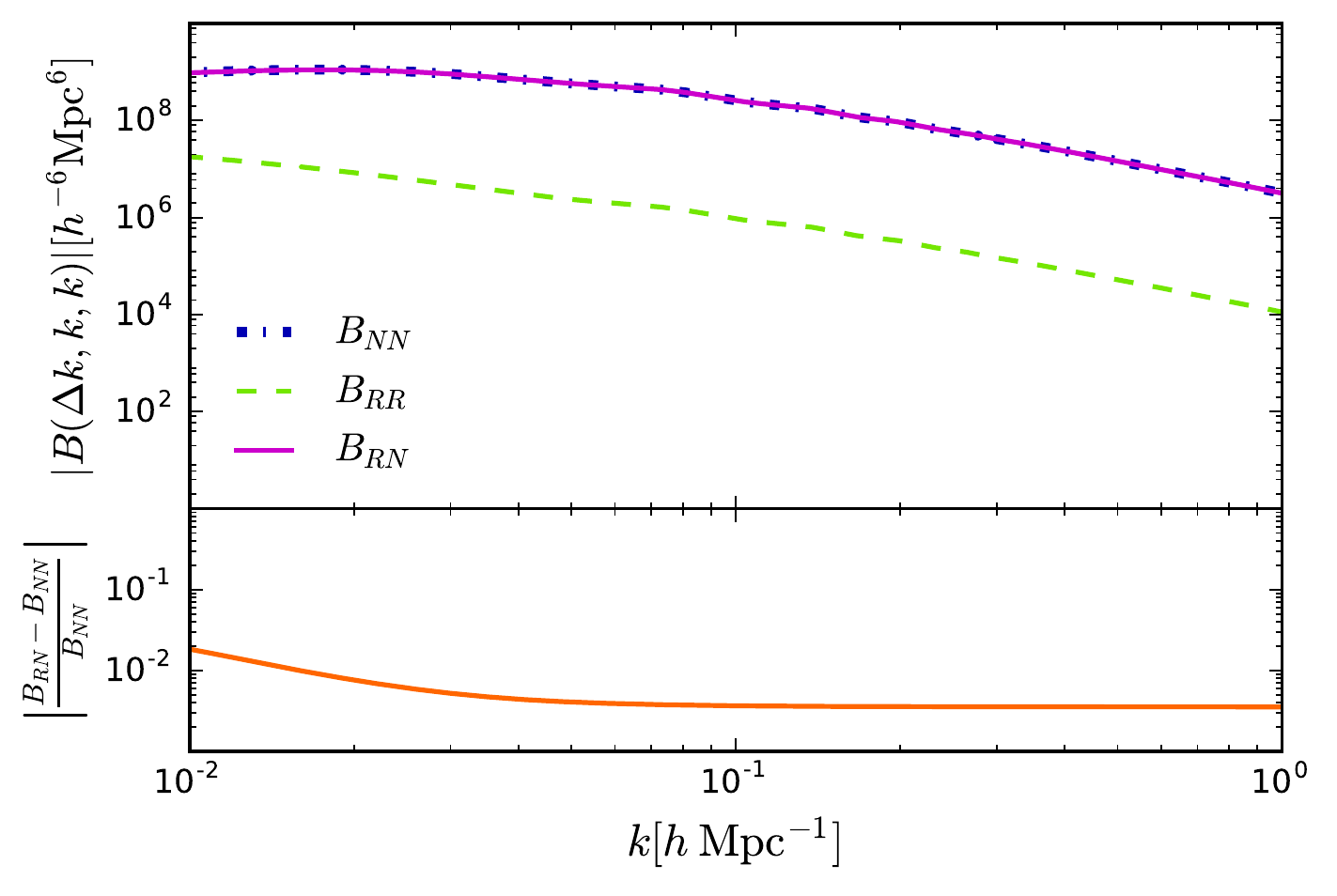}
\caption[Comparison of the Newtonian, relativistic and total tree-level bispectrum corrections in the squeezed limit for $f_{\textrm{NL}}=0$ at redshift $z=0$.]{Upper panel: Comparison of the Newtonian, relativistic and total tree-level bispectrum corrections in the squeezed limit with $\Delta k=0.013\:h\textrm{Mpc}^{-1}$ for $f_{\textrm{NL}}=0$ at redshift $z=0$. Bottom panel: Relative difference of the total tree-level bispectrum with respect to the Newtonian tree-level bispectrum normalised with the Newtonian tree-level.}
\label{Bi1}
\end{figure}

\begin{figure}[htbp]
\centering
\includegraphics[width=100mm]{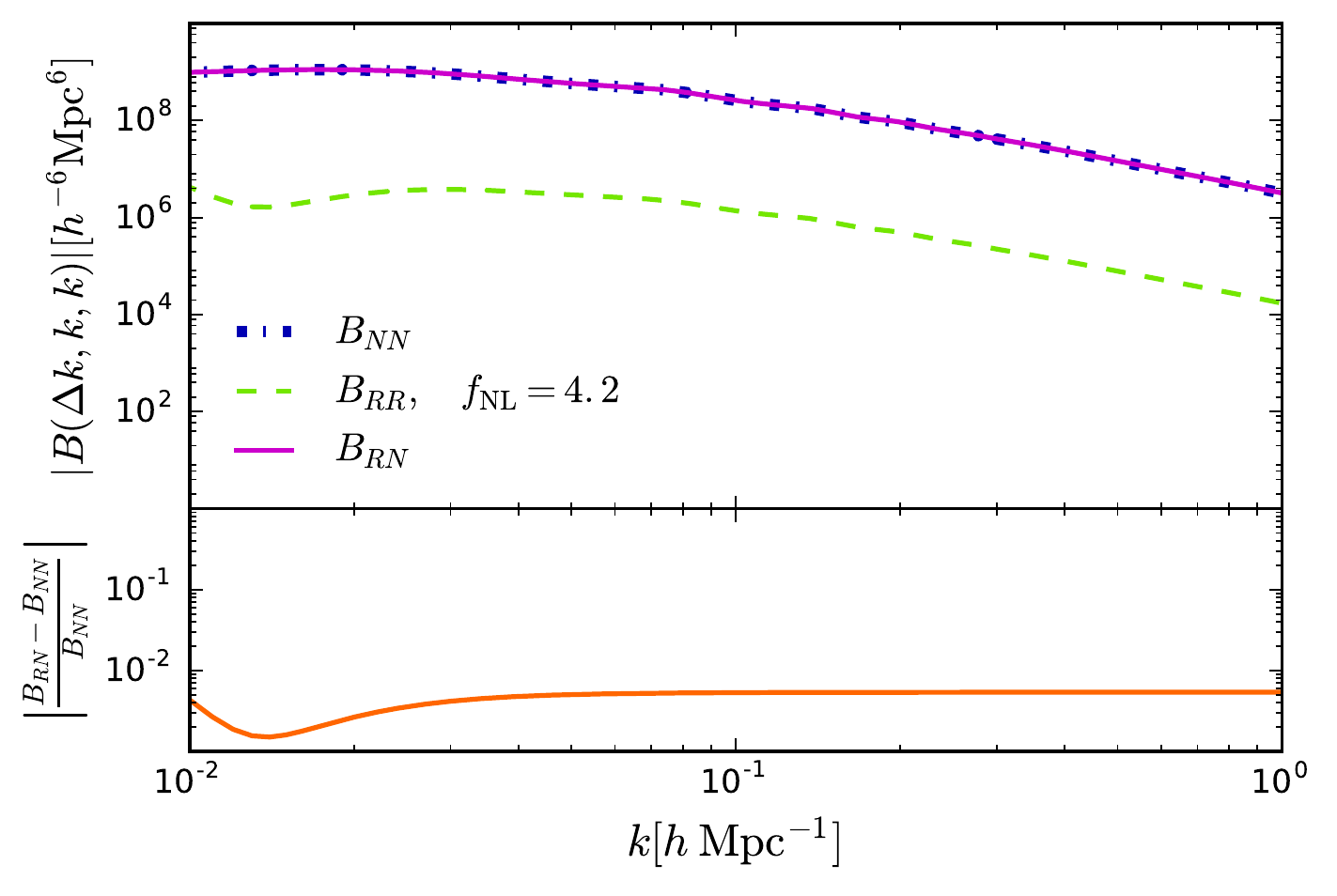}
\caption[Comparison of the Newtonian, relativistic and total tree-level bispectrum corrections in the squeezed limit for $f_{\textrm{NL}}=4.2$ at redshift $z=0$.]{Upper panel: Comparison of the Newtonian, relativistic and total tree-level bispectrum corrections in the squeezed limit with $\Delta k=0.013\:h\textrm{Mpc}^{-1}$ for $f_{\textrm{NL}}=4.2$ at redshift $z=0$. Bottom panel: Relative difference of the total tree-level bispectrum with respect to the Newtonian tree-level bispectrum normalised with the Newtonian tree-level.}
\label{Bi2}
\end{figure}

\begin{figure}[htbp]
\centering
\includegraphics[width=100mm]{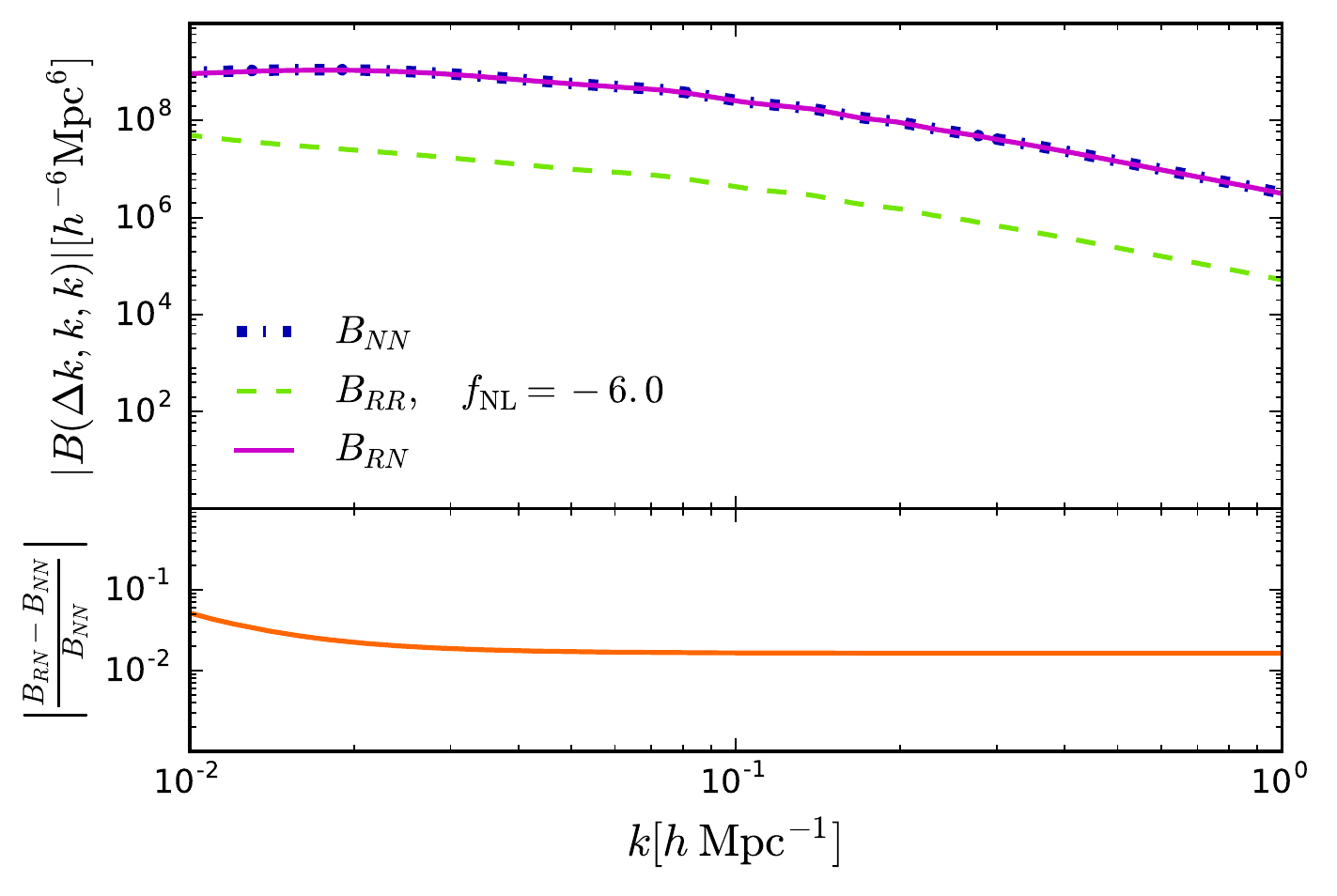}
\caption[Comparison of the Newtonian, relativistic and total tree-level bispectrum corrections in the squeezed limit for $f_{\textrm{NL}}=-6.0$ at redshift $z=0$.]{Upper panel: Comparison of the Newtonian, relativistic and total tree-level bispectrum corrections in the squeezed limit with $\Delta k=0.013\:h\textrm{Mpc}^{-1}$ for $f_{\textrm{NL}}=-6.0$ at redshift $z=0$. Bottom panel: Relative difference of the total tree-level bispectrum with respect to the Newtonian tree-level bispectrum normalised with the Newtonian tree-level.}
\label{Bi3}
\end{figure}

In Figures \ref{Bi1}, \ref{Bi2} and \ref{Bi3} we present a comparison of the Newtonian tree-level bispectrum $B_{NN}(k_1,k_2,k_3,\eta)$, the  relativistic tree-level bispectrum given by $B_{RR}(k_1,k_2,k_3,\eta)$ and the total tree-level bispectrum $B_{RN}(k_1,k_2,k_3,\eta)$, all in the squeezed limit, with $\Delta k=0.013\:h\textrm{Mpc}^{-1}$ when $f_{\textrm{NL}}=0$  and for the limiting values of $f_{\textrm{NL}}$ given by Ref.~\cite{2020A&A...641A...9P}, the relative difference of the total tree-level bispectrum  with respect to the Newtonian bispectrum is shown in the bottom panels. The relativistic corrections at this level are subdominant with respect to the Newtonian tree-level bispectrum.

\section{Discussion}
\label{sec:conclusions}

We calculated purely general relativistic corrections to the density power spectrum at one-loop. For the synchronous-comoving gauge the primordial non-Gaussianity of the local type can be added naturally and we have also computed the contribution of these parameters. 
The modifications that relativistic contributions bring to the density power spectrum are below  $0.01\%$ except at very large scales  where we find a $1\%$ pure relativistic contribution (see Fig.~\ref{GROne}). 
On the other hand, the primordial non-Gaussianity values allowed by the latest Cosmic Microwave Background observations in the local configuration yield significant contributions mostly from the $g_{\textrm{NL}}$ parameter (see Fig.~\ref{Combinationsz1}), since $g_{\textrm{NL}}$ is not highly constrained, opposite to $f_{\textrm{NL}}$ for which we have tighter constrains.

The relativistic terms contributing to the higher order amplitude of the density contrast have been derived from a long-wavelength approximation and do not account for effects at all scales. However, it is expected that at small scales the weak field and therefore the Newtonian regime describe best the matter structure. As mentioned above, it is precisely at the large scales where primordial non-Gaussianity contributes to the density power spectrum. Therefore, the formalism employed here to derive relativistic contributions is naturally extended to include the dominant PNG contributions to the density contrast and its polispectra.

The actual corrections from General Relativity to the non-linear bias of galaxies seem to remove the General Relativity effects presented here, however the form of the volume distortions reproduce the form of those expressed in our non-linear prescriptions for the density contrast---at least at second order \cite{2019JCAP...05..020U} but decoupling large and short scales---.  Yet the local primordial non-Gaussianity terms cannot be removed by local coordinate transformations \cite{2019JCAP...12..048U}, and terms with such factors are precisely what dominates the signal in the one-loop spectra. It is also important to mention that the GR effects removed by the coordinate transformation are carried at second order and the third order effects described here might survive the coordinate changes. The details of adapting the expressions coming from the volume distortions as galaxies evolve from an initial time, and the precise consequences for a non-linear galaxy power spectrum are left for future work.

Our results show that pure relativistic corrections $P_{RR}(k,\eta)$ have a too small contribution at too large scales to be observed in the present or future large scale structure probes. On the other hand, the primordial non-Gaussianity contributions, corresponding to values within the 1-$\sigma$ amplitudes of $g_{\textrm{NL}}$ allowed by Planck \cite{2020A&A...641A...9P}, yield a significant contribution to $\delta^{(3)}$, and to the one-loop power spectrum observable in the next generation of galaxy surveys. While the deviations from the linear prescription lie within the cosmic variance errors, it may be possible to probe these values through cross-correlations of the future surveys with the measurements of anisotropies in the Cosmic Microwave Background. We shall explore the implications of this effect in order to constrain primordial non-Gaussianity through this and other methods in a future work. 

\chapter{Contributions from primordial non-Gaussianity and General Relativity to the galaxy power spectrum}
\chaptermark{Contributions from PNG and GR to the galaxy power spectrum}
\label{paper2}

\section{Introduction}

The work presented in this chapter is based on the paper in Ref.~\cite{2021arXiv210710815M}.
In this chapter we extend our work from chapter~\ref{chapter:Relativistic}, by not only including relativistic corrections to the power spectrum at one-loop and the input from primordial non-Gaussianity but also  introducing a bias prescription guided by the parametrization in Refs.~\cite{2006PhRvD..74j3512M, 2008PhRvD..78l3519M}, and adopt a set of two parameters, a linear $b_{\delta}$ and a non-linear parameter $b_{\mathrm{NL}}$ to compute the galaxy power spectrum in the synchronous-comoving gauge.\footnote{The synchronous-comoving gauge is an adequate gauge choice to express the Lagrangian frame and the simplest to specify a galaxy bias at second order \cite{2015CQGra..32q5019B}.} We calculate the source galaxy power spectrum (for brevity we omit the ``source'' term in rest of the chapter) for a range of reasonable values of the primordial non-Gaussianity and bias parameters and compare results with the forecasted measurements from Stage-IV experiments, specifically from a $15,000$ $\mathrm{deg^2}$ and a $40,000$ $\mathrm{deg^2}$ (all-sky) galaxy survey. 

In addition, we show that suitable values of the bias parameters  constitute an effective strategy to renormalize the (divergent) relativistic contributions at large scales without affecting the dominant primordial non-Gaussianity effect.

This chapter is structured as follows: in section~\ref{galaxy-galaxy-section} we introduce the bias model and the mathematical expression for the galaxy power spectrum. In section~\ref{Results-section} we present our results in a series of plots of the galaxy power spectrum with a range of values for the primordial non-Gaussianity parameters and for the non-linear bias.
 The spectra include the observational uncertainties from Stage-IV galaxy surveys. We also give a combination of bias parameter values which serve as an effective renormalization of the relativistic contributions, which are otherwise divergent at large scales. Finally in section~\ref{Discussion-section} we discuss our results.
\section{Galaxy power spectrum}
\label{galaxy-galaxy-section}
\subsection{Bias model}
To compute the galaxy power spectrum we follow the renormalized perturbative bias model of Refs.~\cite{2006PhRvD..74j3512M, 2008PhRvD..78l3519M}. This approach assumes the Taylor expansion of a general function of the galaxy density contrast, given by
\begin{equation}
\delta_g=c_\delta\delta+\frac{1}{2}c_{\delta^2}(\delta^2-\sigma^2)+\frac{1}{3!}c_{\delta^3}\delta^3+ \epsilon+\mathcal{O}(\delta^4),
\label{MCdensity}
\end{equation}
where $\sigma^2=\langle\delta^2\rangle$ is the variance of $\delta$, that ensures $\langle\delta_g\rangle=0$. The coefficients of the Taylor expansion $c_{\delta^n}$ constitute the bias parameters, while $\epsilon$ is a random noise variable that allows for stochasticity, which means that the relation between $\delta_g$ and $\delta$ is not deterministic, and we need to allow for some noise, e.g.~from shot noise. This noise appears as white noise in the large scales\cite{1999ApJ...520...24D}, and it is important to model it, since it has a contribution to the measurement errors of the statistics. The variance of $\epsilon$ is given as $\langle\epsilon^2\rangle=N_0$, we assume this noise is uncorrelated with the density fluctuations $\langle\epsilon\delta\rangle=0$ and also $\langle\epsilon\rangle=0$ \cite{2010PhDT.........4J}. 

In this model large-higher order perturbative corrections are eliminated through the redefinition of the bias parameters, unlike the approach where the galaxy density is defined with a function of a smoothed mass-density field, in order to be consistent with a Taylor expansion (see e.g.~Refs.~\cite{1998MNRAS.301..797H,1993ApJ...413..447F}), the renormalization bias approach avoids an arbitrary modification of the large scales and allows to use a small number of bias parameters in the description. In this work we focus on large scales and we do not take into account small scales effects like baryonic effects. This prescription was originally defined for matter density in the Eulerian frame, though we adopt it for the Lagrangian density expressed earlier. Our parameters should thus be interpreted as Lagrangian variables.\footnote{The Lagrangian bias parameters should coincide with the Eulerian set at large scales, because the coordinate change to the Eulerian frame alters quantities at small scales only \cite{2014ApJ...785....2B,2015CQGra..32q5019B}.}

If we expand the density contrast in perturbative orders up to the leading non-linear contribution to the power spectrum, then Eq.~\eqref{MCdensity} is expanded as
\begin{equation}
\delta_g=c_\delta\left(\delta^{(1)}+\frac{\delta^{(2)}}{2}+\frac{\delta^{(3)}}{6}\right)+\frac{c_{\delta^2}}{2}\left(\delta^{(1)2}+\delta^{(1)}\delta^{(2)}\right)+...,
\label{galaxy-density}
\end{equation}
where $\delta^{(1)}$, $\delta^{(2)}$ and $\delta^{(3)}$ are defined in Eqs.~\eqref{delta1N}, \eqref{Reldelta2} and \eqref{Reldelta3} respectively.

\subsection{Relativistic galaxy power spectrum }
Analogously to the definition of the power spectrum for the density contrast in Eq.~\eqref{powerspectdef}, the galaxy power spectrum is defined as
\begin{equation}
\langle\delta_g(k,\eta)\delta_g(k',\eta)\rangle=(2\pi)^3P_{gg}(k,\eta)\delta_D(k-k'),
\label{galaxypower}
\end{equation}
where $k$ is the comoving wavenumber in Fourier space. Since $\delta^{(1)}$ is Gaussian, the non-vanishing leading order non-linear corrections to the galaxy power spectrum are of order $\delta^{(4)}$. Using Eq.~\eqref{galaxy-density} and Eq.~\eqref{galaxypower}, the galaxy power spectrum $P_{gg}(k,\eta)$ is given by
\begin{align}
\nonumber
\label{eq:g-spectrum1}
P_{gg}(k,\eta)=&(c_\delta)^2\left[P_L(k,\eta)+2P_R^{(1,3)}(k,\eta)+P_R^{(2,2)}(k,\eta)\right]\\
+&(2c_{\delta}c_{\delta^2})\left[P_{R1}(k,\eta)+P_{R2}(k,\eta)\right],
\end{align}
where $P_L(k,\eta)$ represents the linear power spectrum, $P_R^{(1,3)}(k,\eta)$ and $P_R^{(2,2)}(k,\eta)$ are the relativistic contributions to the one-loop matter power spectrum in Eqs.~\eqref{integral22} and \eqref{integral13}, and are given by 
\begin{align}
\nonumber
P_R^{(1,3)}(k,\eta)=&\frac{k^3}{(2\pi)^2}P_L(k,\eta)\int_{0}^{\infty} drP_L(kr,\eta)\Bigg\{81\left(\frac{\mathcal{H}^4}{k^4}\right)
\bigg[\Big(-g_{\mathrm{NL}}+\frac{5}{9}f_{\mathrm{NL}}+\frac{25}{54}\Big)\\
&+\frac{1+2r^2}{6r^2}\Big(g_{\mathrm{NL}}-\frac{10}{3}f_{\mathrm{NL}}+\frac{50}{27}\Big) \bigg]\Bigg\},\\\nonumber
P_R^{(2,2)}(k,\eta)=&\frac{k^3}{2\pi^2}\int_{0}^{\infty}dr  P_L(kr,\eta)\int_{-1}^{1}dxP_L(k\sqrt{1+r^2-2rx},\eta)\\
  &\times\Bigg\{\left(\frac{{\mathcal{H}^4}}{k^4}\right)\bigg[\frac{6f_{\mathrm {NL}}-10-25r^2+25rx}{4r(1-2rx+r^2)}\bigg]^2\Bigg\}.
\end{align}
The  contributions $P_{R1}(k,\eta)$ and $P_{R2}(k,\eta)$, which are not present in the one-loop matter spectrum due to the bias expansion that we are using for $\delta_g$ in Eq.~\eqref{galaxy-density}, which includes more non-linear terms, these are given by
\begin{align}
\nonumber
P_{R1}(k,\eta)=&\frac{k^3}{(2\pi)^2}P_L(k,\eta)\int_{0}^{\infty} drP_L(kr,\eta)\\
&\times\int_{-1}^{1} dx\left(\frac{\mathcal{H}^2}{k^2}\right)\left[3f_{\mathrm{NL}}(1+r^2+2rx)+\frac{5}{6}(3rx-6+6r^2)\right],\\
\nonumber
P_{R2}(k,\eta)=&\frac{k^3}{(2\pi)^2}\int_{0}^{\infty} drP_L(kr,\eta)\int_{-1}^{1} dxP_L(k\sqrt{1+r^2-2rx},\eta)\\
&\times\left(\frac{\mathcal{H}^2}{k^2}\right)\left[\frac{6f_{\mathrm{NL}}-10-25r^2+25rx}{4(1-2rx+r^2)}\right].
\end{align}

We now re-express the bias factors in terms of the linear and non-linear bias parameters for $P_{gg}(k,\eta)$ in terms of the $c_{\delta}$ and $c_{\delta^2}$. As mentioned above, we are following the renormalized perturbative bias model of Refs.~\cite{2006PhRvD..74j3512M, 2008PhRvD..78l3519M}, where the author redefines the bias parameters $c_{\delta^n}$ in such a way that divergences are removed, managing to group together terms that have the same type of contribution to the galaxy power spectrum. At leading order the correspondence is straightforward and is reduced to parameters
\begin{align}
b_{\delta}=&c_{\delta}\,,\quad
b_{\mathrm{NL}}=2c_{\delta^2}\,,
\end{align}

\noindent and we get
\begin{align}
\nonumber
\label{eq:g-spectrum2}
P_{gg}(k,\eta)=&(b_\delta)^2\left(P_L(k,\eta)+2P_R^{(1,3)}(k,\eta)+P_R^{(2,2)}(k,\eta)\right)\\
+&(b_{\delta}b_{\mathrm{NL}})\Big(P_{R1}(k,\eta)+P_{R2}(k,
\eta)\Big).
\end{align}

The  bias parameter $b_{\mathrm{NL}}$ corresponds to $b_{\phi}$ introduced in Ref.~\cite{2008PhRvD..78l3519M}, the latter accounts for the dominant contributions from primordial non-Gaussianity. We use a different label to emphasise that $b_{\mathrm{NL}}$ includes primordial non-Gaussianity as well as relativistic terms. Note, that the terms multiplied by $b_{\mathrm{NL}}$ include those of the \textit{scale-dependent bias}, which is the focus of previous studies (see e.g.~Refs.~\cite{2008PhRvD..77l3514D,2012PhRvD..85d1301B}). 

As can be seen from Eq.~\eqref{eq:g-spectrum2}, with the parametrization chosen above, the one-loop matter power spectrum is included in the terms multiplied by the square of the linear bias, $(b_\delta)^2$. This is consistent with the fact that the relativistic solutions are part of the linear correspondence between matter density and spatial curvature (the lowest order solution in the gradient expansion).  This characteristic feature of General Relativity and primordial non-Gaussianity plays a crucial role in the bias parameter fitting as we will see below. 
 
\section{Results}
\label{Results-section}

In this section we present the galaxy power spectrum for a set of values of interest of the galaxy bias parameters. This is best appreciated in plots of the galaxy power spectrum itself. All our calculations were performed numerically in Python, using as an input a linear power spectrum generated with the \texttt{CLASS} Boltzmann solver \cite{2011arXiv1104.2932L,2011JCAP...07..034B}, for realisations of the $\Lambda$CDM cosmology, taking parameter values from  the Planck collaboration results \cite{2016A&A...594A..13P}.

In all our plots of the galaxy power spectrum, the value for the linear bias parameter is fixed to $b_{\delta}=1.41$ in a survey bin with mean redshift $z=1$, following the parametrisation \cite{2020A&A...642A.191E,2008arXiv0810.0003R}
\begin{equation}
b_\delta=\sqrt{1+z} \, .
\label{def_lin_bias}
\end{equation}

\begin{figure}[tbp]
\centering
\includegraphics[width=145mm]{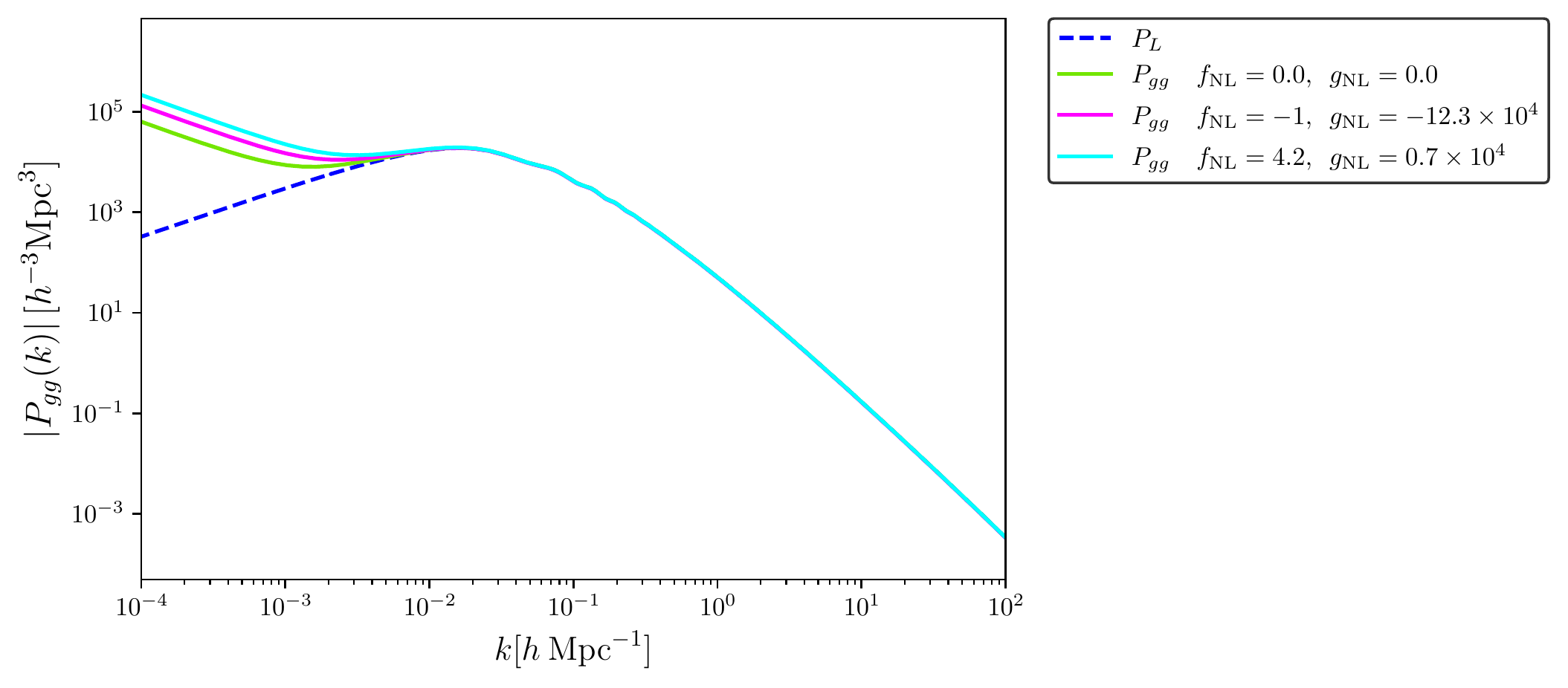}
\caption[Galaxy power spectrum in a wide scale range, at redshift $z=1$, $b_\delta=1.41$ and $b_{\mathrm{NL}}=0.2$, for different limiting  values of $f_{\mathrm {NL}}$ and $g_{\mathrm{NL}}$.]{Galaxy power spectrum in a wide scale range, at redshift $z=1$, $b_\delta=1.41$ and $b_{\mathrm{NL}}=0.2$, for different limiting  values of $f_{\mathrm {NL}}$ and $g_{\mathrm{NL}}$ reported by Planck \cite{2020A&A...641A...9P}. No divergence is observed in any of the cases at small scales (large $k$-modes). The divergences at the other end are discussed in section~\ref{sec:renorma}.}
\label{allscales}
\end{figure}

First,  we show that there is no divergence of the non-linear contributions at small scales. We plot in Fig.~\ref{allscales} the galaxy power spectrum of Eq.~\eqref{eq:g-spectrum2} at $z=1$. This includes the non-Gaussianity contributions with the limiting values of parameters $f_{\mathrm {NL}}$ and $g_{\mathrm {NL}}$ reported by the Planck collaboration \cite{2020A&A...641A...9P} (the value of $b_{\mathrm{NL}}$ for this Figure is arbitrary within the perturbative expansion hierarchy).

\begin{figure}[tbp]
\centering
\includegraphics[width=145mm]{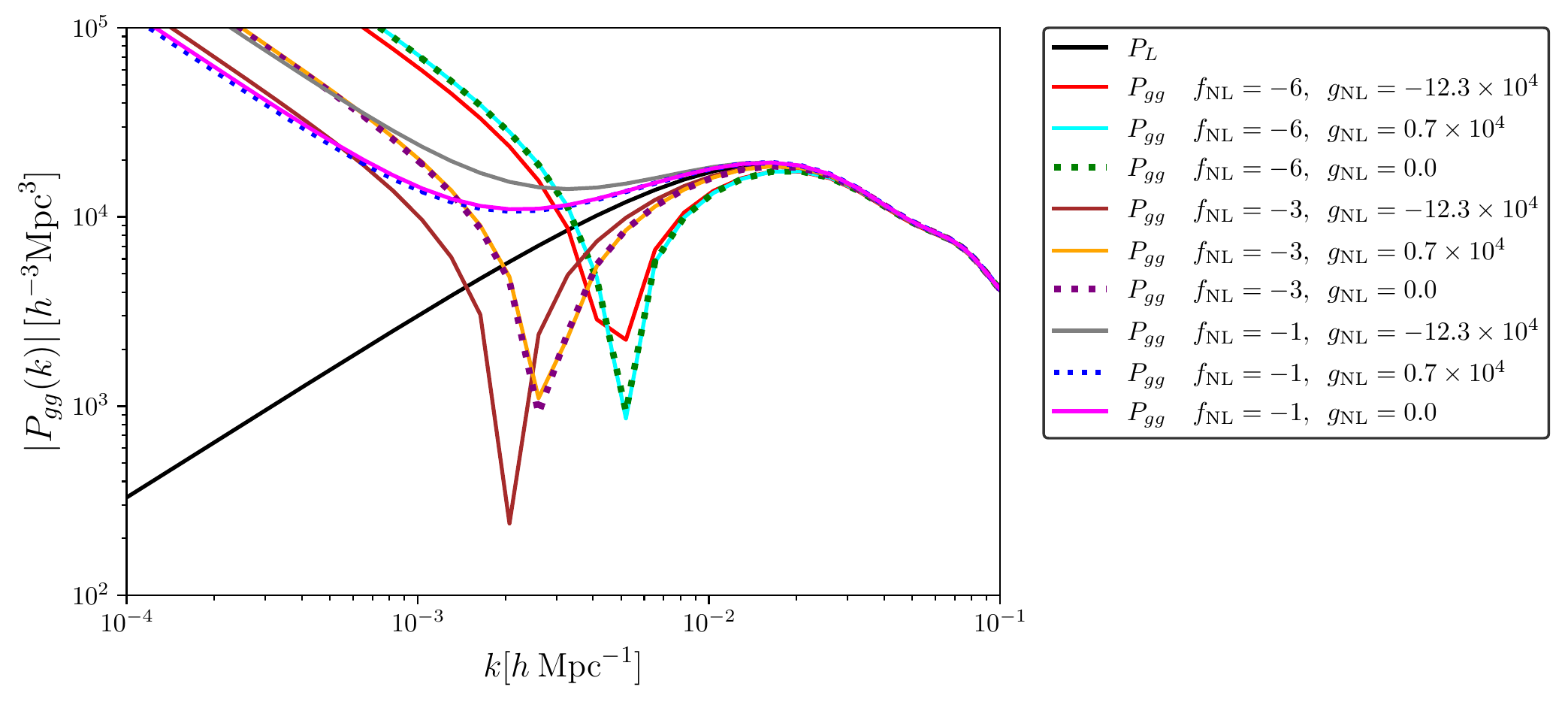}
\caption[Galaxy power spectrum, at redshift $z=1$, $b_\delta=1.41$ and $b_{\mathrm{NL}}=1.0$, for different limiting  values of $f_{\mathrm {NL}}$ and $g_{\mathrm{NL}}$.]{Galaxy power spectrum, at redshift $z=1$, $b_\delta=1.41$ and $b_{\mathrm{NL}}=1.0$, for different limiting  values of $f_{\mathrm {NL}}$ and $g_{\mathrm{NL}}$ reported by Planck \cite{2020A&A...641A...9P}. We observe negative contributions to the galaxy power spectrum for $f_{\mathrm{NL}}<-1$.}
\label{Pggfnlcompare}
\end{figure}

Let us stress at this point, a technical but important issue, relevant for all plots in this chapter. We show throughout the minimum value of $f_{\mathrm{NL}}=-1$, since smaller values yield dominant (negative) contributions to the matter power spectrum on large scales as shown in Fig.~\ref{Pggfnlcompare}. In that sense, in the non-linear contributions to the matter power spectrum presented in chapter~\ref{chapter:Relativistic}  we discarded the possibility of using  values of $g_{\mathrm{NL}}>7$ (see Fig. \ref{Differentgnl}). We note, however, that such a restriction is not necessary in the galaxy power spectrum in general. This is because the two new contributions $P_{R1}(k,\eta)$ and $P_{R2}(k,\eta)$, balance the original one-loop terms as dictated by the values of the $b_{\mathrm{NL}}$ parameter as exemplified below. This shows that after taking into account more corrections, it is possible to allow more values of $g_{\mathrm{NL}}$ and we expect something similar for $f_{
\mathrm{NL}}$ after accounting more contributions to the galaxy power spectrum, however this is left to future work.

\begin{figure}[htbp]
\centering
\includegraphics[width=130mm]{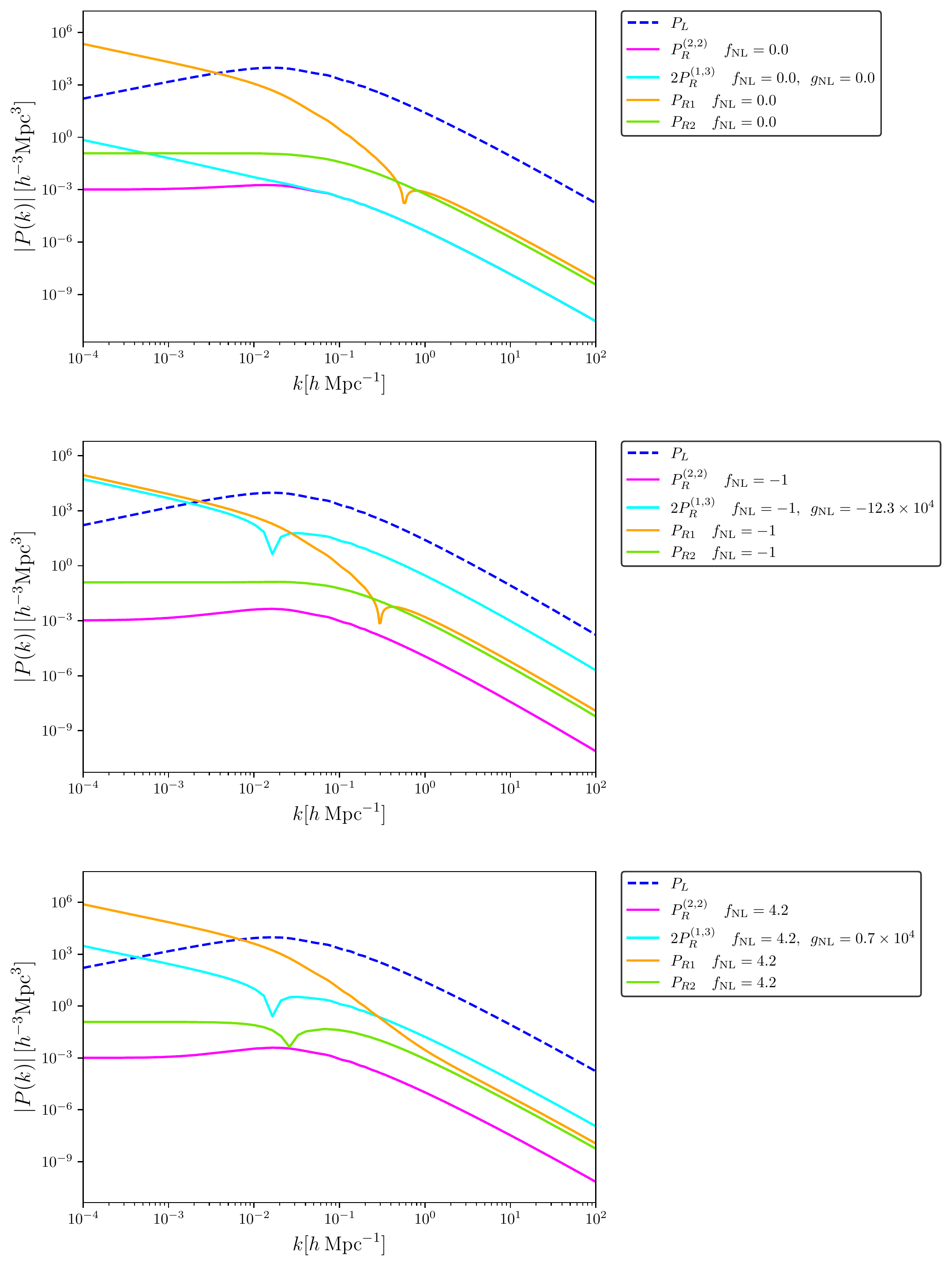} 
  
  \caption[Contributions from each term to the galaxy power spectrum at redshift $z=1$, in Eq.~\eqref{eq:g-spectrum2}, for Gaussian initial conditions and for the limiting values of $f_{\mathrm{NL}}$ and $g_{\mathrm{NL}}$.]{Contributions from each term to the galaxy power spectrum at redshift $z=1$, in Eq.~\eqref{eq:g-spectrum2}, for Gaussian initial conditions (upper plot), and for the limiting values of $f_{\mathrm{NL}}$ and $g_{\mathrm{NL}}$ reported by Planck \cite{2020A&A...641A...9P} (middle and lower plot).}
  \label{separate-contributions}
\end{figure}

In Fig.~\ref{separate-contributions} we plot separately the contributions to the galaxy power spectrum by each term in Eq.~\eqref{eq:g-spectrum2} (setting both bias parameters to unity). The dominant contribution at large scales comes from $P_{R1}(k,\eta)$, an additional term to the one-loop matter spectrum,  followed by $P^{(1,3)}_R(k,\eta)$, which is the only term with $g_{\mathrm{NL}}$ dependence.

 Using Eq.~\eqref{def_lin_bias} for $b_\delta$, in the rest of this section we present the galaxy power spectrum for values for $b_{\mathrm{NL}}$ following two criteria: First we compute values for which the spectrum deviates from the linear prescription beyond the uncertainty in current and future surveys, and thus show observable relativistic or primordial non-Gaussianity contributions, and subsequently we compute values which cancel the divergent part of the relativistic contribution at large scales, thus showing observable features at large scales only in the presence of primordial non-Gaussianity.  

\subsection{Viable bias parameter values}
\label{sec:results1}

\begin{figure}[htbp]
\centering
\includegraphics[width=140mm]{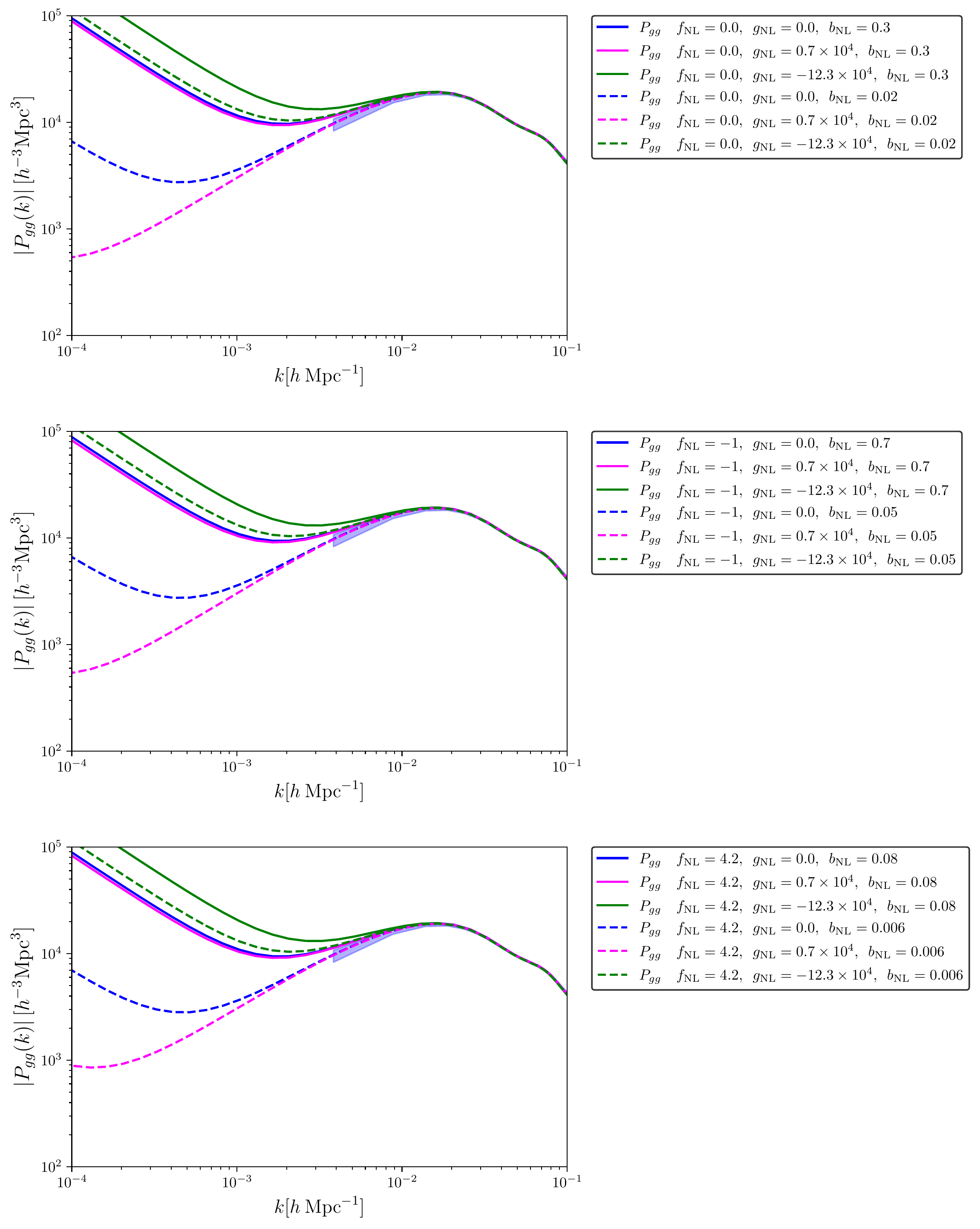}
 \caption[Galaxy power spectrum at redshift $z=1$, with $b_\delta=1.41$, for the limiting values of $f_{\mathrm{NL}}$ and $g_{\mathrm{NL}}$, including the forecasted $1\sigma$ uncertainties of a cosmic variance limited survey of $15,000$ $\mathrm{deg}^2$.]{Galaxy power spectrum at redshift $z=1$, with $b_\delta=1.41$ (a choice justified in the text), for the limiting values of $f_{\mathrm{NL}}$ (zero for the top plot, $-1$ for the middle plot, and 4.2 for the bottom) each with limiting  $g_{\mathrm{NL}}$ values reported by Planck \cite{2020A&A...641A...9P}. The blue shaded area corresponds to the forecasted $1\sigma$ uncertainties of a cosmic variance limited survey of $15,000$ $\mathrm{deg}^2$. See text for more details.}
  \label{Survey15000more}
\end{figure}

\begin{figure}[htbp]
  \centering
    \includegraphics[width=140mm]{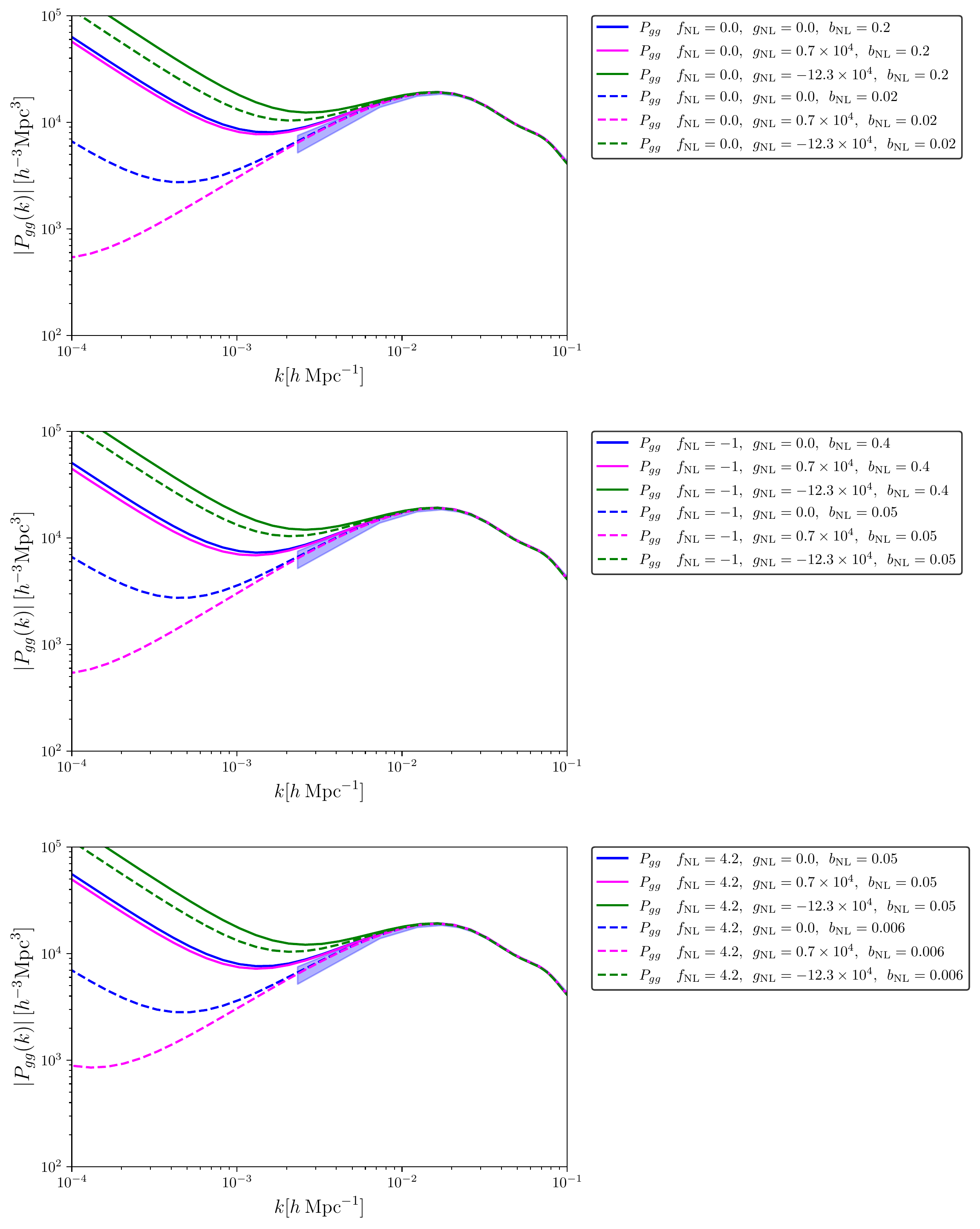}
  \caption[Galaxy power spectrum at redshift $z=1$, with $b_\delta=1.41$, for the limiting values of $f_{\mathrm{NL}}$ and $g_{\mathrm{NL}}$, including the forecasted $1\sigma$ uncertainties of a cosmic variance limited survey of $40,000$ $\mathrm{deg}^2$.]{Galaxy power spectrum at redshift $z=1$, with $b_\delta=1.41$ (a choice justified in the text), for the limiting values of $f_{\mathrm{NL}}$ (zero for the top plot, $-1$ for the middle plot, and 4.2 for the bottom) each with limiting  $g_{\mathrm{NL}}$ values reported by Planck \cite{2020A&A...641A...9P}. The blue shaded area corresponds to the forecasted $1\sigma$ uncertainties of a cosmic variance limited survey of $40,000$ $\mathrm{deg}^2$. See text for more details.}
  \label{Survey40000more}
\end{figure}

In Fig.~\ref{Survey15000more} and Fig.~\ref{Survey40000more} we present the galaxy spectrum at redshift $z=1$ for combinations of the limiting values of $f_{\mathrm {NL}}$ and $g_{\mathrm {NL}}$. In each plot we display a shaded area, corresponding to the predicted $1\sigma$ measurement errors, as presented in chapter~\ref{chapter:Relativistic} (see Eqs.~\eqref{error} to \eqref{fisher}) for a survey of $15,000$ $\mathrm{deg^2}$ (Euclid-like, Fig.~\ref{Survey15000more}), and a survey of $40,000$ $\mathrm{deg^2}$ in Fig.~\ref{Survey40000more}.

Our forecasts assume idealised cosmic variance limited surveys, i.e., with negligible shot noise \cite{1997PhRvL..79.3806T}. Other specifications are the redshift bin width $\Delta z=1.0$ at a central redshift $z=1$, and the largest measured scale $k_{\mathrm {min}}\simeq2\pi/V_{\mathrm{ bin}}^{1/3}=0.001h\mathrm{Mpc}^{-1}$ for a $40,000 \, {\mathrm{deg}}^2$ survey and $k_{\mathrm {min}}=0.002h\mathrm{Mpc}^{-1}$ for a $15,000 \, {\mathrm {deg}}^2$ survey. We note that shot noise contributions and, most importantly, large scale systematic effects (see e.g.~Ref.~\cite{2021arXiv210613725M}) are expected to increase the error budget. Note that we are not taking into account wide-angle effects (see e.g.~Refs.~\cite{2021arXiv210608857C,2021arXiv210813424Y,2015MNRAS.447.1789Y,2012JCAP...10..025B} for related works). We fix the value of $b_{\mathrm{NL}}$ to show in solid lines the minimum value required to have a galaxy power spectrum that could be distinguished in forthcoming surveys considering the forecasted $1\sigma$ measurement errors.

Additionally, we show in dashed lines the smallest $b_{\mathrm{NL}}$ values that result in a well behaved galaxy power spectrum. As already stated, the fiducial value for $b_\delta$ is chosen to be $b_{\delta}=1.41$ at a mean redshift $z=1$. 

We note that there are alternative bias expansion models for which the linear galaxy bias parameter, expressed in the literature as $b_1$ \cite{2018PhR...733....1D} (in our case defined as $b_{\delta}$) is determined by averaging over the halo bias parameters (see e.g.~Ref.~\cite{2019JCAP...12..048U}). However this approach is not required when determining the non-linear bias parameters, as these can be well approximated from other relations with the linear bias \cite{2019MNRAS.483.2078Y}. We do not use this approach in our calculation. 

\subsection{Avoiding large-scale divergences}
\label{sec:renorma}

As mentioned earlier, a few recent works have argued that a divergent behaviour on large scales of the galaxy power spectrum due to the relativistic corrections is nonphysical \cite{2019JCAP...12..048U,2020JCAP...11..064G,2021arXiv210608857C}. In order to avoid such divergence, a coordinate transformation must be performed, which places the observed spectrum in a ``local frame" where the divergences disappear.

\begin{figure}[htbp]
\centering
\includegraphics[width=145mm]{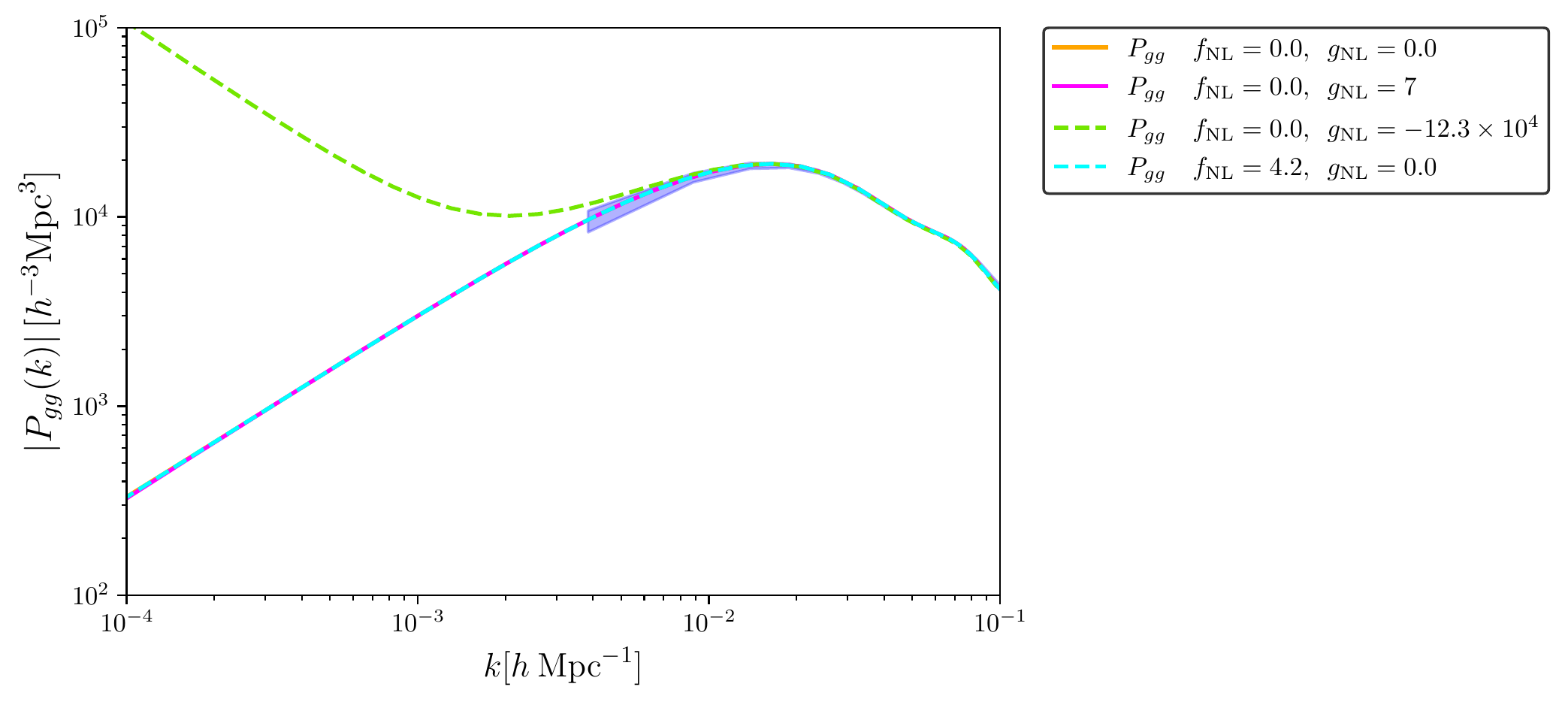}
\caption[Renormalized galaxy power spectrum, at redshift $z=1$, with the specific choice of $b_\delta=1.41$, and $b_{\mathrm{NL}}=2.44\times10^{-6}$.]{Renormalized galaxy power spectrum, at redshift $z=1$, with the specific choice of $b_\delta=1.41$, and $b_{\mathrm{NL}}=2.44\times10^{-6}$. The lines show limiting values of $f_{\mathrm {NL}}$ and $g_{\mathrm{NL}}$ as reported by Planck \cite{2020A&A...641A...9P}. The shaded area is as described in Fig.~\ref{Survey15000more}. Note that only the largest $g_{\mathrm{NL}}$ value yields a distinguishable departure from the linear spectrum. The orange line is hidden behind the magenta line.}
\label{renormalized}
\end{figure}

After studying the behaviour of our expressions in the large scales we have found that there is a possibility to suppress the divergences with a suitable choice of the new $b_{\mathrm{NL}}$ parameter, however, we are aware that this approach might not be suitable for other models. This \textit{renormalization} method sets a parameter value which cancels the relativistic effects of the dominant terms $P_{R1}(k,\eta)$ and $P_{R}^{(1,3)}(k,\eta)$ at large scales, in the absence of primordial non-Gaussianity. For this bias choice, the maximum value of $g_{\mathrm{NL}}$ that keeps the perturbation theory hierarchy is of the order of $g_{\mathrm{NL}}\sim 7$, since the use of higher values, even though are allowed by Planck, leads to negative contributions to the power spectrum. The result of this effective renormalization through parameter fitting is presented in Fig.~\ref{renormalized}. Note that when adding non-Gaussianity contributions the only case that departs significantly from the linear spectrum is when $g_{\mathrm{NL}}$ takes the minimum value allowed by Planck results.

\section{Discussion}
\label{Discussion-section}

In this chapter we have computed the non-linear source galaxy power spectrum including the leading order relativistic and primordial non-Gaussianity contributions, modulated through a set of bias parameters that follow closely the standard prescriptions. 

Besides identifying the well known scale-dependent bias feature, we have non-linear contributions from the relativistic treatment of the matter density. The main result is the galaxy power spectrum including one-loop corrections, expressed in Eq.~\eqref{eq:g-spectrum2}, with the different elements plotted and discussed in detail in section~\ref{Results-section}. In order to assess the detectability of such contributions, our plots include the predicted $1\sigma$ uncertainties of planned Stage-IV galaxy surveys, thus showing which parameter values may be discriminated in future galaxy catalogues. However, we note that we have not included contributions from redshift space distortions (see e.g.~Ref.~\cite{2018MNRAS.478.1341K}). We leave this to future work. 

An important first result is that, even in the absence of PNG, some values of the non-linear bias parameter could be discriminated by future surveys ($b_{\mathrm{NL}} \gtrsim 0.3$), which, conversely means that relativistic contributions could be detected in the galaxy power spectrum. On the other hand, such signal is degenerate with that of a large $f_{\mathrm{NL}}$ since the scale-dependence is identical for relativistic and primordial non-Gaussianity terms at large scales (see e.g.~Fig.~\ref{separate-contributions}). 

The so-called universality relations between bias parameters would fix values of $b_{\phi}$ allowing to debias contributions of primordial non-Gaussianity parameters in the Newtonian formalism \cite{2020JCAP...12..031B,2021arXiv210706887B,2021arXiv210615604M}. However, we find that the corresponding values for $b_{\mathrm{NL}}$ from the universality relation are larger than all the examples here of section~\ref{sec:results1}.
This calls for numerical simulations of galaxy formation and complementary probes of non-Gaussianity in the galaxy or lensing maps, in order to reanalyse the correspondence between bias parameters and also to disentangle relativistic and primordial non-Gaussianity contributions. 

Alternatively, a suitable value of the non-linear bias parameter $b_{\mathrm{NL}}$ can be chosen in order to cancel the divergences of the different contributions at the largest scales, as we show in section~\ref{sec:renorma}. This represents an effective renormalization of the general relativistic corrections to the galaxy power spectrum, so they remain convergent at all scales. In this case, and as shown in Fig.~\ref{renormalized}, the non-Gaussianity contributions can be distinguished from relativistic corrections.

It is important to note that an extreme value of the $g_{\mathrm{NL}}$ parameter, as allowed by Planck constraints, yields the maximum contribution of the non-linear terms to $P_{gg}(k,\eta)$, even when $b_{\mathrm{NL}} = 0$. In such case, our plots show that values of order $g_{\mathrm{NL}} \sim -10^5$ could be detected (or ruled out) in the planned all-sky surveys. We thus conclude that $g_{\mathrm{NL}}$ should not be ignored in the search for primordial non-Gaussianities imprinted in the galaxy power spectrum.

While our results do not account for the full relativistic effects in the observed galaxy distribution, we are confident to have at hand a tool for incorporating the dominant contributions from primordial non-Gaussianity and relativistic non-linearities into the theoretical galaxy power spectrum. 

\chapter{Discussion and Outlook}
\label{Conclusions}

In this thesis, we have investigated the General Relativity and primordial non-Gaussianity effects on the power spectrum. This study is particularly relevant nowadays, since Stage-IV experiments will explore scales that require a relativistic description and statistical tools as the power spectrum for their analysis.

We computed the one-loop power spectrum, which includes Newtonian and relativistic contributions, as well as primordial non-Gaussianity contributions given in terms of the  parameters $f_{\mathrm{NL}}$ and $g_{\mathrm{NL}}$. In addition, we also computed the source galaxy power spectrum, which also includes the effects of primordial non-Gaussianity and General Relativity. The leading contributions to the source galaxy power spectrum come from one-loop matter power spectrum terms dominant at large scales and from factors of the non-linear bias parameter.

To summarise, in chapter~\ref{Introduction} we begin by reviewing the Standard Cosmology, which provides the basic mathematical formalism for our main calculations. We also presented how cosmological distances are measured in Cosmology and how the Cosmic Microwave Background is measured and analysed to constrain cosmological parameters. 

Followed by chapter~\ref{Perturbation-Theory} where we review in detail cosmological perturbation theory, presenting the relevant gauges for our calculations. We also review the Newtonian standard perturbation theory, presenting the main equations, and non-linear solutions up to third order necessary for the calculation of the one-loop power spectrum. In addition, we provide an introduction to the statistical quantities used in this thesis, the power spectrum and bispectrum. We also provide a brief description of primordial non-Gaussianity in this context. 

Chapters~\ref{chapter:Relativistic} and \ref{paper2} contain the original work of the present thesis. Firstly, in chapter~\ref{chapter:Relativistic} we present the evolution equations in synchronous-comoving gauge, followed by two equivalent methods to obtain the solutions for the relativistic density contrast in the long-wavelength approximation. One is the cosmological perturbation theory which is the standard method, but lengthy nevertheless. And the second method, the gradient expansion approach which is more convenient to obtain high order solutions. These solutions also include contributions from primordial non-Gaussianity, given in terms of the parameters $f_{\mathrm{NL}}$ and $g_{\mathrm{NL}}$. 
We complement these solutions with the well known Newtonian solutions for the density contrast. This allows us to compute our main result, the total one-loop power spectrum, which includes relativistic and Newtonian contributions and is given by Eq.~\eqref{Oneloopexp}. Additionally, we also calculate the total bispectrum at the tree-level.

We then discussed the possibility of these relativistic effects being detectable with the future surveys considering different limiting values for $f_{\mathrm{NL}}$ and $g_{\mathrm{NL}}$. Our results show that pure relativistic corrections do not contribute enough at large scale to be observed by future large scale structure probes. However, if we take into account primordial non-Gaussianity contributions, specially those coming from $g_{\mathrm{NL}}$ parameter, these yield a significant contribution to the one-loop power spectrum which could be observable with the future surveys.

Future direction on this topic would be the computation of the total one-loop bispectrum, which could help us to complement the information obtained from the one-loop power spectrum. In order to do that, we would require fourth order solutions for the density contrast. While the relativistic solution might not be difficult to calculate using the gradient expansion approach, the Newtonian solution although is known, is rather lengthy which suppose an analytical challenge, that should be carefully addressed. It is also worth mentioning that the one-loop bispectrum contributions for the current primordial non-Gaussianity constraints have not been detected yet, therefore we leave this computation for future work. 

Subsequently, in chapter~\ref{paper2} we explore an important extension to the work presented in chapter~\ref{chapter:Relativistic}. Using the solutions for the relativistic density contrast and introducing a bias prescription, we compute the real space source galaxy power spectrum given in Eq.~\eqref{eq:g-spectrum2}.  We found two main contributions, the first one from the relativistic one-loop power spectrum terms, previously calculated in chapter~\ref{chapter:Relativistic}, and the second from newly calculated terms that arose due to the bias expansion employed, these are expressed as factors of the non-linear bias parameter $b_{\mathrm{NL}}$. 

We used our bias model to assess the ability of Stage-IV surveys to constrain primordial non-Gaussianity. This was done evaluating the expression for the source galaxy power spectrum for a range of limiting values of $f_{\mathrm{NL}}$ and $g_{\mathrm{NL}}$ and bias parameters, and then comparing them with the forecasted measurements from Stage-IV experiments, specifically from a $15,000$ $\mathrm{deg}^2$ (Euclid-like) and a $40,000$ $\mathrm{deg}^2$ (all-sky) galaxy survey. We found that even for the cases where do not account for primordial non-Gaussianity, some values of the non-linear bias parameter $b_{\mathrm{NL}}$ could be discriminated by future surveys, which could mean the possibility to detect relativistic effects in the galaxy power spectrum. However, it is important to point out that relativistic and primordial non-Gaussianity at large scales have the same scale-dependance, meaning that signals are degenerate for large $f_{\mathrm{NL}}$. 

Finally, we showed how this non-linear bias parameter can effectively renormalize diverging relativistic contributions at large scales. Such renormalization allows to distinguish relativistic and primordial non-Gaussianity corrections.

We note that our expression for the source galaxy power spectrum does not account for all the relativistic effects observed in the galaxy distribution. Nonetheless, we have presented a strategy to incorporate dominant contributions from primordial non-Gaussianity and relativistic non-linearities.

An obvious step towards computing observables from our formalism is the inclusion of redshift space distortions (see e.g.~Refs.~\cite{2020JCAP...11..064G,2021arXiv210608857C}). This becomes relevant due to the fact that galaxy surveys do not measure the positions in configuration space directly, but these are determined through the measurement of redshifts instead, which are affected by the peculiar velocities of galaxies, and which impact the observed distances directly, modifying the power spectrum measured in redshift space.

As a first step towards accounting for redshift space distortions we could consider the so-called Kaiser model, this is a linear model, that is valid for galaxies that are sufficiently far away, such that their separations are small respect to the distances between them and the observer. In this model the overdensities in redshift space and therefore the power spectrum, depend not only in the magnitude of the wavevector $\mathbf{k}$, but also in the cosine of the angle between the line of sight and the wavevector $\mathbf{k}$ \cite{2012MNRAS.420.2102S}. Moreover, for the scales of interest in our work, we must take on account wide-angle effects, where the distance between galaxies is comparable to that of the observer. The implementation of the latter is non-trivial, and we leave the task for future work. 

In addition to redshift space distortions, another relativistic effect relevant on large scales is lensing. Gravitational lensing distorts the images of high-redshift sources due to the gravitational potential on the line of sight that deflects the light rays of distant sources, this effect need to be taken into account in the galaxy power spectrum for the correct analysis of future surveys. Other relativistic effects that have been considered in the literature and are important since they modify the galaxy power spectrum on ultra large scales are, for example, Doppler, Sachs-Wolfe, Integrated Sachs-Wolfe (ISW) effect and time-delay terms \cite{2014JCAP...09..037B,2014JCAP...11..013B,2017JCAP...09..040J,2018JCAP...03..036J}.
These additional effects are important not only to improve the accuracy but also contain additional information, however including these effects is beyond the scope of this work and it is also left for future work.

Besides including more effects, another extension to this work would be perform numerical simulations, using relativistic N-body codes (e.g.~gevolution~\cite{2016JCAP...07..053A} and GRAMSES~\cite{2020JCAP...01..007B}). The aim of this  would be to study the relation in between bias parameters used in our approach, and being able to disentangle primordial non-Gaussianities from the relativistic contributions. This could also be important to take into account in the $f_{\mathrm{NL}}$ and $g_{\mathrm{NL}}$ constraint analysis.

\begin{appendices}

\chapter{One-loop corrections to the density power spectrum }
\label{one-loopderivation}
In this appendix we show the explicit derivation of the one-loop corrections to the density power spectrum presented in section \ref{sec:One-loop}. For simplicity, in this derivation we omit the explicit time dependence in $\delta(\eta,\mathbf{k})$.

The first one-loop order contribution is defined as 
\begin{equation}
\langle\delta^{(2)}(\mathbf{k})\delta^{(2)}(\mathbf{k}')\rangle\equiv(2\pi)^3P^{(2,2)}(\mathbf{k})\delta_D(\mathbf{k}+\mathbf{k}').
\label{P22def}
\end{equation}

As first step we evaluate the left hand side of Eq.~\eqref{P22def}, using the definition of $\delta^{(2)}(\mathbf{k})$ given by
\begin{equation}
{\delta^{(2)}(\mathbf k)}=\int\frac{d^3q_1d^3q_2}{(2\pi)^3}\delta_D(\mathbf k-\mathbf q_1-\mathbf q_2)\mathcal{F}^{(2)}(\mathbf q_1,\mathbf q_2)\delta^{(1)}(\mathbf q_1)\delta^{(1)}({\mathbf q_2}),
\end{equation}

\noindent then we write Eq.~\eqref{P22def} as
\begin{align}
\nonumber
\langle\delta^{(2)}(\mathbf{k})\delta^{(2)}(\mathbf{k}')\rangle=&\int\frac{d^3q_1}{(2\pi)^3}\int\frac{d^3q_1'}{(2\pi)^3}\mathcal{F}^{(2)}(\mathbf{q}_1,\mathbf{k}-\mathbf{q}_1)\mathcal{F}^{(2)}(\mathbf{q}_1',\mathbf{k'}-\mathbf{q}_1')\\
\times&\langle\delta^{(1)}(\mathbf{q}_1)\delta^{(1)}(\mathbf{k}-\mathbf{q}_1)\delta^{(1)}(\mathbf{q}_1')\delta^{(1)}(\mathbf{k'}-\mathbf{q}_1')\rangle.
\label{A2}
\end{align}

\noindent Using the Wick's theorem from Eq.~\eqref{Wickstheorem} we have e.g.
\begin{align}
\nonumber
\langle\delta^{(1)}(\mathbf{q}_1)\delta^{(1)}(\mathbf{q}_2)\delta^{(1)}(\mathbf{q}_3)\delta^{(1)}(\mathbf{q}_4)\rangle=&\langle\delta^{(1)}(\mathbf{q}_1)\delta^{(1)}(\mathbf{q}_2)\rangle \langle\delta^{(1)}(\mathbf{q}_3)\delta^{(1)}(\mathbf{q}_4)\rangle\\
+&\langle\delta^{(1)}(\mathbf{q}_1)\delta^{(1)}(\mathbf{q}_3)\rangle \langle\delta^{(1)}(\mathbf{q}_2)\delta^{(1)}(\mathbf{q}_4)\rangle\\
\nonumber
+&\langle\delta^{(1)}(\mathbf{q}_1)\delta^{(1)}(\mathbf{q}_4)\rangle \langle\delta^{(1)}(\mathbf{q}_3)\delta^{(1)}(\mathbf{q}_2)\rangle,
\end{align}

\noindent which allow us to write the correlator in Eq.~\eqref{A2} in terms of 2-point correlators as follows
\begin{align}
\nonumber
\langle\delta^{(2)}(\mathbf{k})\delta^{(2)}(\mathbf{k}')\rangle=&\int\frac{d^3q_1}{(2\pi)^3}\int\frac{d^3q_1'}{(2\pi)^3}\mathcal{F}^{(2)}(\mathbf{q}_1,\mathbf{k}-\mathbf{q}_1)\mathcal{F}^{(2)}(\mathbf{q}_1',\mathbf{k'}-\mathbf{q}_1')
\\
\nonumber
\times&\{\langle\delta^{(1)}(\mathbf{q}_1)\delta^{(1)}(\mathbf{k}-\mathbf{q}_1)\rangle\langle\delta^{(1)}(\mathbf{q}_1')\delta^{(1)}(\mathbf{k'}-\mathbf{q}_1')\rangle\\
+&\langle\delta^{(1)}(\mathbf{q}_1)\delta^{(1)}(\mathbf{k'}-\mathbf{q}_1')\rangle\langle\delta^{(1)}(\mathbf{q}_1')\delta^{(1)}(\mathbf{k}-\mathbf{q}_1)\rangle\\
\nonumber
+&\langle\delta^{(1)}(\mathbf{q}_1)\delta^{(1)}(\mathbf{q}_1')\rangle\langle\delta^{(1)}(\mathbf{k}-\mathbf{q}_1)\delta^{(1)}(\mathbf{k'}-\mathbf{q}_1')\rangle\},
\end{align}

\noindent finally, using the definition for the power spectrum given in Eq.~\eqref{powerspectdef}, we can rewrite our expression as
\begin{align}
\nonumber
\langle\delta^{(2)}(\mathbf{k})\delta^{(2)}(\mathbf{k}')\rangle=&\int d^3q_1\int d^3q_1'\mathcal{F}^{(2)}(\mathbf{q}_1,\mathbf{k}-\mathbf{q}_1)\mathcal{F}^{(2)}(\mathbf{q}_1',\mathbf{k'}-\mathbf{q}_1')\\
\times&\left[P_L(\mathbf{q}_1)P_L(\mathbf{k}-\mathbf{q}_1)\delta_D(\mathbf{q}_1+\mathbf{q}_1')\delta_D(\mathbf{k}-\mathbf{q}_1+\mathbf{k}'-\mathbf{q}_1')\right.\\
\nonumber
+&\left.P_L(\mathbf{q}_1)P_L(\mathbf{q}_1')\delta_D(\mathbf{k}'+\mathbf{q}_1-\mathbf{q}_1')\delta_D(\mathbf{k}+\mathbf{q}_1'-\mathbf{q}_1)\right]\\
\nonumber
=&(2\pi)^3\left\{2\int\frac{d^3q}{(2\pi^3)}P_L(q)P_L(|\mathbf{k}-\mathbf{q}|)\left[\mathcal{F}^{(2)}(\mathbf{q},\mathbf{k}-\mathbf{q})\right]^2\right\}\delta_D(\mathbf{k}+\mathbf{k}').
\end{align}

As a result we can express
\begin{equation}
P^{(2,2)}(\mathbf{k})=2\int\frac{d^3q}{(2\pi^3)}P_L(q)P_L(|\mathbf{k}-\mathbf{q}|)\left[\mathcal{F}^{(2)}(\mathbf{q},\mathbf{k}-\mathbf{q})\right]^2.
\end{equation}

The second one-loop order contribution to the density power spectrum is defined as 
\begin{equation}
\langle\delta^{(1)}(\mathbf{k}')\delta^{(3)}(\mathbf{k})\rangle\equiv(2\pi)^3P^{(1,3)}(\mathbf{k})\delta_D(\mathbf{k}+\mathbf{k}').
\label{P13def}
\end{equation}

Again, if we evaluate the left hand side of Eq.~\eqref{P13def} using the definition of $\delta^{(1)}(\mathbf{k})$ given by
\begin{equation}
\delta^{(1)}(\mathbf{k})=\mathcal{F}^{(1)}(\mathbf{k})\delta^{(1)}(\mathbf{k}),
\end{equation}

\noindent and the definition for $\delta^{(3)}(\mathbf{k})$
\begin{align}
\delta^{(3)}(\mathbf k)=\int\frac{d^3q_1d^3q_2d^3q_3}{(2\pi)^6}&\delta_D(\mathbf k-\mathbf q_1-\mathbf q_2-\mathbf q_3 )\mathcal{F}^{(3)}(\mathbf q_1,\mathbf q_2,\mathbf q_3)\\
\times&\delta^{(1)}(\mathbf q_1)\delta^{(1)}(\mathbf q_2)\delta^{(1)}({\mathbf q_3}),
\nonumber
\end{align}

\noindent we can write Eq.~\eqref{P13def} as
\begin{align}
\nonumber
\langle\delta^{(1)}(\mathbf{k}')\delta^{(3)}(\mathbf{k})\rangle&=\mathcal{F}^{(1)}(\mathbf{k}')\int\frac{d^3q_1}{(2\pi)^3}\int\frac{d^3q_2}{(2\pi)^3}\int d^3q_3\delta_D(\mathbf{k}-\mathbf{q}_1-\mathbf{q}_2-\mathbf{q}_3)\\
&\times\mathcal{F}^{(3)}(\mathbf{q}_1,\mathbf{q}_2,\mathbf{q}_3)\langle\delta^{(1)}(\mathbf{k}')\delta^{(1)}(\mathbf{q}_1)\delta^{(1)}(\mathbf{q}_2)\delta^{(1)}(\mathbf{q}_3)\rangle.
\label{A7}
\end{align}

\noindent Using the Wick's theorem from Eq.~\eqref{Wickstheorem}, we can write the correlator in Eq.~\eqref{A7} in terms of 2-point correlators as follows
\begin{align}
\nonumber
\langle\delta^{(1)}(\mathbf{k}')\delta^{(3)}(\mathbf{k})\rangle&=\mathcal{F}^{(1)}(\mathbf{k}')\int\frac{d^3q_1}{(2\pi)^3}\int\frac{d^3q_2}{(2\pi)^3}\int d^3q_3\delta_D(\mathbf{k}-\mathbf{q}_1-\mathbf{q}_2-\mathbf{q}_3)\\
\nonumber
&\times\mathcal{F}^{(3)}(\mathbf{q}_1,\mathbf{q}_2,\mathbf{q}_3)\{\langle\delta^{(1)}(\mathbf{k}')\delta^{(1)}(\mathbf{q}_1)\rangle \langle\delta^{(1)}(\mathbf{q}_2)\delta^{(1)}(\mathbf{q}_3)\rangle\\
&+\langle\delta^{(1)}(\mathbf{k}')\delta^{(1)}(\mathbf{q}_2)\rangle \langle\delta^{(1)}(\mathbf{q}_1)\delta^{(1)}(\mathbf{q}_3)\rangle\\
\nonumber
&+\langle\delta^{(1)}(\mathbf{k}')\delta^{(1)}(\mathbf{q}_3)\rangle \langle\delta^{(1)}(\mathbf{q}_1)\delta^{(1)}(\mathbf{q}_2)\rangle\},
\end{align}

\noindent finally, using the definition for the power spectrum given in Eq.~\eqref{powerspectdef}, we can rewrite our expression as
\begin{align}
\nonumber
\langle\delta^{(1)}(\mathbf{k}')\delta^{(3)}(\mathbf{k})\rangle&=3\mathcal{F}^{(1)}(\mathbf{k}')\int d^3q_1\int d^3q_2\int d^3q_3\delta_D(\mathbf{k}-\mathbf{q}_1-\mathbf{q}_2-\mathbf{q}_3)\\
&\times \mathcal{F}^{(3)}(\mathbf{q}_1,\mathbf{q}_2,\mathbf{q}_3)P_L(\mathbf{k}')P_L(\mathbf{q}_2)\delta_D(\mathbf{k}'+\mathbf{q}_1)\delta_D(\mathbf{q}_2+\mathbf{q}_3)\\
\nonumber
&=(2\pi)^3\left\{ 3\mathcal{F}^{(1)}(-\mathbf{k})P_L(\mathbf{k})\int\frac{d^3q}{(2\pi)^3}P_L(\mathbf{q})\mathcal{F}^{(3)}(\mathbf{k},\mathbf{q},-\mathbf{q})\right\}\delta_{D}(\mathbf{k}+\mathbf{k}').
\end{align}

As a result we can express
\begin{equation}
P^{(1,3)}(\mathbf{k})=3\mathcal{F}^{(1)}(\mathbf{k})P_L(\mathbf{k})\int\frac{d^3q}{(2\pi)^3}P_L(\mathbf{q})\mathcal{F}^{(3)}(\mathbf{k},\mathbf{q},-\mathbf{q}),
\end{equation}

where the Hermiticity properties allow us to write $\mathcal{F}^{(1)}(-\mathbf{k})=\mathcal{F}^{(1)}(\mathbf{k})$ \cite{2010PhDT.........4J}.

\chapter{Infrared limits}

\label{IRL}

The infrared (IR) contributions of the one-loop integrals in Eqs.~\eqref{integral22} and \eqref{integral13}
can be computed as the part of the integral from $r=0$ to a small value $\epsilon$. With this consideration, the one-loop power spectrum can be written as:
\begin{eqnarray}
P^{(2,2)}(k,\eta) &=& \frac{k^3}{2\pi^2}\left( \int_0^\epsilon + \int_\epsilon^\infty \right)
r^2 dr 
P_L(kr,\eta)\int_{-1}^{1}dxP_L(k\sqrt{1+r^2-2rx},\eta) \nonumber\\
&&  \times \left[ (\mathcal{F}^{(2)}_N)^2+2 \mathcal{F}^{(2)}_N\mathcal{F}^{(2)}_R+(\mathcal{F}^{(2)}_R)^2 \right]
\,, \\
P^{(1,3)}(k,\eta) &=& \frac{k^3}{4\pi^2}P_L(k,\eta) \left( \int_0^\epsilon + \int_\epsilon^\infty \right) dr  P_L(kr,\eta) \left(\mathcal{F}^{(3)}_N + \mathcal{F}^{(3)}_R \right) \,,
\end{eqnarray}
where the integrals in $r$ have been split between a possible divergent infrared contribution from 0 to $\epsilon$ and a finite contribution from $\epsilon$ to $\infty$ which, in the limit of $\epsilon \to 0$ will correspond to the Cauchy principal value of the integral.
Using $P_L(k,\eta)\propto k^{n_s}$ as $k\to 0$ the  infrared contributions  can be computed analytically, which we will write explicitly in the following expressions. Note that, in the cases where the integrals diverge we will write the expressions as the limit
\begin{equation}
    \int_0^\epsilon = \lim_{\delta \to 0} \int_\delta^\epsilon \,,
\end{equation}
in order to see the divergence rate.
For the three different terms in $P^{(2,2)}(k,\eta)$ we obtain the expressions:
\begin{align}
{}_{\textrm{IR}}P^{(2,2)}_{NN} =& \frac{k^3}{2\pi^2}P_L(k)P_L(k\epsilon) \frac{\epsilon}{3(n_s+1)} \,, \label{ir22}\\
{}_{\textrm{IR}}P^{(2,2)}_{C} =& \frac{\mathcal{H}^2k}{2\pi^2}P_L(k)P_L(k\epsilon) \frac{\epsilon}{n_s+1} 
  \left( \frac{27}{7}f_{\textrm{NL}} - \frac{95}{42} \right)\,,\\
{}_{\textrm{IR}}P^{(2,2)}_{RR}  =&\frac{\mathcal{H}^4}{2\pi^2 k} P_L(k)P_L(k\epsilon)\left\{  \frac{\epsilon A}{(n_s+1)}+ 
\frac{B}{\epsilon(n_s-1)}\lim_{\delta \to 0} \left[ 1- \left( \frac{\delta}{\epsilon}\right)^{n_s-1}\right]
\right\}  \,, \label{divergence22}
\end{align}

where
\begin{eqnarray}
A&=& 9f_{\textrm{NL}}^2 - \frac{35}{2}f_{\textrm{NL}}+\frac{725}{27} \,,\\
B&=& \frac{(3f_{\textrm{NL}}-5)^2}{2}\,.
\end{eqnarray}
In these expressions the value of $\epsilon$ is small but fixed, meaning that the purely relativistic term diverges approximately as $\delta^{-0.03}$ for $\delta \to 0$. Meanwhile the infrared contributions to $P^{(1,3)}(k,\eta)$ read:
\begin{align}
{}_{\textrm{IR}}P^{(1,3)}_{NN} =& - \frac{k^3}{4\pi^2} P_L(k)P_L(k\epsilon) \frac{\epsilon}{3(n_s+1)} \label{ir13} \,,\\
{}_{\textrm{IR}}P^{(1,3)}_{RR}  =&  \frac{\mathcal{H}^4}{4\pi^2 k} P_L(k)P_L(k\epsilon) \left\{  \frac{\epsilon C}{(n_s+1)}+ 
\frac{D}{\epsilon(n_s-1)}
\times\lim_{\delta \to 0} \left[ 1- \left( \frac{\delta}{\epsilon}\right)^{n_s-1}\right]
\right\} \,, 
\label{divergence13}
\end{align}
where
\begin{eqnarray}
C&=& 54 \left( -g_{\textrm{NL}} - \frac{5}{6}f_{\textrm{NL}} -\frac{175}{108} \right) \,,\\
D&=& -\frac{27}{2} \left( -g_{\textrm{NL}} + \frac{10}{3}f_{\textrm{NL}} -\frac{50}{27} \right) \,.
\end{eqnarray}
We see that the second term in \eqref{divergence13} diverges at the same rate as \eqref{divergence22}.
For the purely Newtonian one loop contribution, the possible infrared problems in the different terms get solved as the combination $2{}_{\textrm{IR}}P_{NN}^{(1,3)}(k,\eta)+{}_{\textrm{IR}}P_{NN}^{(2,2)}(k,\eta)$ cancels out, as read from the expressions \eqref{ir22} and \eqref{ir13} (see Ref.~\cite{2016JCAP...09..015M}). However for the relativistic term this does not happen as the expressions \eqref{divergence13} and \eqref{divergence22} do not cancel.

In order to obtain finite results for the relativistic one-loop contribution, we set a lower limit different from zero in the $r$ integrals. The fact that the divergence is very slow allows the results to not be very dependent on this limit, but only as $r_c^{-0.03}$. Moreover, as stated in Ref.~\cite{2019JCAP...07..030C}, the observations have a minimum $k$ accessible to them, corresponding to their maximum observed scale.
Throughout this work we chose this limit to be in the parameter $q = kr$ as $q_c = 10^{-5}h{\textrm{Mpc}^{-1}}$ which is close to the limit chosen in Ref.~\cite{2019JCAP...07..030C} as $q_c = H_0 \approx 3\times 10^{-4}h{\textrm{Mpc}^{-1}}$.

\end{appendices}

\begin{singlespace}

\bibliography{bib/thesis_bib}

\providecommand{\href}[2]{#2}\begingroup\raggedright\begin{thebibliography}{100}

\bibitem{2016PhRvL.116f1102A}
B.~P. {Abbott}, R.~{Abbott}, T.~D. {Abbott}, M.~R. {Abernathy}, F.~{Acernese},
  K.~{Ackley} et~al., \emph{{Observation of Gravitational Waves from a Binary
  Black Hole Merger}},
  \href{https://doi.org/10.1103/PhysRevLett.116.061102}{\emph{\prl} {\bfseries
  116} (2016) 061102} [\href{https://arxiv.org/abs/1602.03837}{{\ttfamily
  1602.03837}}].

\bibitem{1999AIPC..476....1S}
G.~F. {Smoot}, \emph{{COBE observations and results}},  in \emph{3K cosmology},
  L.~{Maiani}, F.~{Melchiorri} and N.~{Vittorio}, eds., vol.~476 of
  \emph{American Institute of Physics Conference Series}, pp.~1--10, May, 1999,
  \href{https://doi.org/10.1063/1.59326}{DOI}
  [\href{https://arxiv.org/abs/astro-ph/9902027}{{\ttfamily
  astro-ph/9902027}}].

\bibitem{2013ApJS..208...19H}
G.~{Hinshaw}, D.~{Larson}, E.~{Komatsu}, D.~N. {Spergel}, C.~L. {Bennett},
  J.~{Dunkley} et~al., \emph{{Nine-year Wilkinson Microwave Anisotropy Probe
  (WMAP) Observations: Cosmological Parameter Results}},
  \href{https://doi.org/10.1088/0067-0049/208/2/19}{\emph{\apjs} {\bfseries
  208} (2013) 19} [\href{https://arxiv.org/abs/1212.5226}{{\ttfamily
  1212.5226}}].

\bibitem{2020A&A...641A...1P}
{Planck Collaboration}, N.~{Aghanim}, Y.~{Akrami}, F.~{Arroja}, M.~{Ashdown},
  J.~{Aumont} et~al., \emph{{Planck 2018 results. I. Overview and the
  cosmological legacy of Planck}},
  \href{https://doi.org/10.1051/0004-6361/201833880}{\emph{\aap} {\bfseries
  641} (2020) A1} [\href{https://arxiv.org/abs/1807.06205}{{\ttfamily
  1807.06205}}].

\bibitem{2016A&A...594A..13P}
{Planck Collaboration}, P.~A.~R. {Ade}, N.~{Aghanim}, M.~{Arnaud},
  M.~{Ashdown}, J.~{Aumont} et~al., \emph{{Planck 2015 results. XIII.
  Cosmological parameters}},
  \href{https://doi.org/10.1051/0004-6361/201525830}{\emph{\aap} {\bfseries
  594} (2016) A13} [\href{https://arxiv.org/abs/1502.01589}{{\ttfamily
  1502.01589}}].

\bibitem{2016A&A...594A..20P}
{Planck Collaboration}, P.~A.~R. {Ade}, N.~{Aghanim}, M.~{Arnaud}, F.~{Arroja},
  M.~{Ashdown} et~al., \emph{{Planck 2015 results. XX. Constraints on
  inflation}}, \href{https://doi.org/10.1051/0004-6361/201525898}{\emph{\aap}
  {\bfseries 594} (2016) A20}
  [\href{https://arxiv.org/abs/1502.02114}{{\ttfamily 1502.02114}}].

\bibitem{2003RvMP...75..559P}
P.~J. {Peebles} and B.~{Ratra}, \emph{{The cosmological constant and dark
  energy}}, \href{https://doi.org/10.1103/RevModPhys.75.559}{\emph{Reviews of
  Modern Physics} {\bfseries 75} (2003) 559}
  [\href{https://arxiv.org/abs/astro-ph/0207347}{{\ttfamily
  astro-ph/0207347}}].

\bibitem{2018RvMP...90d5002B}
G.~{Bertone} and D.~{Hooper}, \emph{{History of dark matter}},
  \href{https://doi.org/10.1103/RevModPhys.90.045002}{\emph{Reviews of Modern
  Physics} {\bfseries 90} (2018) 045002}
  [\href{https://arxiv.org/abs/1605.04909}{{\ttfamily 1605.04909}}].

\bibitem{2005PhR...405..279B}
G.~{Bertone}, D.~{Hooper} and J.~{Silk}, \emph{{Particle dark matter: evidence,
  candidates and constraints}},
  \href{https://doi.org/10.1016/j.physrep.2004.08.031}{\emph{\physrep}
  {\bfseries 405} (2005) 279}
  [\href{https://arxiv.org/abs/hep-ph/0404175}{{\ttfamily hep-ph/0404175}}].

\bibitem{2005MNRAS.362..505C}
S.~{Cole}, W.~J. {Percival}, J.~A. {Peacock}, P.~{Norberg}, C.~M. {Baugh},
  C.~S. {Frenk} et~al., \emph{{The 2dF Galaxy Redshift Survey: power-spectrum
  analysis of the final data set and cosmological implications}},
  \href{https://doi.org/10.1111/j.1365-2966.2005.09318.x}{\emph{\mnras}
  {\bfseries 362} (2005) 505}
  [\href{https://arxiv.org/abs/astro-ph/0501174}{{\ttfamily
  astro-ph/0501174}}].

\bibitem{2017AJ....154...28B}
M.~R. {Blanton}, M.~A. {Bershady}, B.~{Abolfathi}, F.~D. {Albareti},
  C.~{Allende Prieto}, A.~{Almeida} et~al., \emph{{Sloan Digital Sky Survey IV:
  Mapping the Milky Way, Nearby Galaxies, and the Distant Universe}},
  \href{https://doi.org/10.3847/1538-3881/aa7567}{\emph{\aj} {\bfseries 154}
  (2017) 28} [\href{https://arxiv.org/abs/1703.00052}{{\ttfamily 1703.00052}}].

\bibitem{2003moco.book.....D}
S.~{Dodelson}, \emph{{Modern cosmology}}. ``Academic Press", 2003.

\bibitem{2003imc..book.....L}
A.~R. {Liddle}, \emph{{An introduction to modern cosmology}}. ``Wiley", 2003.

\bibitem{1992PhRvD..46..585M}
N.~{Makino}, M.~{Sasaki} and Y.~{Suto}, \emph{{Analytic approach to the
  perturbative expansion of nonlinear gravitational fluctuations in
  cosmological density and velocity fields}},
  \href{https://doi.org/10.1103/PhysRevD.46.585}{\emph{\prd} {\bfseries 46}
  (1992) 585}.

\bibitem{1994ApJ...431..495J}
B.~{Jain} and E.~{Bertschinger}, \emph{{Second-Order Power Spectrum and
  Nonlinear Evolution at High Redshift}},
  \href{https://doi.org/10.1086/174502}{\emph{\apj} {\bfseries 431} (1994) 495}
  [\href{https://arxiv.org/abs/astro-ph/9311070}{{\ttfamily
  astro-ph/9311070}}].

\bibitem{2002PhR...367....1B}
F.~{Bernardeau}, S.~{Colombi}, E.~{Gazta{\~n}aga} and R.~{Scoccimarro},
  \emph{{Large-scale structure of the Universe and cosmological perturbation
  theory}},
  \href{https://doi.org/10.1016/S0370-1573(02)00135-7}{\emph{\physrep}
  {\bfseries 367} (2002) 1}
  [\href{https://arxiv.org/abs/astro-ph/0112551}{{\ttfamily
  astro-ph/0112551}}].

\bibitem{2011arXiv1110.3193L}
R.~{Laureijs}, J.~{Amiaux}, S.~{Arduini}, J.~L. {Augu{\`e}res},
  J.~{Brinchmann}, R.~{Cole} et~al., \emph{{Euclid Definition Study Report}},
  {\emph{arXiv e-prints} (2011) arXiv:1110.3193}
  [\href{https://arxiv.org/abs/1110.3193}{{\ttfamily 1110.3193}}].

\bibitem{2016arXiv161100036D}
{DESI Collaboration}, A.~{Aghamousa}, J.~{Aguilar}, S.~{Ahlen}, S.~{Alam},
  L.~E. {Allen} et~al., \emph{{The DESI Experiment Part I: Science,Targeting,
  and Survey Design}}, {\emph{arXiv e-prints} (2016) arXiv:1611.00036}
  [\href{https://arxiv.org/abs/1611.00036}{{\ttfamily 1611.00036}}].

\bibitem{2019ApJS..242....2C}
N.~E. {Chisari}, D.~{Alonso}, E.~{Krause}, C.~D. {Leonard}, P.~{Bull},
  J.~{Neveu} et~al., \emph{{Core Cosmology Library: Precision Cosmological
  Predictions for LSST}},
  \href{https://doi.org/10.3847/1538-4365/ab1658}{\emph{\apjs} {\bfseries 242}
  (2019) 2} [\href{https://arxiv.org/abs/1812.05995}{{\ttfamily 1812.05995}}].

\bibitem{2005JCAP...10..010B}
N.~{Bartolo}, S.~{Matarrese} and A.~{Riotto}, \emph{{Signatures of primordial
  non-Gaussianity in the large-scale structure of the universe}},
  \href{https://doi.org/10.1088/1475-7516/2005/10/010}{\emph{\jcap} {\bfseries
  2005} (2005) 010} [\href{https://arxiv.org/abs/astro-ph/0501614}{{\ttfamily
  astro-ph/0501614}}].

\bibitem{2015JCAP...10..024D}
R.~{de Putter}, O.~{Dor{\'e}} and D.~{Green}, \emph{{Is there scale-dependent
  bias in single-field inflation?}},
  \href{https://doi.org/10.1088/1475-7516/2015/10/024}{\emph{\jcap} {\bfseries
  2015} (2015) 024} [\href{https://arxiv.org/abs/1504.05935}{{\ttfamily
  1504.05935}}].

\bibitem{2010CQGra..27l4011D}
V.~{Desjacques} and U.~{Seljak}, \emph{{Primordial non-Gaussianity from the
  large-scale structure}},
  \href{https://doi.org/10.1088/0264-9381/27/12/124011}{\emph{Classical and
  Quantum Gravity} {\bfseries 27} (2010) 124011}
  [\href{https://arxiv.org/abs/1003.5020}{{\ttfamily 1003.5020}}].

\bibitem{2004PhR...402..103B}
N.~{Bartolo}, E.~{Komatsu}, S.~{Matarrese} and A.~{Riotto},
  \emph{{Non-Gaussianity from inflation: theory and observations}},
  \href{https://doi.org/10.1016/j.physrep.2004.08.022}{\emph{\physrep}
  {\bfseries 402} (2004) 103}
  [\href{https://arxiv.org/abs/astro-ph/0406398}{{\ttfamily
  astro-ph/0406398}}].

\bibitem{2012MNRAS.422.2854G}
T.~{Giannantonio}, C.~{Porciani}, J.~{Carron}, A.~{Amara} and A.~{Pillepich},
  \emph{{Constraining primordial non-Gaussianity with future galaxy surveys}},
  \href{https://doi.org/10.1111/j.1365-2966.2012.20604.x}{\emph{\mnras}
  {\bfseries 422} (2012) 2854}
  [\href{https://arxiv.org/abs/1109.0958}{{\ttfamily 1109.0958}}].

\bibitem{2008cmbg.book.....D}
R.~{Durrer}, \emph{{The Cosmic Microwave Background}}. ``Cambridge University
  Press", 2008.

\bibitem{1986ApJ...304...15B}
J.~M. {Bardeen}, J.~R. {Bond}, N.~{Kaiser} and A.~S. {Szalay}, \emph{{The
  Statistics of Peaks of Gaussian Random Fields}},
  \href{https://doi.org/10.1086/164143}{\emph{\apj} {\bfseries 304} (1986) 15}.

\bibitem{2008ApJ...684L...1C}
C.~{Carbone}, L.~{Verde} and S.~{Matarrese}, \emph{{Non-Gaussian Halo Bias and
  Future Galaxy Surveys}}, \href{https://doi.org/10.1086/592020}{\emph{\apjl}
  {\bfseries 684} (2008) L1} [\href{https://arxiv.org/abs/0806.1950}{{\ttfamily
  0806.1950}}].

\bibitem{2008JCAP...08..031S}
A.~{Slosar}, C.~{Hirata}, U.~{Seljak}, S.~{Ho} and N.~{Padmanabhan},
  \emph{{Constraints on local primordial non-Gaussianity from large scale
  structure}},
  \href{https://doi.org/10.1088/1475-7516/2008/08/031}{\emph{\jcap} {\bfseries
  2008} (2008) 031} [\href{https://arxiv.org/abs/0805.3580}{{\ttfamily
  0805.3580}}].

\bibitem{2011PhRvD..84h3509H}
N.~{Hamaus}, U.~{Seljak} and V.~{Desjacques}, \emph{{Optimal constraints on
  local primordial non-Gaussianity from the two-point statistics of large-scale
  structure}}, \href{https://doi.org/10.1103/PhysRevD.84.083509}{\emph{\prd}
  {\bfseries 84} (2011) 083509}
  [\href{https://arxiv.org/abs/1104.2321}{{\ttfamily 1104.2321}}].

\bibitem{2008PhRvD..78l3534T}
A.~{Taruya}, K.~{Koyama} and T.~{Matsubara}, \emph{{Signature of primordial
  non-Gaussianity on the matter power spectrum}},
  \href{https://doi.org/10.1103/PhysRevD.78.123534}{\emph{\prd} {\bfseries 78}
  (2008) 123534} [\href{https://arxiv.org/abs/0808.4085}{{\ttfamily
  0808.4085}}].

\bibitem{2009MNRAS.396...85D}
V.~{Desjacques}, U.~{Seljak} and I.~T. {Iliev}, \emph{{Scale-dependent bias
  induced by local non-Gaussianity: a comparison to N-body simulations}},
  \href{https://doi.org/10.1111/j.1365-2966.2009.14721.x}{\emph{\mnras}
  {\bfseries 396} (2009) 85} [\href{https://arxiv.org/abs/0811.2748}{{\ttfamily
  0811.2748}}].

\bibitem{2013MNRAS.428.1116R}
A.~J. {Ross}, W.~J. {Percival}, A.~{Carnero}, G.-b. {Zhao}, M.~{Manera},
  A.~{Raccanelli} et~al., \emph{{The clustering of galaxies in the SDSS-III DR9
  Baryon Oscillation Spectroscopic Survey: constraints on primordial
  non-Gaussianity}}, \href{https://doi.org/10.1093/mnras/sts094}{\emph{\mnras}
  {\bfseries 428} (2013) 1116}
  [\href{https://arxiv.org/abs/1208.1491}{{\ttfamily 1208.1491}}].

\bibitem{2015PhRvD..91d3506F}
S.~{Ferraro} and K.~M. {Smith}, \emph{{Using large scale structure to measure
  f$_{NL}$ , g$_{NL}$ and {\ensuremath{\tau}}$_{NL}$}},
  \href{https://doi.org/10.1103/PhysRevD.91.043506}{\emph{\prd} {\bfseries 91}
  (2015) 043506} [\href{https://arxiv.org/abs/1408.3126}{{\ttfamily
  1408.3126}}].

\bibitem{2017PhRvD..95l3513D}
R.~{de Putter} and O.~{Dor{\'e}}, \emph{{Designing an inflation galaxy survey:
  How to measure {\ensuremath{\sigma}} (f$_{NL}$){\ensuremath{\sim}}1 using
  scale-dependent galaxy bias}},
  \href{https://doi.org/10.1103/PhysRevD.95.123513}{\emph{\prd} {\bfseries 95}
  (2017) 123513} [\href{https://arxiv.org/abs/1412.3854}{{\ttfamily
  1412.3854}}].

\bibitem{2018MNRAS.478.1341K}
D.~{Karagiannis}, A.~{Lazanu}, M.~{Liguori}, A.~{Raccanelli}, N.~{Bartolo} and
  L.~{Verde}, \emph{{Constraining primordial non-Gaussianity with bispectrum
  and power spectrum from upcoming optical and radio surveys}},
  \href{https://doi.org/10.1093/mnras/sty1029}{\emph{\mnras} {\bfseries 478}
  (2018) 1341} [\href{https://arxiv.org/abs/1801.09280}{{\ttfamily
  1801.09280}}].

\bibitem{2019JCAP...09..010C}
E.~{Castorina}, N.~{Hand}, U.~{Seljak}, F.~{Beutler}, C.-H. {Chuang}, C.~{Zhao}
  et~al., \emph{{Redshift-weighted constraints on primordial non-Gaussianity
  from the clustering of the eBOSS DR14 quasars in Fourier space}},
  \href{https://doi.org/10.1088/1475-7516/2019/09/010}{\emph{\jcap} {\bfseries
  2019} (2019) 010} [\href{https://arxiv.org/abs/1904.08859}{{\ttfamily
  1904.08859}}].

\bibitem{2020JCAP...11..052K}
D.~{Karagiannis}, A.~{Slosar} and M.~{Liguori}, \emph{{Forecasts on primordial
  non-Gaussianity from 21 cm intensity mapping experiments}},
  \href{https://doi.org/10.1088/1475-7516/2020/11/052}{\emph{\jcap} {\bfseries
  2020} (2020) 052} [\href{https://arxiv.org/abs/1911.03964}{{\ttfamily
  1911.03964}}].

\bibitem{2020JCAP...12..031B}
A.~{Barreira}, \emph{{On the impact of galaxy bias uncertainties on primordial
  non-Gaussianity constraints}},
  \href{https://doi.org/10.1088/1475-7516/2020/12/031}{\emph{\jcap} {\bfseries
  2020} (2020) 031} [\href{https://arxiv.org/abs/2009.06622}{{\ttfamily
  2009.06622}}].

\bibitem{2021arXiv210706887B}
A.~{Barreira}, \emph{{Predictions for local PNG bias in the galaxy power
  spectrum and bispectrum and the consequences for $f_{\mathrm {NL}}$
  constraints}}, {\emph{arXiv e-prints} (2021) arXiv:2107.06887}
  [\href{https://arxiv.org/abs/2107.06887}{{\ttfamily 2107.06887}}].

\bibitem{2021arXiv210208315P}
A.~{Pezzotta}, M.~{Crocce}, A.~{Eggemeier}, A.~G. {S{\'a}nchez} and
  R.~{Scoccimarro}, \emph{{Testing one-loop galaxy bias: cosmological
  constraints from the power spectrum}}, {\emph{arXiv e-prints} (2021)
  arXiv:2102.08315} [\href{https://arxiv.org/abs/2102.08315}{{\ttfamily
  2102.08315}}].

\bibitem{2009ApJ...703.1230J}
D.~{Jeong} and E.~{Komatsu}, \emph{{Primordial Non-Gaussianity, Scale-dependent
  Bias, and the Bispectrum of Galaxies}},
  \href{https://doi.org/10.1088/0004-637X/703/2/1230}{\emph{\apj} {\bfseries
  703} (2009) 1230} [\href{https://arxiv.org/abs/0904.0497}{{\ttfamily
  0904.0497}}].

\bibitem{2015JCAP...07..004T}
M.~{Tellarini}, A.~J. {Ross}, G.~{Tasinato} and D.~{Wands}, \emph{{Non-local
  bias in the halo bispectrum with primordial non-Gaussianity}},
  \href{https://doi.org/10.1088/1475-7516/2015/07/004}{\emph{\jcap} {\bfseries
  2015} (2015) 004} [\href{https://arxiv.org/abs/1504.00324}{{\ttfamily
  1504.00324}}].

\bibitem{2016JCAP...06..014T}
M.~{Tellarini}, A.~J. {Ross}, G.~{Tasinato} and D.~{Wands}, \emph{{Galaxy
  bispectrum, primordial non-Gaussianity and redshift space distortions}},
  \href{https://doi.org/10.1088/1475-7516/2016/06/014}{\emph{\jcap} {\bfseries
  2016} (2016) 014} [\href{https://arxiv.org/abs/1603.06814}{{\ttfamily
  1603.06814}}].

\bibitem{2017JCAP...03..006D}
E.~{Di Dio}, H.~{Perrier}, R.~{Durrer}, G.~{Marozzi}, A.~{Moradinezhad Dizgah},
  J.~{Nore{\~n}a} et~al., \emph{{Non-Gaussianities due to relativistic
  corrections to the observed galaxy bispectrum}},
  \href{https://doi.org/10.1088/1475-7516/2017/03/006}{\emph{\jcap} {\bfseries
  2017} (2017) 006} [\href{https://arxiv.org/abs/1611.03720}{{\ttfamily
  1611.03720}}].

\bibitem{2018JCAP...07..050K}
K.~{Koyama}, O.~{Umeh}, R.~{Maartens} and D.~{Bertacca}, \emph{{The observed
  galaxy bispectrum from single-field inflation in the squeezed limit}},
  \href{https://doi.org/10.1088/1475-7516/2018/07/050}{\emph{\jcap} {\bfseries
  2018} (2018) 050} [\href{https://arxiv.org/abs/1805.09189}{{\ttfamily
  1805.09189}}].

\bibitem{2021JCAP...04..013M}
R.~{Maartens}, S.~{Jolicoeur}, O.~{Umeh}, E.~M. {De Weerd} and C.~{Clarkson},
  \emph{{Local primordial non-Gaussianity in the relativistic galaxy
  bispectrum}},
  \href{https://doi.org/10.1088/1475-7516/2021/04/013}{\emph{\jcap} {\bfseries
  2021} (2021) 013} [\href{https://arxiv.org/abs/2011.13660}{{\ttfamily
  2011.13660}}].

\bibitem{2008PhRvD..77l3514D}
N.~{Dalal}, O.~{Dor{\'e}}, D.~{Huterer} and A.~{Shirokov}, \emph{{Imprints of
  primordial non-Gaussianities on large-scale structure: Scale-dependent bias
  and abundance of virialized objects}},
  \href{https://doi.org/10.1103/PhysRevD.77.123514}{\emph{\prd} {\bfseries 77}
  (2008) 123514} [\href{https://arxiv.org/abs/0710.4560}{{\ttfamily
  0710.4560}}].

\bibitem{2010PhRvD..81f3530G}
T.~{Giannantonio} and C.~{Porciani}, \emph{{Structure formation from
  non-Gaussian initial conditions: Multivariate biasing, statistics, and
  comparison with N-body simulations}},
  \href{https://doi.org/10.1103/PhysRevD.81.063530}{\emph{\prd} {\bfseries 81}
  (2010) 063530} [\href{https://arxiv.org/abs/0911.0017}{{\ttfamily
  0911.0017}}].

\bibitem{2015JCAP...09..029A}
R.~{Angulo}, M.~{Fasiello}, L.~{Senatore} and Z.~{Vlah}, \emph{{On the
  statistics of biased tracers in the Effective Field Theory of Large Scale
  Structures}},
  \href{https://doi.org/10.1088/1475-7516/2015/09/029}{\emph{\jcap} {\bfseries
  2015} (2015) 029} [\href{https://arxiv.org/abs/1503.08826}{{\ttfamily
  1503.08826}}].

\bibitem{2015JCAP...12..043A}
V.~{Assassi}, D.~{Baumann} and F.~{Schmidt}, \emph{{Galaxy bias and primordial
  non-Gaussianity}},
  \href{https://doi.org/10.1088/1475-7516/2015/12/043}{\emph{\jcap} {\bfseries
  2015} (2015) 043} [\href{https://arxiv.org/abs/1510.03723}{{\ttfamily
  1510.03723}}].

\bibitem{2020PhRvD.102j3530E}
A.~{Eggemeier}, R.~{Scoccimarro}, M.~{Crocce}, A.~{Pezzotta} and A.~G.
  {S{\'a}nchez}, \emph{{Testing one-loop galaxy bias: Power spectrum}},
  \href{https://doi.org/10.1103/PhysRevD.102.103530}{\emph{\prd} {\bfseries
  102} (2020) 103530} [\href{https://arxiv.org/abs/2006.09729}{{\ttfamily
  2006.09729}}].

\bibitem{2020JCAP...08..007D}
J.-P. {Dai}, L.~{Verde} and J.-Q. {Xia}, \emph{{What can we learn by combining
  the skew spectrum and the power spectrum?}},
  \href{https://doi.org/10.1088/1475-7516/2020/08/007}{\emph{\jcap} {\bfseries
  2020} (2020) 007} [\href{https://arxiv.org/abs/2002.09904}{{\ttfamily
  2002.09904}}].

\bibitem{2018PhR...733....1D}
V.~{Desjacques}, D.~{Jeong} and F.~{Schmidt}, \emph{{Large-scale galaxy bias}},
  \href{https://doi.org/10.1016/j.physrep.2017.12.002}{\emph{\physrep}
  {\bfseries 733} (2018) 1} [\href{https://arxiv.org/abs/1611.09787}{{\ttfamily
  1611.09787}}].

\bibitem{1994ARA&A..32..319W}
M.~{White}, D.~{Scott} and J.~{Silk}, \emph{{Anisotropies in the Cosmic
  Microwave Background}},
  \href{https://doi.org/10.1146/annurev.astro.32.1.319}{\emph{\araa} {\bfseries
  32} (1994) 319}.

\bibitem{1995Sci...268..829S}
D.~{Scott}, J.~{Silk} and M.~{White}, \emph{{From Microwave Anisotropies to
  Cosmology}},
  \href{https://doi.org/10.1126/science.268.5212.829}{\emph{Science} {\bfseries
  268} (1995) 829} [\href{https://arxiv.org/abs/astro-ph/9505015}{{\ttfamily
  astro-ph/9505015}}].

\bibitem{2002PhRvD..66j3508T}
M.~{Tegmark} and M.~{Zaldarriaga}, \emph{{Separating the early universe from
  the late universe: Cosmological parameter estimation beyond the black box}},
  \href{https://doi.org/10.1103/PhysRevD.66.103508}{\emph{\prd} {\bfseries 66}
  (2002) 103508} [\href{https://arxiv.org/abs/astro-ph/0207047}{{\ttfamily
  astro-ph/0207047}}].

\bibitem{2004ApJ...606..702T}
M.~{Tegmark}, M.~R. {Blanton}, M.~A. {Strauss}, F.~{Hoyle}, D.~{Schlegel},
  R.~{Scoccimarro} et~al., \emph{{The Three-Dimensional Power Spectrum of
  Galaxies from the Sloan Digital Sky Survey}},
  \href{https://doi.org/10.1086/382125}{\emph{\apj} {\bfseries 606} (2004) 702}
  [\href{https://arxiv.org/abs/astro-ph/0310725}{{\ttfamily
  astro-ph/0310725}}].

\bibitem{2014ApJ...794L..11B}
M.~{Bruni}, J.~C. {Hidalgo} and D.~{Wands}, \emph{{Einstein's Signature in
  Cosmological Large-scale Structure}},
  \href{https://doi.org/10.1088/2041-8205/794/1/L11}{\emph{\apjl} {\bfseries
  794} (2014) L11} [\href{https://arxiv.org/abs/1405.7006}{{\ttfamily
  1405.7006}}].

\bibitem{2016PDU....13...30B}
N.~{Bartolo}, D.~{Bertacca}, M.~{Bruni}, K.~{Koyama}, R.~{Maartens},
  S.~{Matarrese} et~al., \emph{{A relativistic signature in large-scale
  structure}}, \href{https://doi.org/10.1016/j.dark.2016.04.002}{\emph{Physics
  of the Dark Universe} {\bfseries 13} (2016) 30}
  [\href{https://arxiv.org/abs/1506.00915}{{\ttfamily 1506.00915}}].

\bibitem{2009PhRvD..80h3514Y}
J.~{Yoo}, A.~L. {Fitzpatrick} and M.~{Zaldarriaga}, \emph{{New perspective on
  galaxy clustering as a cosmological probe: General relativistic effects}},
  \href{https://doi.org/10.1103/PhysRevD.80.083514}{\emph{\prd} {\bfseries 80}
  (2009) 083514} [\href{https://arxiv.org/abs/0907.0707}{{\ttfamily
  0907.0707}}].

\bibitem{2010PhRvD..82h3508Y}
J.~{Yoo}, \emph{{General relativistic description of the observed galaxy power
  spectrum: Do we understand what we measure?}},
  \href{https://doi.org/10.1103/PhysRevD.82.083508}{\emph{\prd} {\bfseries 82}
  (2010) 083508} [\href{https://arxiv.org/abs/1009.3021}{{\ttfamily
  1009.3021}}].

\bibitem{2011PhRvD..84d3516C}
A.~{Challinor} and A.~{Lewis}, \emph{{Linear power spectrum of observed source
  number counts}},
  \href{https://doi.org/10.1103/PhysRevD.84.043516}{\emph{\prd} {\bfseries 84}
  (2011) 043516} [\href{https://arxiv.org/abs/1105.5292}{{\ttfamily
  1105.5292}}].

\bibitem{2011PhRvD..84f3505B}
C.~{Bonvin} and R.~{Durrer}, \emph{{What galaxy surveys really measure}},
  \href{https://doi.org/10.1103/PhysRevD.84.063505}{\emph{\prd} {\bfseries 84}
  (2011) 063505} [\href{https://arxiv.org/abs/1105.5280}{{\ttfamily
  1105.5280}}].

\bibitem{2012PhRvD..85b3504J}
D.~{Jeong}, F.~{Schmidt} and C.~M. {Hirata}, \emph{{Large-scale clustering of
  galaxies in general relativity}},
  \href{https://doi.org/10.1103/PhysRevD.85.023504}{\emph{\prd} {\bfseries 85}
  (2012) 023504} [\href{https://arxiv.org/abs/1107.5427}{{\ttfamily
  1107.5427}}].

\bibitem{2014PhRvD..90l3507Y}
J.~{Yoo}, \emph{{Proper-time hypersurface of nonrelativistic matter flows:
  Galaxy bias in general relativity}},
  \href{https://doi.org/10.1103/PhysRevD.90.123507}{\emph{\prd} {\bfseries 90}
  (2014) 123507} [\href{https://arxiv.org/abs/1408.5137}{{\ttfamily
  1408.5137}}].

\bibitem{2014JCAP...09..037B}
D.~{Bertacca}, R.~{Maartens} and C.~{Clarkson}, \emph{{Observed galaxy number
  counts on the lightcone up to second order: I. Main result}},
  \href{https://doi.org/10.1088/1475-7516/2014/09/037}{\emph{\jcap} {\bfseries
  2014} (2014) 037} [\href{https://arxiv.org/abs/1405.4403}{{\ttfamily
  1405.4403}}].

\bibitem{2014JCAP...11..013B}
D.~{Bertacca}, R.~{Maartens} and C.~{Clarkson}, \emph{{Observed galaxy number
  counts on the lightcone up to second order: II. Derivation}},
  \href{https://doi.org/10.1088/1475-7516/2014/11/013}{\emph{\jcap} {\bfseries
  2014} (2014) 013} [\href{https://arxiv.org/abs/1406.0319}{{\ttfamily
  1406.0319}}].

\bibitem{2014PhRvD..90b3513Y}
J.~{Yoo} and M.~{Zaldarriaga}, \emph{{Beyond the linear-order relativistic
  effect in galaxy clustering: Second-order gauge-invariant formalism}},
  \href{https://doi.org/10.1103/PhysRevD.90.023513}{\emph{\prd} {\bfseries 90}
  (2014) 023513} [\href{https://arxiv.org/abs/1406.4140}{{\ttfamily
  1406.4140}}].

\bibitem{2014JCAP...12..017D}
E.~{Di Dio}, R.~{Durrer}, G.~{Marozzi} and F.~{Montanari}, \emph{{Galaxy number
  counts to second order and their bispectrum}},
  \href{https://doi.org/10.1088/1475-7516/2014/12/017}{\emph{\jcap} {\bfseries
  2014} (2014) 017} [\href{https://arxiv.org/abs/1407.0376}{{\ttfamily
  1407.0376}}].

\bibitem{2015PhRvD..91d3507C}
S.~{Chen} and D.~J. {Schwarz}, \emph{{Fluctuations of differential number
  counts of radio continuum sources}},
  \href{https://doi.org/10.1103/PhysRevD.91.043507}{\emph{\prd} {\bfseries 91}
  (2015) 043507} [\href{https://arxiv.org/abs/1407.4682}{{\ttfamily
  1407.4682}}].

\bibitem{2016JCAP...09..046Y}
J.~{Yoo} and F.~{Scaccabarozzi}, \emph{{Unified treatment of the luminosity
  distance in cosmology}},
  \href{https://doi.org/10.1088/1475-7516/2016/09/046}{\emph{\jcap} {\bfseries
  2016} (2016) 046} [\href{https://arxiv.org/abs/1606.08453}{{\ttfamily
  1606.08453}}].

\bibitem{2017JCAP...03..010T}
J.~{Tr{\o}st Nielsen} and R.~{Durrer}, \emph{{Higher order relativistic galaxy
  number counts: dominating terms}},
  \href{https://doi.org/10.1088/1475-7516/2017/03/010}{\emph{\jcap} {\bfseries
  2017} (2017) 010} [\href{https://arxiv.org/abs/1606.02113}{{\ttfamily
  1606.02113}}].

\bibitem{2017JCAP...03..034U}
O.~{Umeh}, S.~{Jolicoeur}, R.~{Maartens} and C.~{Clarkson}, \emph{{A general
  relativistic signature in the galaxy bispectrum: the local effects of
  observing on the lightcone}},
  \href{https://doi.org/10.1088/1475-7516/2017/03/034}{\emph{\jcap} {\bfseries
  2017} (2017) 034} [\href{https://arxiv.org/abs/1610.03351}{{\ttfamily
  1610.03351}}].

\bibitem{2017JCAP...09..040J}
S.~{Jolicoeur}, O.~{Umeh}, R.~{Maartens} and C.~{Clarkson}, \emph{{Imprints of
  local lightcone \textbackslash projection effects on the galaxy bispectrum.
  Part II}}, \href{https://doi.org/10.1088/1475-7516/2017/09/040}{\emph{\jcap}
  {\bfseries 2017} (2017) 040}
  [\href{https://arxiv.org/abs/1703.09630}{{\ttfamily 1703.09630}}].

\bibitem{2018JCAP...09..037F}
G.~{Fanizza}, J.~{Yoo} and S.~G. {Biern}, \emph{{Non-linear general
  relativistic effects in the observed redshift}},
  \href{https://doi.org/10.1088/1475-7516/2018/09/037}{\emph{\jcap} {\bfseries
  2018} (2018) 037} [\href{https://arxiv.org/abs/1805.05959}{{\ttfamily
  1805.05959}}].

\bibitem{2019arXiv190808400F}
J.~L. {Fuentes}, J.~C. {Hidalgo} and K.~A. {Malik}, \emph{{Galaxy number counts
  at second order: an independent approach}}, {\emph{arXiv e-prints} (2019)
  arXiv:1908.08400} [\href{https://arxiv.org/abs/1908.08400}{{\ttfamily
  1908.08400}}].

\bibitem{2019MNRAS.486L.101C}
C.~{Clarkson}, E.~M. {de Weerd}, S.~{Jolicoeur}, R.~{Maartens} and O.~{Umeh},
  \emph{{The dipole of the galaxy bispectrum}},
  \href{https://doi.org/10.1093/mnrasl/slz066}{\emph{\mnras} {\bfseries 486}
  (2019) L101} [\href{https://arxiv.org/abs/1812.09512}{{\ttfamily
  1812.09512}}].

\bibitem{2019arXiv190501293M}
E.~{Mitsou}, J.~{Yoo}, R.~{Durrer}, F.~{Scaccabarozzi} and V.~{Tansella},
  \emph{{Angular $N$-point spectra and cosmic variance on the light-cone}},
  {\emph{arXiv e-prints} (2019) arXiv:1905.01293}
  [\href{https://arxiv.org/abs/1905.01293}{{\ttfamily 1905.01293}}].

\bibitem{2020JCAP...03..065M}
R.~{Maartens}, S.~{Jolicoeur}, O.~{Umeh}, E.~M. {De Weerd}, C.~{Clarkson} and
  S.~{Camera}, \emph{{Detecting the relativistic galaxy bispectrum}},
  \href{https://doi.org/10.1088/1475-7516/2020/03/065}{\emph{\jcap} {\bfseries
  2020} (2020) 065} [\href{https://arxiv.org/abs/1911.02398}{{\ttfamily
  1911.02398}}].

\bibitem{2018PhRvD..97b3531B}
D.~{Bertacca}, A.~{Raccanelli}, N.~{Bartolo}, M.~{Liguori}, S.~{Matarrese} and
  L.~{Verde}, \emph{{Relativistic wide-angle galaxy bispectrum on the light
  cone}}, \href{https://doi.org/10.1103/PhysRevD.97.023531}{\emph{\prd}
  {\bfseries 97} (2018) 023531}
  [\href{https://arxiv.org/abs/1705.09306}{{\ttfamily 1705.09306}}].

\bibitem{2020arXiv201215326F}
J.~L. {Fuentes}, J.~C. {Hidalgo} and K.~A. {Malik}, \emph{{Galaxy number counts
  at second order in perturbation theory: a leading-order term comparison}},
  {\emph{arXiv e-prints} (2020) arXiv:2012.15326}
  [\href{https://arxiv.org/abs/2012.15326}{{\ttfamily 2012.15326}}].

\bibitem{2016JCAP...07..053A}
J.~{Adamek}, D.~{Daverio}, R.~{Durrer} and M.~{Kunz}, \emph{{gevolution: a
  cosmological N-body code based on General Relativity}},
  \href{https://doi.org/10.1088/1475-7516/2016/07/053}{\emph{\jcap} {\bfseries
  2016} (2016) 053} [\href{https://arxiv.org/abs/1604.06065}{{\ttfamily
  1604.06065}}].

\bibitem{2016PhRvL.116y1302B}
E.~{Bentivegna} and M.~{Bruni}, \emph{{Effects of Nonlinear Inhomogeneity on
  the Cosmic Expansion with Numerical Relativity}},
  \href{https://doi.org/10.1103/PhysRevLett.116.251302}{\emph{\prl} {\bfseries
  116} (2016) 251302} [\href{https://arxiv.org/abs/1511.05124}{{\ttfamily
  1511.05124}}].

\bibitem{2020JCAP...01..007B}
C.~{Barrera-Hinojosa} and B.~{Li}, \emph{{GRAMSES: a new route to general
  relativistic N-body simulations in cosmology. Part I. Methodology and code
  description}},
  \href{https://doi.org/10.1088/1475-7516/2020/01/007}{\emph{\jcap} {\bfseries
  2020} (2020) 007} [\href{https://arxiv.org/abs/1905.08890}{{\ttfamily
  1905.08890}}].

\bibitem{2011JCAP...10..031B}
T.~{Baldauf}, U.~{Seljak}, L.~{Senatore} and M.~{Zaldarriaga}, \emph{{Galaxy
  bias and non-linear structure formation in general relativity}},
  \href{https://doi.org/10.1088/1475-7516/2011/10/031}{\emph{\jcap} {\bfseries
  2011} (2011) 031} [\href{https://arxiv.org/abs/1106.5507}{{\ttfamily
  1106.5507}}].

\bibitem{2011JCAP...04..011B}
N.~{Bartolo}, S.~{Matarrese} and A.~{Riotto}, \emph{{Relativistic effects and
  primordial non-Gaussianity in the galaxy bias}},
  \href{https://doi.org/10.1088/1475-7516/2011/04/011}{\emph{\jcap} {\bfseries
  2011} (2011) 011} [\href{https://arxiv.org/abs/1011.4374}{{\ttfamily
  1011.4374}}].

\bibitem{2012PhRvD..85d1301B}
M.~{Bruni}, R.~{Crittenden}, K.~{Koyama}, R.~{Maartens}, C.~{Pitrou} and
  D.~{Wands}, \emph{{Disentangling non-Gaussianity, bias, and general
  relativistic effects in the galaxy distribution}},
  \href{https://doi.org/10.1103/PhysRevD.85.041301}{\emph{\prd} {\bfseries 85}
  (2012) 041301} [\href{https://arxiv.org/abs/1106.3999}{{\ttfamily
  1106.3999}}].

\bibitem{2015CQGra..32q5019B}
D.~{Bertacca}, N.~{Bartolo}, M.~{Bruni}, K.~{Koyama}, R.~{Maartens},
  S.~{Matarrese} et~al., \emph{{Galaxy bias and gauges at second order in
  general relativity}},
  \href{https://doi.org/10.1088/0264-9381/32/17/175019}{\emph{Classical and
  Quantum Gravity} {\bfseries 32} (2015) 175019}
  [\href{https://arxiv.org/abs/1501.03163}{{\ttfamily 1501.03163}}].

\bibitem{2019JCAP...05..020U}
O.~{Umeh}, K.~{Koyama}, R.~{Maartens}, F.~{Schmidt} and C.~{Clarkson},
  \emph{{General relativistic effects in the galaxy bias at second order}},
  \href{https://doi.org/10.1088/1475-7516/2019/05/020}{\emph{\jcap} {\bfseries
  2019} (2019) 020} [\href{https://arxiv.org/abs/1901.07460}{{\ttfamily
  1901.07460}}].

\bibitem{2019JCAP...12..048U}
O.~{Umeh} and K.~{Koyama}, \emph{{The galaxy bias at second order in general
  relativity with non-Gaussian initial conditions}},
  \href{https://doi.org/10.1088/1475-7516/2019/12/048}{\emph{\jcap} {\bfseries
  2019} (2019) 048} [\href{https://arxiv.org/abs/1907.08094}{{\ttfamily
  1907.08094}}].

\bibitem{2019JCAP...07..030C}
L.~{Castiblanco}, R.~{Gannouji}, J.~{Nore{\~n}a} and C.~{Stahl},
  \emph{{Relativistic cosmological large scale structures at one-loop}},
  \href{https://doi.org/10.1088/1475-7516/2019/07/030}{\emph{\jcap} {\bfseries
  2019} (2019) 030} [\href{https://arxiv.org/abs/1811.05452}{{\ttfamily
  1811.05452}}].

\bibitem{2019arXiv191213034C}
J.~{Calles}, L.~{Castiblanco}, J.~{Nore{\~n}a} and C.~{Stahl}, \emph{{From
  matter to galaxies: General relativistic bias for the one-loop bispectrum}},
  {\emph{arXiv e-prints} (2019) arXiv:1912.13034}
  [\href{https://arxiv.org/abs/1912.13034}{{\ttfamily 1912.13034}}].

\bibitem{2020JCAP...11..064G}
N.~{Grimm}, F.~{Scaccabarozzi}, J.~{Yoo}, S.~G. {Biern} and J.-O. {Gong},
  \emph{{Galaxy power spectrum in general relativity}},
  \href{https://doi.org/10.1088/1475-7516/2020/11/064}{\emph{\jcap} {\bfseries
  2020} (2020) 064} [\href{https://arxiv.org/abs/2005.06484}{{\ttfamily
  2005.06484}}].

\bibitem{2021arXiv210608857C}
E.~{Castorina} and E.~{di Dio}, \emph{{The observed galaxy power spectrum in
  General Relativity}}, {\emph{arXiv e-prints} (2021) arXiv:2106.08857}
  [\href{https://arxiv.org/abs/2106.08857}{{\ttfamily 2106.08857}}].

\bibitem{2012reco.book.....E}
G.~F.~R. {Ellis}, R.~{Maartens} and M.~A.~H. {MacCallum}, \emph{{Relativistic
  Cosmology}}. ``Cambridge University Press, Cambridge U.K.", 2012.

\bibitem{2009fcgr.book.....S}
B.~{Schutz}, \emph{{A First Course in General Relativity}}. ``Cambridge
  University Press", 2009.

\bibitem{2008cosm.book.....W}
S.~{Weinberg}, \emph{{Cosmology}}. ``Oxford University Press", 2008.

\bibitem{2016inco.book.....R}
B.~{Ryden}, \emph{{Introduction to Cosmology}}. ``Cambridge University Press",
  2016.

\bibitem{1965ApJ...142..419P}
A.~A. {Penzias} and R.~W. {Wilson}, \emph{{A Measurement of Excess Antenna
  Temperature at 4080 Mc/s.}},
  \href{https://doi.org/10.1086/148307}{\emph{\apj} {\bfseries 142} (1965)
  419}.

\bibitem{2013arXiv1302.4640L}
J.~{Lesgourgues}, \emph{{TASI Lectures on Cosmological Perturbations}},
  {\emph{arXiv e-prints} (2013) arXiv:1302.4640}
  [\href{https://arxiv.org/abs/1302.4640}{{\ttfamily 1302.4640}}].

\bibitem{2014A&A...571A...1P}
{Planck Collaboration}, P.~A.~R. {Ade}, N.~{Aghanim}, M.~I.~R. {Alves},
  C.~{Armitage-Caplan}, M.~{Arnaud} et~al., \emph{{Planck 2013 results. I.
  Overview of products and scientific results}},
  \href{https://doi.org/10.1051/0004-6361/201321529}{\emph{\aap} {\bfseries
  571} (2014) A1} [\href{https://arxiv.org/abs/1303.5062}{{\ttfamily
  1303.5062}}].

\bibitem{Baumann}
D.~{Baumann}, ``Advanced cosmology (lecture notes).''
  \url{http://cosmology.amsterdam/education/advanced-cosmology/}, 2021.

\bibitem{2002astro.ph..9504R}
D.~D. {Reid}, D.~W. {Kittell}, E.~E. {Arsznov} and G.~B. {Thompson}, \emph{{The
  picture of our universe: A view from modern cosmology}}, {\emph{arXiv
  e-prints} (2002) astro}
  [\href{https://arxiv.org/abs/astro-ph/0209504}{{\ttfamily
  astro-ph/0209504}}].

\bibitem{2008arXiv0802.3688H}
W.~{Hu}, \emph{{Lecture Notes on CMB Theory: From Nucleosynthesis to
  Recombination}}, {\emph{arXiv e-prints} (2008) arXiv:0802.3688}
  [\href{https://arxiv.org/abs/0802.3688}{{\ttfamily 0802.3688}}].

\bibitem{2016ASSP...45....3W}
D.~{Wands}, O.~F. {Piattella} and L.~{Casarini}, \emph{{Physics of the Cosmic
  Microwave Background Radiation}},  in \emph{The Cosmic Microwave Background},
  vol.~45 of \emph{Astrophysics and Space Science Proceedings}, p.~3, Jan.,
  2016, \href{https://doi.org/K35-72414}{DOI}
  [\href{https://arxiv.org/abs/1504.06335}{{\ttfamily 1504.06335}}].

\bibitem{2017GReGr..49...18L}
E.~{Lifshitz}, \emph{{Republication of: On the gravitational stability of the
  expanding universe}},
  \href{https://doi.org/10.1007/s10714-016-2165-8}{\emph{General Relativity and
  Gravitation} {\bfseries 49} (2017) 18}.

\bibitem{1963AdPhy..12..185L}
E.~M. {Lifshitz} and I.~M. {Khalatnikov}, \emph{{Investigations in relativistic
  cosmology{\textdagger}}},
  \href{https://doi.org/10.1080/00018736300101283}{\emph{Advances in Physics}
  {\bfseries 12} (1963) 185}.

\bibitem{1967PThPh..37..831T}
K.~{Tomita}, \emph{{Non-Linear Theory of Gravitational Instability in the
  Expanding Universe}},
  \href{https://doi.org/10.1143/PTP.37.831}{\emph{Progress of Theoretical
  Physics} {\bfseries 37} (1967) 831}.

\bibitem{1980PhRvD..22.1882B}
J.~M. {Bardeen}, \emph{{Gauge-invariant cosmological perturbations}},
  \href{https://doi.org/10.1103/PhysRevD.22.1882}{\emph{\prd} {\bfseries 22}
  (1980) 1882}.

\bibitem{1984PThPS..78....1K}
H.~{Kodama} and M.~{Sasaki}, \emph{{Cosmological Perturbation Theory}},
  \href{https://doi.org/10.1143/PTPS.78.1}{\emph{Progress of Theoretical
  Physics Supplement} {\bfseries 78} (1984) 1}.

\bibitem{1992PhR...215..203M}
V.~F. {Mukhanov}, H.~A. {Feldman} and R.~H. {Brandenberger}, \emph{{Theory of
  cosmological perturbations}},
  \href{https://doi.org/10.1016/0370-1573(92)90044-Z}{\emph{\physrep}
  {\bfseries 215} (1992) 203}.

\bibitem{2009PhR...475....1M}
K.~A. {Malik} and D.~{Wands}, \emph{{Cosmological perturbations}},
  \href{https://doi.org/10.1016/j.physrep.2009.03.001}{\emph{\physrep}
  {\bfseries 475} (2009) 1} [\href{https://arxiv.org/abs/0809.4944}{{\ttfamily
  0809.4944}}].

\bibitem{1995ApJ...455....7M}
C.-P. {Ma} and E.~{Bertschinger}, \emph{{Cosmological Perturbation Theory in
  the Synchronous and Conformal Newtonian Gauges}},
  \href{https://doi.org/10.1086/176550}{\emph{\apj} {\bfseries 455} (1995) 7}
  [\href{https://arxiv.org/abs/astro-ph/9506072}{{\ttfamily
  astro-ph/9506072}}].

\bibitem{1997CQGra..14.2585B}
M.~{Bruni}, S.~{Matarrese}, S.~{Mollerach} and S.~{Sonego},
  \emph{{Perturbations of spacetime: gauge transformations and gauge invariance
  at second order and beyond}},
  \href{https://doi.org/10.1088/0264-9381/14/9/014}{\emph{Classical and Quantum
  Gravity} {\bfseries 14} (1997) 2585}
  [\href{https://arxiv.org/abs/gr-qc/9609040}{{\ttfamily gr-qc/9609040}}].

\bibitem{1990CQGra...7.1169S}
J.~M. {Stewart}, \emph{{Perturbations of Friedmann-Robertson-Walker
  cosmological models}},
  \href{https://doi.org/10.1088/0264-9381/7/7/013}{\emph{Classical and Quantum
  Gravity} {\bfseries 7} (1990) 1169}.

\bibitem{2016JCAP...01..030V}
E.~{Villa} and C.~{Rampf}, \emph{{Relativistic perturbations in
  {\ensuremath{\Lambda}}CDM: Eulerian \& Lagrangian approaches}},
  \href{https://doi.org/10.1088/1475-7516/2016/01/030}{\emph{\jcap} {\bfseries
  2016} (2016) 030} [\href{https://arxiv.org/abs/1505.04782}{{\ttfamily
  1505.04782}}].

\bibitem{Cosper}
H.~{Kurki-Suonio}, ``Cosmological perturbation theory, part 1 (lecture
  notes).'' \url{https://www.mv.helsinki.fi/home/hkurkisu/CosPer.pdf}, 2020.

\bibitem{2011PhRvD..83l3512C}
A.~J. {Christopherson}, K.~A. {Malik} and D.~R. {Matravers}, \emph{{Estimating
  the amount of vorticity generated by cosmological perturbations in the early
  universe}}, \href{https://doi.org/10.1103/PhysRevD.83.123512}{\emph{\prd}
  {\bfseries 83} (2011) 123512}
  [\href{https://arxiv.org/abs/1008.4866}{{\ttfamily 1008.4866}}].

\bibitem{2014JCAP...09..023N}
E.~{Nalson}, A.~J. {Christopherson} and K.~A. {Malik}, \emph{{Effects of
  non-linearities on magnetic field generation}},
  \href{https://doi.org/10.1088/1475-7516/2014/09/023}{\emph{\jcap} {\bfseries
  2014} (2014) 023} [\href{https://arxiv.org/abs/1312.6504}{{\ttfamily
  1312.6504}}].

\bibitem{2001astro.ph..1009B}
E.~{Bertschinger}, \emph{{Cosmological Perturbation Theory and Structure
  Formation}}, {\emph{arXiv e-prints} (2000) astro}
  [\href{https://arxiv.org/abs/astro-ph/0101009}{{\ttfamily
  astro-ph/0101009}}].

\bibitem{2003gr.qc.....3004M}
P.~G. {Miedema} and W.~A. {van Leeuwen}, \emph{{Density Perturbations in the
  Early Universe}}, {\emph{arXiv e-prints} (2003) gr}
  [\href{https://arxiv.org/abs/gr-qc/0303004}{{\ttfamily gr-qc/0303004}}].

\bibitem{1986ApJ...311....6G}
M.~H. {Goroff}, B.~{Grinstein}, S.~J. {Rey} and M.~B. {Wise}, \emph{{Coupling
  of modes of cosmological mass density fluctuations}},
  \href{https://doi.org/10.1086/164749}{\emph{\apj} {\bfseries 311} (1986) 6}.

\bibitem{1981MNRAS.197..931J}
R.~{Juszkiewicz}, \emph{{On the evolution of cosmological adiabatic
  perturbations in the weakly non-linear regime}},
  \href{https://doi.org/10.1093/mnras/197.4.931}{\emph{\mnras} {\bfseries 197}
  (1981) 931}.

\bibitem{1983MNRAS.203..345V}
E.~T. {Vishniac}, \emph{{Why weakly non-linear effects are small in a
  zero-pressure cosmology}},
  \href{https://doi.org/10.1093/mnras/203.2.345}{\emph{\mnras} {\bfseries 203}
  (1983) 345}.

\bibitem{2010PhDT.........4J}
D.~{Jeong}, \emph{{Cosmology with high $(z>1)$ redshift galaxy surveys}}, Ph.D.
  thesis, University of Texas at Austin, Aug., 2010.

\bibitem{Peebles1980}
P.~J.~E. {Peebles}, \emph{{The Large-Scale Structure of the Universe}}.
  ``Princeton University Press", 1980.

\bibitem{Baldauf}
T.~{Baldauf}, ``Advanced cosmology statistics, non-gaussianity and
  non-linearity (lecture notes).''
  \url{http://www.damtp.cam.ac.uk/user/tb561/AdvCosmo/AdvCosmo19_notes0803.pdf},
  2019.

\bibitem{2015daen.book.....A}
L.~{Amendola} and S.~{Tsujikawa}, \emph{{Dark Energy}}. ``Cambridge University
  Press, Cambridge U.K.", 2015.

\bibitem{1984ApJ...279..499F}
J.~N. {Fry}, \emph{{The Galaxy correlation hierarchy in perturbation theory}},
  \href{https://doi.org/10.1086/161913}{\emph{\apj} {\bfseries 279} (1984)
  499}.

\bibitem{2011JCAP...10..026L}
A.~{Lewis}, \emph{{The real shape of non-Gaussianities}},
  \href{https://doi.org/10.1088/1475-7516/2011/10/026}{\emph{\jcap} {\bfseries
  2011} (2011) 026} [\href{https://arxiv.org/abs/1107.5431}{{\ttfamily
  1107.5431}}].

\bibitem{2018arXiv180300070P}
O.~F. {Piattella}, \emph{{Lecture Notes in Cosmology}}, {\emph{arXiv e-prints}
  (2018) arXiv:1803.00070} [\href{https://arxiv.org/abs/1803.00070}{{\ttfamily
  1803.00070}}].

\bibitem{2016A&A...594A..17P}
{Planck Collaboration}, P.~A.~R. {Ade}, N.~{Aghanim}, M.~{Arnaud}, F.~{Arroja},
  M.~{Ashdown} et~al., \emph{{Planck 2015 results. XVII. Constraints on
  primordial non-Gaussianity}},
  \href{https://doi.org/10.1051/0004-6361/201525836}{\emph{\aap} {\bfseries
  594} (2016) A17} [\href{https://arxiv.org/abs/1502.01592}{{\ttfamily
  1502.01592}}].

\bibitem{2003JHEP...05..013M}
J.~{Maldacena}, \emph{{Non-gaussian features of primordial fluctuations in
  single field inflationary models}},
  \href{https://doi.org/10.1088/1126-6708/2003/05/013}{\emph{Journal of High
  Energy Physics} {\bfseries 2003} (2003) 013}
  [\href{https://arxiv.org/abs/astro-ph/0210603}{{\ttfamily
  astro-ph/0210603}}].

\bibitem{2003NuPhB.667..119A}
V.~{Acquaviva}, N.~{Bartolo}, S.~{Matarrese} and A.~{Riotto},
  \emph{{Gauge-invariant second-order perturbations and non-Gaussianity from
  inflation}},
  \href{https://doi.org/10.1016/S0550-3213(03)00550-9}{\emph{Nuclear Physics B}
  {\bfseries 667} (2003) 119}
  [\href{https://arxiv.org/abs/astro-ph/0209156}{{\ttfamily
  astro-ph/0209156}}].

\bibitem{2007JCAP...01..002C}
X.~{Chen}, M.-x. {Huang}, S.~{Kachru} and G.~{Shiu}, \emph{{Observational
  signatures and non-Gaussianities of general single-field inflation}},
  \href{https://doi.org/10.1088/1475-7516/2007/01/002}{\emph{\jcap} {\bfseries
  2007} (2007) 002} [\href{https://arxiv.org/abs/hep-th/0605045}{{\ttfamily
  hep-th/0605045}}].

\bibitem{2018arXiv181208197C}
M.~{Celoria} and S.~{Matarrese}, \emph{{Primordial Non-Gaussianity}},
  {\emph{arXiv e-prints} (2018) arXiv:1812.08197}
  [\href{https://arxiv.org/abs/1812.08197}{{\ttfamily 1812.08197}}].

\bibitem{2000MNRAS.313..141V}
L.~{Verde}, L.~{Wang}, A.~F. {Heavens} and M.~{Kamionkowski},
  \emph{{Large-scale structure, the cosmic microwave background and primordial
  non-Gaussianity}},
  \href{https://doi.org/10.1046/j.1365-8711.2000.03191.x}{\emph{\mnras}
  {\bfseries 313} (2000) 141}
  [\href{https://arxiv.org/abs/astro-ph/9906301}{{\ttfamily
  astro-ph/9906301}}].

\bibitem{2001PhRvD..63f3002K}
E.~{Komatsu} and D.~N. {Spergel}, \emph{{Acoustic signatures in the primary
  microwave background bispectrum}},
  \href{https://doi.org/10.1103/PhysRevD.63.063002}{\emph{\prd} {\bfseries 63}
  (2001) 063002} [\href{https://arxiv.org/abs/astro-ph/0005036}{{\ttfamily
  astro-ph/0005036}}].

\bibitem{2010CQGra..27l4010K}
E.~{Komatsu}, \emph{{Hunting for primordial non-Gaussianity in the cosmic
  microwave background}},
  \href{https://doi.org/10.1088/0264-9381/27/12/124010}{\emph{Classical and
  Quantum Gravity} {\bfseries 27} (2010) 124010}
  [\href{https://arxiv.org/abs/1003.6097}{{\ttfamily 1003.6097}}].

\bibitem{2020A&A...641A...9P}
{Planck Collaboration}, Y.~{Akrami}, F.~{Arroja}, M.~{Ashdown}, J.~{Aumont},
  C.~{Baccigalupi} et~al., \emph{{Planck 2018 results. IX. Constraints on
  primordial non-Gaussianity}},
  \href{https://doi.org/10.1051/0004-6361/201935891}{\emph{\aap} {\bfseries
  641} (2020) A9} [\href{https://arxiv.org/abs/1905.05697}{{\ttfamily
  1905.05697}}].

\bibitem{2020JCAP...04..028M}
R.~{Martinez-Carrillo}, J.~{De-Santiago}, J.~C. {Hidalgo} and K.~A. {Malik},
  \emph{{Relativistic and non-Gaussianity contributions to the one-loop power
  spectrum}}, \href{https://doi.org/10.1088/1475-7516/2020/04/028}{\emph{\jcap}
  {\bfseries 2020} (2020) 028}
  [\href{https://arxiv.org/abs/1911.04359}{{\ttfamily 1911.04359}}].

\bibitem{2018JCAP...06..016G}
H.~A. {Gressel} and M.~{Bruni}, \emph{{f$_{NL}$-g$_{NL}$ mixing in the matter
  density field at higher orders}},
  \href{https://doi.org/10.1088/1475-7516/2018/06/016}{\emph{\jcap} {\bfseries
  2018} (2018) 016} [\href{https://arxiv.org/abs/1712.08687}{{\ttfamily
  1712.08687}}].

\bibitem{2014ApJ...785....2B}
M.~{Bruni}, J.~C. {Hidalgo}, N.~{Meures} and D.~{Wands}, \emph{{Non-Gaussian
  Initial Conditions in {\ensuremath{\Lambda}}CDM: Newtonian, Relativistic, and
  Primordial Contributions}},
  \href{https://doi.org/10.1088/0004-637X/785/1/2}{\emph{\apj} {\bfseries 785}
  (2014) 2} [\href{https://arxiv.org/abs/1307.1478}{{\ttfamily 1307.1478}}].

\bibitem{Meures}
N.~Meures, \emph{{General Relativistic Effects on Cosmological Observations}},
  Ph.D. thesis, University of Portsmouth, 2012.

\bibitem{2016JCAP...02..021C}
P.~{Carrilho} and K.~A. {Malik}, \emph{{Vector and tensor contributions to the
  curvature perturbation at second order}},
  \href{https://doi.org/10.1088/1475-7516/2016/02/021}{\emph{\jcap} {\bfseries
  2016} (2016) 021} [\href{https://arxiv.org/abs/1507.06922}{{\ttfamily
  1507.06922}}].

\bibitem{1990PhRvD..42.3936S}
D.~S. {Salopek} and J.~R. {Bond}, \emph{{Nonlinear evolution of long-wavelength
  metric fluctuations in inflationary models}},
  \href{https://doi.org/10.1103/PhysRevD.42.3936}{\emph{\prd} {\bfseries 42}
  (1990) 3936}.

\bibitem{2005JCAP...05..004L}
D.~H. {Lyth}, K.~A. {Malik} and M.~{Sasaki}, \emph{{A general proof of the
  conservation of the curvature perturbation}},
  \href{https://doi.org/10.1088/1475-7516/2005/05/004}{\emph{\jcap} {\bfseries
  2005} (2005) 004} [\href{https://arxiv.org/abs/astro-ph/0411220}{{\ttfamily
  astro-ph/0411220}}].

\bibitem{2009pdp..book.....L}
D.~H. {Lyth} and A.~R. {Liddle}, \emph{{The Primordial Density Perturbation}}.
  ``Cambridge University Press, Cambridge U.K.", 2009.

\bibitem{1975PThPh..54..730T}
K.~{Tomita}, \emph{{Evolution of Irregularities in a Chaotic Early Universe}},
  \href{https://doi.org/10.1143/PTP.54.730}{\emph{Progress of Theoretical
  Physics} {\bfseries 54} (1975) 730}.

\bibitem{2013MNRAS.430L..54R}
C.~{Rampf} and G.~{Rigopoulos}, \emph{{Zel'dovich approximation and general
  relativity.}}, \href{https://doi.org/10.1093/mnrasl/sls049}{\emph{\mnras}
  {\bfseries 430} (2013) L54}
  [\href{https://arxiv.org/abs/1210.5446}{{\ttfamily 1210.5446}}].

\bibitem{2013PhRvD..87l3525R}
C.~{Rampf} and G.~{Rigopoulos}, \emph{{Initial conditions for cold dark matter
  particles and general relativity}},
  \href{https://doi.org/10.1103/PhysRevD.87.123525}{\emph{\prd} {\bfseries 87}
  (2013) 123525} [\href{https://arxiv.org/abs/1305.0010}{{\ttfamily
  1305.0010}}].

\bibitem{1995PhRvD..52.2007D}
N.~{Deruelle} and D.~{Langlois}, \emph{{Long wavelength iteration of Einstein's
  equations near a spacetime singularity}},
  \href{https://doi.org/10.1103/PhysRevD.52.2007}{\emph{\prd} {\bfseries 52}
  (1995) 2007} [\href{https://arxiv.org/abs/gr-qc/9411040}{{\ttfamily
  gr-qc/9411040}}].

\bibitem{1984ucp..book.....W}
R.~M. {Wald}, \emph{{General relativity}}. ``The University of Chicago Press",
  1984.

\bibitem{2010CQGra..27l4002W}
D.~{Wands}, \emph{{Local non-Gaussianity from inflation}},
  \href{https://doi.org/10.1088/0264-9381/27/12/124002}{\emph{Classical and
  Quantum Gravity} {\bfseries 27} (2010) 124002}
  [\href{https://arxiv.org/abs/1004.0818}{{\ttfamily 1004.0818}}].

\bibitem{2016PhRvD..93d3539C}
A.~J. {Christopherson}, J.~C. {Hidalgo}, C.~{Rampf} and K.~A. {Malik},
  \emph{{Second-order cosmological perturbation theory and initial conditions
  for N -body simulations}},
  \href{https://doi.org/10.1103/PhysRevD.93.043539}{\emph{\prd} {\bfseries 93}
  (2016) 043539} [\href{https://arxiv.org/abs/1511.02220}{{\ttfamily
  1511.02220}}].

\bibitem{2009PhRvD..80d3531C}
J.~{Carlson}, M.~{White} and N.~{Padmanabhan}, \emph{{Critical look at
  cosmological perturbation theory techniques}},
  \href{https://doi.org/10.1103/PhysRevD.80.043531}{\emph{\prd} {\bfseries 80}
  (2009) 043531} [\href{https://arxiv.org/abs/0905.0479}{{\ttfamily
  0905.0479}}].

\bibitem{1996ApJ...473..620S}
R.~{Scoccimarro} and J.~A. {Frieman}, \emph{{Loop Corrections in Nonlinear
  Cosmological Perturbation Theory. II. Two-Point Statistics and
  Self-Similarity}}, \href{https://doi.org/10.1086/178177}{\emph{\apj}
  {\bfseries 473} (1996) 620}
  [\href{https://arxiv.org/abs/astro-ph/9602070}{{\ttfamily
  astro-ph/9602070}}].

\bibitem{2011arXiv1104.2932L}
J.~{Lesgourgues}, \emph{{The Cosmic Linear Anisotropy Solving System (CLASS) I:
  Overview}}, {\emph{arXiv e-prints} (2011) arXiv:1104.2932}
  [\href{https://arxiv.org/abs/1104.2932}{{\ttfamily 1104.2932}}].

\bibitem{2011JCAP...07..034B}
D.~{Blas}, J.~{Lesgourgues} and T.~{Tram}, \emph{{The Cosmic Linear Anisotropy
  Solving System (CLASS). Part II: Approximation schemes}},
  \href{https://doi.org/10.1088/1475-7516/2011/07/034}{\emph{\jcap} {\bfseries
  2011} (2011) 034} [\href{https://arxiv.org/abs/1104.2933}{{\ttfamily
  1104.2933}}].

\bibitem{2019arXiv191009273E}
{Euclid Collaboration}, A.~{Blanchard}, S.~{Camera}, C.~{Carbone}, V.~F.
  {Cardone}, S.~{Casas} et~al., \emph{{Euclid preparation: VII. Forecast
  validation for Euclid cosmological probes}}, {\emph{arXiv e-prints} (2019)
  arXiv:1910.09273} [\href{https://arxiv.org/abs/1910.09273}{{\ttfamily
  1910.09273}}].

\bibitem{2020A&A...642A.191E}
{Euclid Collaboration}, A.~{Blanchard}, S.~{Camera}, C.~{Carbone}, V.~F.
  {Cardone}, S.~{Casas} et~al., \emph{{Euclid preparation. VII. Forecast
  validation for Euclid cosmological probes}},
  \href{https://doi.org/10.1051/0004-6361/202038071}{\emph{\aap} {\bfseries
  642} (2020) A191} [\href{https://arxiv.org/abs/1910.09273}{{\ttfamily
  1910.09273}}].

\bibitem{2008arXiv0810.0003R}
A.~{Rassat}, A.~{Amara}, L.~{Amendola}, F.~J. {Castander}, T.~{Kitching},
  M.~{Kunz} et~al., \emph{{Deconstructing Baryon Acoustic Oscillations: A
  Comparison of Methods}}, {\emph{arXiv e-prints} (2008) arXiv:0810.0003}
  [\href{https://arxiv.org/abs/0810.0003}{{\ttfamily 0810.0003}}].

\bibitem{1997PhRvL..79.3806T}
M.~{Tegmark}, \emph{{Measuring Cosmological Parameters with Galaxy Surveys}},
  \href{https://doi.org/10.1103/PhysRevLett.79.3806}{\emph{\prl} {\bfseries 79}
  (1997) 3806} [\href{https://arxiv.org/abs/astro-ph/9706198}{{\ttfamily
  astro-ph/9706198}}].

\bibitem{2007ApJ...665...14S}
H.-J. {Seo} and D.~J. {Eisenstein}, \emph{{Improved Forecasts for the Baryon
  Acoustic Oscillations and Cosmological Distance Scale}},
  \href{https://doi.org/10.1086/519549}{\emph{\apj} {\bfseries 665} (2007) 14}
  [\href{https://arxiv.org/abs/astro-ph/0701079}{{\ttfamily
  astro-ph/0701079}}].

\bibitem{2011PhRvD..83j3527T}
A.~{Taruya}, S.~{Saito} and T.~{Nishimichi}, \emph{{Forecasting the
  cosmological constraints with anisotropic baryon acoustic oscillations from
  multipole expansion}},
  \href{https://doi.org/10.1103/PhysRevD.83.103527}{\emph{\prd} {\bfseries 83}
  (2011) 103527} [\href{https://arxiv.org/abs/1101.4723}{{\ttfamily
  1101.4723}}].

\bibitem{2019OJAp....2E..13M}
K.~{Markovic}, B.~{Bose} and A.~{Pourtsidou}, \emph{{Assessing non-linear
  models for galaxy clustering I: unbiased growth forecasts from multipole
  expansion}}, \href{https://doi.org/10.21105/astro.1904.11448}{\emph{The Open
  Journal of Astrophysics} {\bfseries 2} (2019) 13}
  [\href{https://arxiv.org/abs/1904.11448}{{\ttfamily 1904.11448}}].

\bibitem{2021arXiv210710815M}
R.~{Martinez-Carrillo}, J.~C. {Hidalgo}, K.~A. {Malik} and A.~{Pourtsidou},
  \emph{{Contributions from primordial non-Gaussianity and general relativity
  to the galaxy power spectrum}},
  \href{https://doi.org/10.1088/1475-7516/2021/12/025}{\emph{\jcap} {\bfseries
  2021} (2021) 025} [\href{https://arxiv.org/abs/2107.10815}{{\ttfamily
  2107.10815}}].

\bibitem{2006PhRvD..74j3512M}
P.~{McDonald}, \emph{{Clustering of dark matter tracers: Renormalizing the bias
  parameters}}, \href{https://doi.org/10.1103/PhysRevD.74.103512}{\emph{\prd}
  {\bfseries 74} (2006) 103512}
  [\href{https://arxiv.org/abs/astro-ph/0609413}{{\ttfamily
  astro-ph/0609413}}].

\bibitem{2008PhRvD..78l3519M}
P.~{McDonald}, \emph{{Primordial non-Gaussianity: Large-scale structure
  signature in the perturbative bias model}},
  \href{https://doi.org/10.1103/PhysRevD.78.123519}{\emph{\prd} {\bfseries 78}
  (2008) 123519} [\href{https://arxiv.org/abs/0806.1061}{{\ttfamily
  0806.1061}}].

\bibitem{1999ApJ...520...24D}
A.~{Dekel} and O.~{Lahav}, \emph{{Stochastic Nonlinear Galaxy Biasing}},
  \href{https://doi.org/10.1086/307428}{\emph{\apj} {\bfseries 520} (1999) 24}
  [\href{https://arxiv.org/abs/astro-ph/9806193}{{\ttfamily
  astro-ph/9806193}}].

\bibitem{1998MNRAS.301..797H}
A.~F. {Heavens}, S.~{Matarrese} and L.~{Verde}, \emph{{The non-linear
  redshift-space power spectrum of galaxies}},
  \href{https://doi.org/10.1046/j.1365-8711.1998.02052.x}{\emph{\mnras}
  {\bfseries 301} (1998) 797}
  [\href{https://arxiv.org/abs/astro-ph/9808016}{{\ttfamily
  astro-ph/9808016}}].

\bibitem{1993ApJ...413..447F}
J.~N. {Fry} and E.~{Gaztanaga}, \emph{{Biasing and Hierarchical Statistics in
  Large-Scale Structure}}, \href{https://doi.org/10.1086/173015}{\emph{\apj}
  {\bfseries 413} (1993) 447}
  [\href{https://arxiv.org/abs/astro-ph/9302009}{{\ttfamily
  astro-ph/9302009}}].

\bibitem{2021arXiv210613725M}
E.-M. {Mueller}, M.~{Rezaie}, W.~J. {Percival}, A.~J. {Ross}, R.~{Ruggeri},
  H.-J. {Seo} et~al., \emph{{The clustering of galaxies in the completed
  SDSS-IV extended Baryon Oscillation Spectroscopic Survey: Primordial
  non-Gaussianity in Fourier Space}}, {\emph{arXiv e-prints} (2021)
  arXiv:2106.13725} [\href{https://arxiv.org/abs/2106.13725}{{\ttfamily
  2106.13725}}].

\bibitem{2021arXiv210813424Y}
M.~{Yousry Elkhashab}, C.~{Porciani} and D.~{Bertacca}, \emph{{The large-scale
  monopole of the power spectrum in a Euclid-like survey: wide-angle effects,
  lensing, and the `finger of the observer'}}, {\emph{arXiv e-prints} (2021)
  arXiv:2108.13424} [\href{https://arxiv.org/abs/2108.13424}{{\ttfamily
  2108.13424}}].

\bibitem{2015MNRAS.447.1789Y}
J.~{Yoo} and U.~{Seljak}, \emph{{Wide-angle effects in future galaxy surveys}},
  \href{https://doi.org/10.1093/mnras/stu2491}{\emph{\mnras} {\bfseries 447}
  (2015) 1789} [\href{https://arxiv.org/abs/1308.1093}{{\ttfamily 1308.1093}}].

\bibitem{2012JCAP...10..025B}
D.~{Bertacca}, R.~{Maartens}, A.~{Raccanelli} and C.~{Clarkson}, \emph{{Beyond
  the plane-parallel and Newtonian approach: wide-angle redshift distortions
  and convergence in general relativity}},
  \href{https://doi.org/10.1088/1475-7516/2012/10/025}{\emph{\jcap} {\bfseries
  2012} (2012) 025} [\href{https://arxiv.org/abs/1205.5221}{{\ttfamily
  1205.5221}}].

\bibitem{2019MNRAS.483.2078Y}
V.~{Yankelevich} and C.~{Porciani}, \emph{{Cosmological information in the
  redshift-space bispectrum}},
  \href{https://doi.org/10.1093/mnras/sty3143}{\emph{\mnras} {\bfseries 483}
  (2019) 2078} [\href{https://arxiv.org/abs/1807.07076}{{\ttfamily
  1807.07076}}].

\bibitem{2021arXiv210615604M}
M.~{Martinelli}, R.~{Dalal}, F.~{Majidi}, Y.~{Akrami}, S.~{Camera} and
  E.~{Sellentin}, \emph{{Ultra-large-scale approximations and galaxy
  clustering: debiasing constraints on cosmological parameters}}, {\emph{arXiv
  e-prints} (2021) arXiv:2106.15604}
  [\href{https://arxiv.org/abs/2106.15604}{{\ttfamily 2106.15604}}].

\bibitem{2012MNRAS.420.2102S}
L.~{Samushia}, W.~J. {Percival} and A.~{Raccanelli}, \emph{{Interpreting
  large-scale redshift-space distortion measurements}},
  \href{https://doi.org/10.1111/j.1365-2966.2011.20169.x}{\emph{\mnras}
  {\bfseries 420} (2012) 2102}
  [\href{https://arxiv.org/abs/1102.1014}{{\ttfamily 1102.1014}}].

\bibitem{2018JCAP...03..036J}
S.~{Jolicoeur}, O.~{Umeh}, R.~{Maartens} and C.~{Clarkson}, \emph{{Imprints of
  local lightcone projection effects on the galaxy bispectrum. Part III.
  Relativistic corrections from nonlinear dynamical evolution on
  large-scales}},
  \href{https://doi.org/10.1088/1475-7516/2018/03/036}{\emph{\jcap} {\bfseries
  2018} (2018) 036} [\href{https://arxiv.org/abs/1711.01812}{{\ttfamily
  1711.01812}}].

\bibitem{2016JCAP...09..015M}
J.~E. {McEwen}, X.~{Fang}, C.~M. {Hirata} and J.~A. {Blazek}, \emph{{FAST-PT: a
  novel algorithm to calculate convolution integrals in cosmological
  perturbation theory}},
  \href{https://doi.org/10.1088/1475-7516/2016/09/015}{\emph{\jcap} {\bfseries
  2016} (2016) 015} [\href{https://arxiv.org/abs/1603.04826}{{\ttfamily
  1603.04826}}].

\end{thebibliography}\endgroup

%
\bibliographystyle{JHEP.bst}



\end{singlespace}

\listoffigures 

\end{document}